\documentclass[useAMS,usenatbib]{mnras}

\usepackage{acronym}

\usepackage{amsmath,array,amsfonts}
\usepackage{amssymb}
\usepackage{graphicx}
\usepackage{grffile}
\usepackage{natbib}
\usepackage{pstricks}
\usepackage{verbatim}
\usepackage{multirow}
\usepackage{caption}
\usepackage{subcaption}

\newcommand{\ltsima} {$\; \buildrel < \over \sim \;$}
\newcommand{\gtsima} {$\; \buildrel > \over \sim \;$}
\newcommand{\lta} {\lower.5ex\hbox{\ltsima}}
\newcommand{\gta} {\lower.5ex\hbox{\gtsima}}

\usepackage{float}
\usepackage{amsmath}
\usepackage{commath}
\usepackage{mleftright}
\usepackage{bm}
\usepackage{natbib}
\usepackage{booktabs}
\usepackage{xcolor}
\definecolor{columbiablue}{rgb}{0.61, 0.87, 1.0}

\usepackage{cleveref}

\usepackage{scalerel}
\usepackage{tikz}
\usetikzlibrary{svg.path}

\definecolor{orcidlogocol}{HTML}{A6CE39}
\tikzset{
  orcidlogo/.pic={
    \fill[orcidlogocol] svg{M256,128c0,70.7-57.3,128-128,128C57.3,256,0,198.7,0,128C0,57.3,57.3,0,128,0C198.7,0,256,57.3,256,128z};
    \fill[white] svg{M86.3,186.2H70.9V79.1h15.4v48.4V186.2z}
                 svg{M108.9,79.1h41.6c39.6,0,57,28.3,57,53.6c0,27.5-21.5,53.6-56.8,53.6h-41.8V79.1z M124.3,172.4h24.5c34.9,0,42.9-26.5,42.9-39.7c0-21.5-13.7-39.7-43.7-39.7h-23.7V172.4z}
                 svg{M88.7,56.8c0,5.5-4.5,10.1-10.1,10.1c-5.6,0-10.1-4.6-10.1-10.1c0-5.6,4.5-10.1,10.1-10.1C84.2,46.7,88.7,51.3,88.7,56.8z};
  }
}

\newcommand\orcidicon[1]{\href{https://orcid.org/#1}{\resizebox{.35cm}{.35cm}{
\begin{tikzpicture}[yscale=-1,transform shape]
\pic{orcidlogo};
\end{tikzpicture}
}}}

\usepackage{hyperref} 

\makeatletter
\newcommand\notsotiny{\@setfontsize\notsotiny\@vipt\@viipt}
\makeatother
\newcommand{\mymatrix}[1]{{\bm{\mathsf{#1}}}}
\newcommand{\myset}[1]{{#1}}
\newcommand{\polarization}{}
\newcommand{\myvector}[1]{{\boldsymbol{#1}}}

\usepackage{soul} 

\title[QUIJOTE-MFI Diffuse polarized foregrounds]{QUIJOTE scientific results -- VIII. Diffuse polarized foregrounds from component separation with QUIJOTE-MFI}

\author[E.~de~la~Hoz]{E.~de~la~Hoz,$^{1,2}$\thanks{e-mail:delahoz@ifca.unican.es}
R.~B.~Barreiro,$^{1}$
P.~Vielva,$^{1}$
E.~Mart\'inez-Gonz\'alez,$^{1}$
\newauthor
J.~A.~Rubi{\~n}o-Mart\'in,$^{3,4}$
B.~Casaponsa,$^{1}$
F.~Guidi,$^{3,4,5}$
M.~Ashdown,$^{6,7}$
\newauthor
R.~T. G\'enova-Santos,$^{3,4}$
E.~Artal,$^{8}$
F.~J. Casas,$^{1}$
R.~Fern\'andez-Cobos,$^{9}$
\newauthor
M.~Fern\'andez-Torreiro,$^{3,4}$
D.~Herranz,$^{1}$
R.~J. Hoyland,$^{3,4}$
A.~N. Lasenby,$^{6,7}$
\newauthor
M.~L\'opez-Caniego,$^{10,11}$
C.~H. L\'opez-Caraballo,$^{3,4}$
M.~W. Peel,$^{3,4}$ 
L.~Piccirillo,$^{12}$
\newauthor
F.~Poidevin,$^{3,4}$
R.~Rebolo,$^{3,4,13}$
B.~Ruiz-Granados,$^{3,4,14}$
D.~Tramonte,$^{15,16,3,4}$
\newauthor
F.~Vansyngel,$^{3,4}$
R.~A. Watson.$^{12}$ 
\\
\\
\textit{\large{Affiliations are listed at the end of the paper}}
}

\date{Accepted 2022 October 2014. Received 2022 October 14; in original form 2022 July 27}

\begin{document}

\maketitle

\begin{abstract}
    We derive linearly polarized astrophysical component maps in the Northern Sky from the QUIJOTE-MFI data at 11 and 13\,GHz in combination with the WMAP K and Ka bands (23 and 33\,GHz) and all \textit{Planck} polarized channels (30-353\,GHz), using the parametric component separation method \textsc{B-SeCRET}. The addition of QUIJOTE-MFI data significantly improves the parameter estimation of the low-frequency foregrounds, especially the estimation of the synchrotron spectral index, $\beta_s$. We present the first detailed $\beta_s$ map of the Northern Celestial Hemisphere at a smoothing scale of 2$^{\circ}$.  We find statistically significant spatial variability across the sky. We obtain an average value of $-3.08$ and a dispersion of $0.13$, considering only pixels with reliable goodness-of-fit. The power law model of the synchrotron emission provides a good fit to the data outside the Galactic plane but fails to track the complexity within this region. Moreover, when we assume a synchrotron model  with uniform curvature, $c_s$, we find a value of  $c_s = -0.0797 \pm 0.0012$. However, there is insufficient statistical significance to determine which model is favoured, either the power law or the power law with uniform curvature. Furthermore, we estimate the thermal dust spectral parameters in polarization. Our CMB, synchrotron, and thermal dust maps are highly correlated with the corresponding products of the PR4 \textit{Planck} release, although some large-scale differences are observed in the synchrotron emission. Finally, we find that the $\beta_s$ estimation in the high signal-to-noise synchrotron emission areas is prior-independent while, outside these regions, the prior governs the $\beta_s$ estimation.
    
\end{abstract}


\section{Introduction}
\label{sec:introduction}

Currently, most of the efforts of the CMB community are devoted to the search for primordial $B$-modes. These predicted $B$-modes at large scales can only be produced by tensor modes, and their detection would constitute compelling evidence of an inflationary phase. The intensity of this primordial signal is determined by the tensor-to-scalar ratio $r$, the relative amplitude between the tensor and scalar modes at a given pivot scale. The current best upper bound on the tensor-to-scalar ratio is: $r<0.032$ at 95\% CL, set by the combination of \textit{Planck},  BICEP2/KeckArray and baryon-acoustic-oscillation data  \citep{tristram_r_constraint}.

The weakness of the primordial $B$-modes makes its detection a tremendous experimental challenge, requiring high-sensitivity experiments as well as an exquisite control of systematics. Indeed, a large effort is currently on-going with the aim to detect, or at least to constrain, $r$ with a sensitivity $\sigma_r (r=0) \leq 10^{-3}$. This includes many planned ground-based experiments, e.g., GroundBIRD \citep{groundBIRD}, LSPE-Strip \citep{LSPE}, CMB-S4 \citep{abazajian2016cmb}, Simons Observatory \citep{ade2019simons} and BICEP array \citep{hui2018bicep}, as well as satellite missions, e.g., LiteBIRD \citep{LiteBIRDPTEP} and PICO \citep{hanany2019pico}. 

The detectability of the primordial $B$-modes could be improved by removing the secondary $B$-mode component induced by weak gravitational lensing. Several delensing procedures have been proposed in the literature \citep{delensing_planck_method,bayesian_delensing} and have been applied to data from current CMB experiments \citep{delensing_planck_method,delensing_planck,delensing_bicep}, and in forecasts of future CMB experiments \citep{patricia_delensing,delensing_so}.

It is necessary to disentangle the CMB polarization signal from those coming from other microwave emissions, such as Galactic synchrotron, thermal dust and extragalactic point sources. Thus, the problem of component separation is a crucial step in order to detect the primordial $B$-mode of CMB polarization. This process benefits from the characterization of foreground emissions using complementary frequency ranges that provide unique information about the contaminants. 

The main diffuse polarized contaminants are the synchrotron emission (at low-frequencies) and the thermal dust emission (at high-frequencies). The best characterization of these diffuse foregrounds  has been done by \textit{Planck} \citep{Planck2018_IV} using a data set covering frequencies from 30 to 353\,GHz. This frequency range limited strongly the estimation of the synchrotron spectral parameters. In \citet{Planck2018_IV} it is shown that, with \textit{Planck} data only, one cannot test the spatial variability of the synchrotron spectral index due to limited sensitivity and frequency coverage. The data only allows a measurement of a global spectral index of $\beta_s = -3.1 \pm 0.1$. The synchrotron spectral index has also been estimated using other datasets, e.g., \citet{Fuskeland2014,Nicoletta2018,Fuskeland2019}. 

The Q-U-I JOint Tenerife Experiment (QUIJOTE) \citep{QUIJOTE} is a polarimetric ground-based CMB experiment whose main scientific goal is the characterization of the polarization of the cosmic microwave background (CMB) and other Galactic and extragalactic physical processes in the frequency range 10–40 GHz and at large angular scales ($\gtrsim 1^{\circ}$). The experiment is located at the Teide Observatory (at $\sim 2400$\,m above sea level) in Tenerife. It is composed of two telescopes equipped with three instruments: the Multi-Frequency Instrument (MFI), the Thirty-GHz Instrument (TGI), and the Forty-GHz Instrument (FGI), operating at 10–20~GHz, 26–36 GHz and 39–49 GHz respectively.

The MFI instrument has been operating from November 2012 to October 2018. It conducted two different surveys: i) a shallow Galactic survey (called ``wide survey'') covering all the visible sky from Tenerife at elevations larger than 30$^{\circ}$, and ii) a deep cosmological survey covering approximately 3000 deg$^2$ in three separated sky patches in the northern sky. In this work we use the QUIJOTE-MFI wide survey maps. This survey provides an average sensitivity in polarization of $\sim$ 35–40 $\mu$K per 1-degree beam in four bands centred around 11, 13, 17
and 19\,GHz \citep{mfiwidesurvey}. Those frequencies are crucial to achieving a better characterization of the low-frequency foregrounds. In intensity, this additional information helps breaking degeneracies between the synchrotron, free-free and anomalous microwave emissions while, in polarization, the QUIJOTE-MFI channels are key to characterize the synchrotron spectral dependence. 

In the present work we perform a component separation analysis to obtain more information about the polarized sky using the QUIJOTE-MFI data\footnote{This is one of the papers which are part of the MFI wide survey data release.} \citep{mfiwidesurvey} in combination with the publicly available \textit{Planck} \citep{NPIPE,Planck2018_I} and Nine-Year WMAP \citep{WMAP2013} data. To perform component separation analysis we use \textsc{B-SeCRET} (Bayesian-Separation of Components and Residual Estimation Tool), a parametric maximum-likelihood method described in \citet{de2020detection}. 

The paper is organized as follows: in Section~\ref{sec:sky_model} we provide details of the main components in the polarized microwave sky and the corresponding parametric models used to characterize them. Section~\ref{sec:cs_method} describes briefly the \textsc{B-SeCRET} method. The data used in the analysis are presented in Section~\ref{sec:data}. Then, the main component separation results obtained are shown in Section~\ref{sec:results}. Finally, the main conclusions from the analysis are given in Section~\ref{sec:conclusions}. In Appendix~\ref{sec:appendix_independent_Q_U} we provide maps of the synchrotron spectral index obtained from independent fits in linear Stokes parameters $Q$ and $U$. Appendix~\ref{sec:appendix_FR} compares the variations on the synchrotron spectral index due to rotations of the polarized angle with  Faraday rotation.

\section{The Microwave Sky Model}
\label{sec:sky_model}

The polarized microwave sky is composed primarily of photons from the CMB, synchrotron and thermal dust. As stated before, the synchrotron emission dominates at low-frequencies while the thermal dust is the principal component at higher frequencies. The contribution from other components, discussed in Section~\ref{subsec:other_contributions}, is expected to be insignificant and not taken into account.  Apart from these astronomical signals, the measured sky signal maps have another contribution from the instrumental noise. The characteristics of this noise depend on the specifications of the experiment. Furthermore, contaminants such as the atmosphere and artificial signals from satellites also contribute to the microwave sky, see \citet{mfiwidesurvey} for more details. Thus, the measured polarized sky signal for a given $\nu$ channel can be expressed as the following sum:

\begin{equation}
    \begin{pmatrix}
        Q \\
        U
    \end{pmatrix}_{\nu}
	=
	\begin{pmatrix}
        Q_{\notsotiny{\rm cmb}} \\
        U_{\notsotiny{\rm cmb}}
    \end{pmatrix}_{\nu}
    +
    \begin{pmatrix}
        Q_s \\
        U_s
    \end{pmatrix}_{\nu}
    +
    \begin{pmatrix}
        Q_d \\
        U_d
    \end{pmatrix}_{\nu}
    +
    \begin{pmatrix}
        Q_n \\
        U_n
    \end{pmatrix}_{\nu}
    \, ,
\label{eq:sky_model}
\end{equation}
where $X_{\rm cmb}$, $X_{s}$, and $X_{d}$ are the CMB, synchrotron and thermal dust signals respectively, and $X_n$ is the instrumental noise ($X \in \{Q,U\}$). In the subsequent subsections we describe the main physical components that encompass the sky signal as well as some effects that alter this signal. Moreover, we present the parametric models that we use in the component separation analysis for each polarized astronomical component.

\subsection{Synchrotron}
\label{subsec:synchrotron}
The synchrotron emission arises from relativistic particles (cosmic rays) passing through the Galactic magnetic field. Its emissivity depends both on the magnetic field strength and  energy distribution of the relativistic particles (generally electrons). These quantities are not uniform in the Galactic disc. For instance, the free electrons are more predominant in compact regions as supernovae remnants. On the other hand, the magnetic field is amplified in some compact regions and can have different strength and direction across the sky. 

The synchrotron spectral energy distribution (SED) is generally described as a power law \citep{rybicki2008radiative}:
\begin{equation}
\begin{pmatrix}
        Q_s \\
        U_s
    \end{pmatrix}_{\nu}
    =
    \begin{pmatrix}
        A_s^Q \\
        A_s^U
    \end{pmatrix}
    \left(\frac{\nu}{\nu_s}\right)^{\beta_s} \, ,
\label{eq:synchrotron_model}
\end{equation}
where $A_s$ is the amplitude in brightness temperature at the pivot frequency $\nu_s = 30$\,GHz and $\beta_s$ is the spectral index which is assumed to be equal for both $Q$ and $U$ Stokes parameters. 

Previous works dedicated to the estimation of the spectral index, found values around $\beta_s\simeq-3.1$ \citep{Planck2018_IV}.  However, the spectral index is expected to vary spatially due to its dependence on the energy distribution of the cosmic rays $N(E)$. Studies such as \citet{Fuskeland2014,Vidal2015,Nicoletta2018,felice_synchrotron,weiland_beta_s} indicate that different polarized regions present different spectral indices. Here, we conduct a more detailed analysis of the $\beta_s$ spatial variations in the Northern Hemisphere by performing a pixel-by-pixel component separation analysis using the QUIJOTE MFI polarized maps.

The S-PASS survey \citep{SPASS}, has provided the most sensitive reconstruction of the $\beta_s$ variations of the South Celestial Hemisphere \citep{Nicoletta2018}. They found large variability over the sky, and a mean value of $-3.22\pm0.08$. Those results were further confirmed in the analysis of \citet{Fuskeland2019} that estimated the spectral index  taking into account the Faraday Rotation effect. They also studied the Galactic plane and found compatible results to those where only WMAP data were used, finding a flatter index in the Galactic plane than at high Galactic latitudes. 

We have also considered an extension of equation~\eqref{eq:synchrotron_model} where we include a possible curvature in the synchrotron's SED:
\begin{equation}
\begin{pmatrix}
        Q_s \\
        U_s
    \end{pmatrix}_{\nu}
    =
    \begin{pmatrix}
        A_s^Q \\
        A_s^U
    \end{pmatrix}_{\nu}
    \left(\frac{\nu}{\nu_s}\right)^{\beta_s + c_s\log{\left(\frac{\nu}{\nu_s}\right)}} \, ,
\label{eq:synchrotron_model_curvature}
\end{equation}
where $c_s$ is the parameter that represents the curvature. This extension is worth studying since a curved spectrum can account for steepening or flattening of the SED due to different effects, e.g., cosmic ray aging effect or multiple synchrotron components along the line of sight. This model could also account for the presence of polarized anomalous microwave emission (AME).

\subsection{Thermal Dust}
\label{subsec:thermal_dust}
The thermal dust radiation comes from dust grains present in the interstellar medium. Those grains absorb ultraviolet light and re-emit as a grey body. In general, these dust grains are not perfectly spherical and typically have their minor axis aligned with the direction of the local magnetic field. This effect yields polarized thermal dust emission. The SED of this radiation is often described as a modified black body with emissivity index $\beta_d$ and dust temperature $T_d$:
\begin{equation}
\begin{pmatrix}
        Q_d \\
        U_d
    \end{pmatrix}_{\nu}
    =
    \begin{pmatrix}
        A_d^Q \\
        A_d^U
    \end{pmatrix}
    \left(\frac{\nu}{\nu_d}\right)^{\beta_d+1}\frac{e^{\gamma\nu_d}-1}{e^{\gamma\nu}-1} \, ,
\label{eq:dust_model}
\end{equation}
where $A_d$ is the amplitude of the dust in brightness temperature evaluated at the pivot frequency $\nu_d= 143$\,GHz and $\gamma=\frac{h}{k_B T_d}$\footnote{$h$ and $k_B$ are Planck and Boltzmann constants respectively.}. The amplitude is well characterized by the higher frequency channels where the other components are clearly sub-dominant. The current temperature map of the dust grains ($T_d$) is obtained from temperature analysis and has values mostly between 14 K and 26 K. The polarized dust emissivity evaluated with \textit{Planck} data is $\beta_d = 1.55\pm0.05$ \citep{Planck2018_IV}. 

Several works support the idea that a single component dust model is too simplistic and more components might be required to fully characterize this emission (e.g., \citealt{McBride2022,Ritacco2022}). Nonetheless, since this paper is focused on the low frequency foregrounds, we keep the model used in \citet{Planck2018_IV} which seems to provide a good description at the \textit{Planck} polarized frequencies (30\,GHz $<\nu<$ 353\,GHz).

\subsection{CMB}
\label{subsec:cmb}

The CMB radiation has a thermal black body spectrum with a temperature of \textbf{$T_{o}=2.7255\pm0.0006$}\,K \citep{T_cmb}.
CMB photons are linearly polarized due to the Thomson scattering experienced with the hot electron gas at the last scattering surface.
Unlike in intensity, where the CMB can be the dominant contribution at intermediate frequencies (70-150\,GHz) and high Galactic latitudes, in polarization, the foreground contribution cannot be overlooked. Thus, in order to detect the primordial $B$-mode,  experiments with very high sensitivity, exquisite control of systematics and a careful removal of foregrounds are mandatory.

The CMB signal at each pixel is given by its amplitude $A_{\mathrm{cmb}}$,  which is the only free parameter for this component. Since the rest of the components are given in brightness temperature we convert the thermodynamic temperature of the CMB to the same units:
\begin{equation}
\begin{pmatrix}
        Q_{\notsotiny{\mathrm{cmb}}} \\
        U_{\notsotiny{\mathrm{cmb}}}
    \end{pmatrix}_{\nu}
    =
    \begin{pmatrix}
        A_{\notsotiny{\mathrm{cmb}}}^Q \\
        A_{\notsotiny{\mathrm{cmb}}}^U
    \end{pmatrix}
    \dfrac{x^2e^x}{(e^x-1)^2} \, ,
\label{eq:cmb_model}
\end{equation}
where $x=\frac{h\nu}{k_B T_o}$. 

\subsection{Faraday Rotation}
\label{subsec:faraday_rotation}

Another issue intrinsic to the polarization signal is the Faraday rotation effect, i.e., the rotation of the plane of polarization that occurs when light passes through the interstellar medium in the presence of a magnetic field. The magnitude of this effect scales with the square of the wavelength, hence its repercussions are more significant at low-frequencies. To properly account for this effect we require a broad knowledge of the Galactic magnetic field as well as the interstellar medium, in order to recognise the regions where the effect is more significant. Moreover, since the instrumental beam has a finite size, the measured signal is an average of the emission from various directions within the beam with slightly different rotation angles. This results in a ``beam depolarization'' of the signal. 

\citet{Faraday_Rotation} show that the possible Faraday Rotation effects at the QUIJOTE-MFI frequencies (10-20/,GHz) are very small in most of the sky, and particularly at high Galactic latitudes. Thus, in this work we have not considered any Faraday Rotation effect. Nevertheless, in Appendix~\ref{sec:appendix_FR} we study variations on the synchrotron spectral index due to rotations of the polarized angle and compare it to Faraday Rotation models such as the one proposed in \citet{Faraday_Rotation}.

\subsection{Other contributions}
\label{subsec:other_contributions}
It is well known that there are other foreground components whose emissions are important for intensity analyses. In particular, at low frequencies, one needs to consider two additional Galactic emission components: the bremsstrahlung radiation generated from electron–ion scattering in interstellar plasma (free-free), and AME, whose physical origin still is not fully clear. At high frequencies, in addition to thermal dust, we find an isotropic extragalactic emission called the cosmic infrared background (CIB), coming from different sources, e.g., dusty star-forming galaxies, quasars, intergalactic stars, inter-cluster dust in the Local group, etc. We also have other contributions such as CO line emission or Sunyaev-Zeldovich effect (SZ) from clusters of galaxies \citep{SZ_effect} that should be taken into account in intensity analyses \citep{Planck2016_X}. In addition, emission from extragalactic point sources, both at radio and infrared frequencies is an important contaminant at small scales. In polarization the problem is simplified since several of these emissions (free-free, CIB, SZ, etc.) are not expected to be polarized (at least significantly), therefore we do not consider them.

The polarization of the anomalous microwave emission is still under study because its nature is still uncertain \citep{ame_review}. Several models have been proposed such as spinning dust particles \citep{Haimoud2013}, magnetic dipole emission \citep{Draine1999} or more recently the proposal of spinning nano-diamonds \citep{Greaves2018}. The predicted polarization fraction of the AME emission for most of these models is below 5\%. From the data analysis point of view, no evidence of polarization has been found in compact region studies (the most stringent constraints on the polarization fraction, $\Pi$, have been provided by \citet{Genova2017}, $\Pi < 0.22\%$ at 41\,GHz). Due to this lack of evidence, we do not take into account the AME component in this work. 

On the other hand, point sources present some degree of polarization, which is in general small (a few percent). However, at the resolutions considered in this work, they are subdominant with respect to Galactic foregrounds. Thus, we do not include them in our analysis. We note however that in the data, a few polarized point sources are present that are not taken into account in the component separation analysis, see \citet{sourceswidesurvey}.

\section{Component Separation Methodology}
\label{sec:cs_method}
In this work, we apply the parametric component separation method \textsc{B-SeCRET} to extract the polarized astrophysical signals. Parametric methods are very powerful since they provide physical information of each sky component. However, they require a profound theoretical understanding of the nature of the foregrounds and accurate knowledge of the experiment's characteristics to avoid biases in the analysis.

Below, in Section~\ref{subsec:ml_method}, we outline the component separation technique applied in this work. Then, in Section~\ref{subsec:priors_assumptions}, we describe the prior information  that is used in the Bayesian analyses.

\subsection{Bayesian analyses}
\label{subsec:ml_method}

The \textsc{B-SeCRET} methodology is a parametric pixel-based maximum-likelihood method, which relies on an Affine-Invariant Markov Chain Monte Carlo  Ensemble sampler to draw samples from a posterior distribution \citep{foreman2013emcee}. This methodology has already been applied in previous studies, e.g., \citet{de2020detection,delahoz_angles}.

\textsc{B-SeCRET} applies Bayesian inference to determine the best-fit model parameters given some prior information. In Bayesian statistics, the probability of the model parameters $\myset{\theta}\polarization_{p}$ given the signal data $\myvector{d}\polarization_{p}$ at the pixel $p$ is proportional to the probability of the $\myvector{d}\polarization_{p}$ given $\myset{\theta}\polarization_{p}$ times the  probability of $\myset{\theta}\polarization_{p}$, i.e.,
\begin{equation}
    \mathcal{P}(\myset{\theta}\polarization_{p}|\myvector{d}\polarization_{p}) \propto \mathcal{P}(\myvector{d}\polarization_{p}|\myset{\theta}\polarization_{p}) \mathcal{P}(\myset{\theta}\polarization_{p}) \, .
    \label{eq:posterior}
\end{equation}
$\mathcal{P}(\myset{\theta}\polarization_{p})$ is commonly known as the prior information, whereas $\mathcal{P}(\myvector{d}\polarization_{p}|\myset{\theta}\polarization_{p})$ is usually referred to as the likelihood. Assuming Gaussian noise, the likelihood of the data can be expressed as 
\begin{equation}
    \mathcal{P}(\myvector{d}\polarization_{p}|\myset{\theta}\polarization_{p}) = \dfrac{\exp\mleft(-\dfrac{1}{2}\mleft(\myvector{d}\polarization_{p}-\myvector{S}_{p}\polarization\mright)^{T}\mymatrix{C}^{-1}\mleft(\myvector{d}\polarization_{p}-\myvector{S}_{p}\polarization\mright)\mright)}{\sqrt{(2\pi)^{N}\det (\mymatrix{C})}} \, ,
    \label{eq:likelihood}
\end{equation}
where $\mymatrix{C}$ is the noise covariance matrix, $N$ is the number of elements in the $\myvector{d}_{p}$ array, and $\myvector{S}_{p}$ is the parametric model considered, which has been described in detail in Section~\ref{sec:sky_model}.

To draw samples from the posterior probability we use the \textsc{Python} implementation \textsc{emcee} \citep{foreman2013emcee} of an affine-invariant ensemble sampler for MCMC \citep{goodman2010ensemble}. In each pixel, the best-fit parameters and their uncertainties are obtained as the median and the standard deviation of their respective marginalized posterior probability. 

\subsection{Priors}
\label{subsec:priors_assumptions}

In this work we benefit from prior information about astrophysical foregrounds to help with convergence and computational time reduction. For example, the synchrotron spectral index is known to be around $-3.1$, although experiments such as S-PASS found a more negative value. Here we use the estimated value obtained with \textit{Planck} polarization data by the SMICA method,  $\beta_s=-3.1\pm0.06$ \citep{Planck2018_IV} and use a broad Gaussian distribution $\mathcal{N}\mleft(-3.1,0.3\mright)$\footnote{$\mathcal{N}\mleft(x,\sigma\mright)$ represents a normal distribution with mean $x$ and variance $\sigma^2$.} as a prior on $\beta_s$. When we include a curvature in the synchrotron model we apply a Gaussian prior $\mathcal{N}\mleft(0,0.1\mright)$ on $c_s$. Moreover, we apply Gaussian priors $\mathcal{N}\mleft(1.55,0.1\mright)$ and $\mathcal{N}\mleft(21,3\mright)$ on both $\beta_d$ and $T_d$ respectively. Finally, flat priors are used in the characterization of the amplitude parameters. 

\section{Data}
\label{sec:data}
The aim of this work is to obtain a better characterization of the low-frequency foregrounds by including the newly released QUIJOTE-MFI maps in component separation analyses. In this Section we summarize the basic details of these maps as well as those from the other experiments used in the analysis, i.e., the K and Ka bands from WMAP and \textit{Planck}'s third and fourth public releases (PR3 and PR4, respectively). We also discuss some technical issues related to the instruments such as the estimated noise, RFI, and the colour corrections.

\subsection{Datasets}
\label{subsec:frequency_maps}

In this analysis we have used the data from the following experiments:
\begin{figure}
    \centering
    \includegraphics[width=\linewidth,trim={0 1cm 0  1cm},clip]{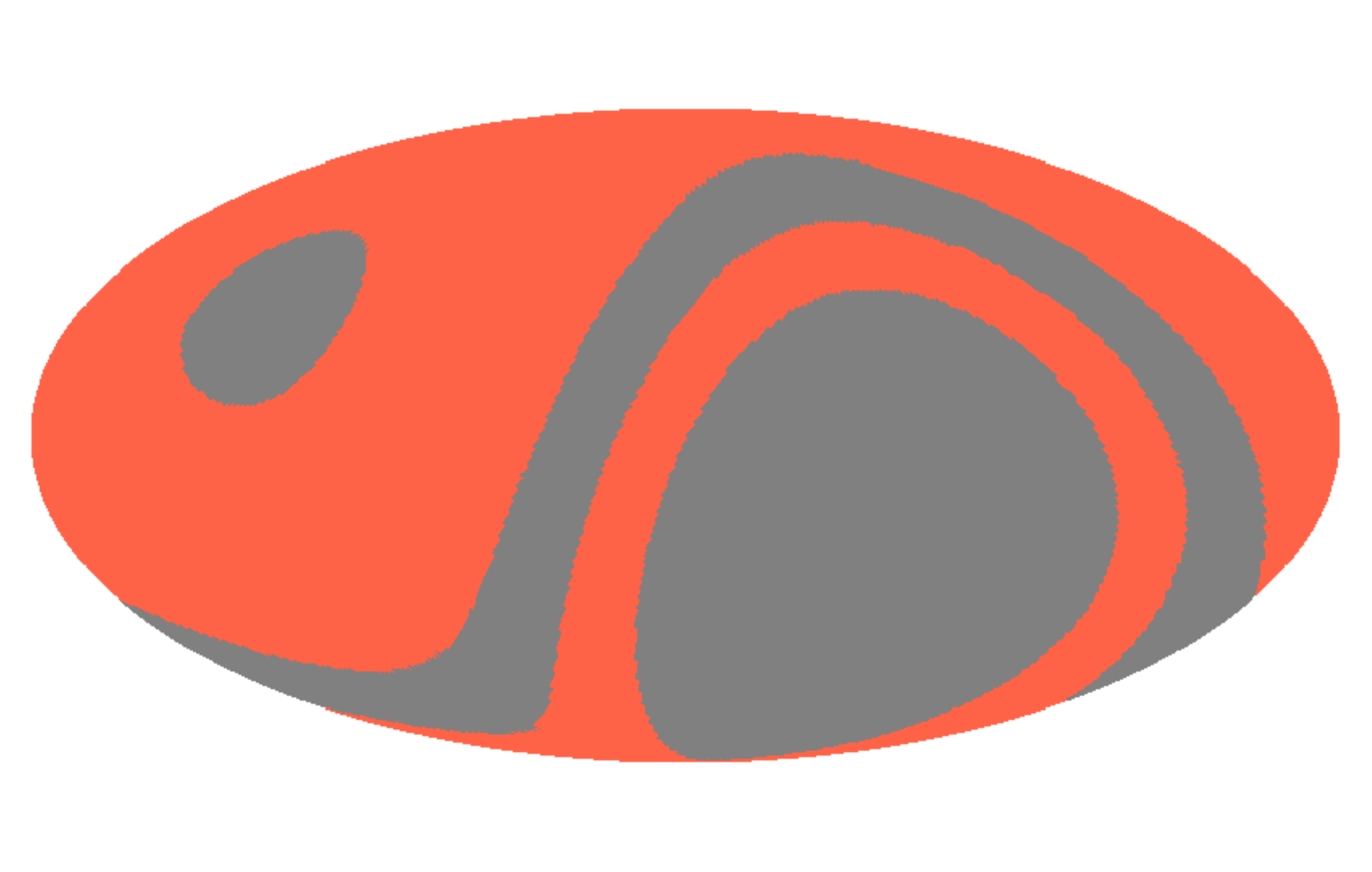}
    \caption{QUIJOTE observed sky after removing the geostationary satellite band and the region around the north celestial pole, which is affected by high atmospheric air-mass ($f_{\mathrm{sky}} = 51$\%, Galactic coordinates centred on (0,0)).}
    \label{fig:mask_satband}
\end{figure}

\begin{itemize}
    \item \textbf{QUIJOTE.} We have used the low frequency QUIJOTE MFI 11 and 13\,GHz channels (MFI) \citep{mfiwidesurvey}  due to their better signal-to-noise ratio. Although QUIJOTE has observed 70\% of the sky there are regions with poorer sensitivity due to the presence of artificial satellites and high atmospheric masses in some directions. Thus, in this analysis we have considered the mask shown in Fig.~\ref{fig:mask_satband}, as the observable sky. This mask (satband+NCP) is described in \citet{mfiwidesurvey}.
    \item \textbf{WMAP.} We have used the low frequency Nine-Year Wilkinson Microwave Anisotropy Probe (WMAP) K  (22.8\,GHz) and Ka (33.1\,GHz) bands \citep{WMAP2013}\footnote{The other bands were not included since they have a much lower synchrotron signal-to-noise ratio and do not contribute to the determination of the synchrotron characteristics.}.
    \item \textbf{\textit{Planck}.} We have used the full set of \textit{Planck} polarization maps i.e., the low frequency instrument (LFI) 30, 44 and 70\,GHz frequency maps and the high frequency instrument (HFI) 100, 143, 217 and 353\,GHz maps. We have obtained results from both PR3\footnote{We used the Planck maps corrected from bandpass leakage.} \citep{Planck2018_II,Planck2018_III} and PR4 \citep{NPIPE} data releases.
\end{itemize}
 
Before component separation analyses, the frequency maps are all convolved (taking appropriately into account the beam window function of each particular frequency map) with a common beam, a Gaussian beam of $\mathrm{FWHM} = 2^\circ$, and downgraded to the same resolution through spherical harmonics, given by the \textsc{HEALPix} parameter $N_{\rm side}=64$. The procedure followed is described below:
\begin{enumerate}
    \item We calculate the spherical harmonics coefficients $(t_{\ell m}, e_{\ell m}, b_{\ell m})$ using the \textsc{healpy} routine \textsc{map2alm}.
    \item To convolve all channels with the same beam we multiply the $(t_{\ell m}, e_{\ell m}, b_{\ell m})$ by $b_{\ell}(2^\circ)\, p_{\ell}(64) / (b_{i,\ell} \, p_{\ell}(N_{\rm{side}}))$, where $b_{\ell}(\alpha)$ is a gaussian beam window function whose FWHM is $\alpha$, $b_{i,\ell}$ is the $i$-th channel beam window function and, $p_{\ell}(N_{\rm{side}})$ is the pixel window function at the resolution $N_{\rm{side}}$.
    \item We obtain the downgraded maps at $N_{\rm{side}} = 64$ applying the \textsc{healpy} routine \textsc{alm2map} to the new $(t_{\ell m}, e_{\ell m}, b_{\ell m})$.
\end{enumerate}

Several combinations of the previous data sets have been tested. Each configuration's name is given by the ``sum'' of the sets of maps included in the analysis.  For example, the configuration composed of PR4 channels in combination with WMAP's K and Ka bands is referred as K/Ka+PR4, or MFI-QUIJOTE low frequency channels in combination with PR4 and WMAP channels is specified as MFI+K/Ka+PR4.

\subsection{Instrumental Effects}
\label{subsec:instrumental_effects}
Real data present different instrumental effects that need to be accounted for. For example, an important contribution to the observed signal is the noise produced by the detectors of each experiment. A proper characterization of the noise levels is key for component separation analyses. In this work, we have calculated the covariance matrix among the frequency channels per pixel, required by the parametric component separation method, using realistic noise simulations specific to each instrument. Each experiment's noise simulations are obtained as follows:
\begin{itemize}
    \item \textbf{QUIJOTE} We have used the correlated noise simulations described in \citet{mfiwidesurvey}. They account for the $1/f$ noise present in the maps, and the correlated noise component between 11 and 13\,GHz.
    \item \textbf{WMAP}  We have generated a set of white noise simulations using the RMS noise per pixel  provided by the WMAP collaboration \citep{wmap_noise_simulations}. The RMS noise $\sigma$ is calculated as $\sigma = \sigma_0/\sqrt{N_{\rm obs}}.$\footnote{$\sigma_0$ and $N_{\rm obs}$ are given in \url{https://lambda.gsfc.nasa.gov/product/wmap/dr5/skymap_info.html}.} 
    \item \textbf{\textit{Planck}.} For PR3 we have used the FFP10 simulations generated by the \textit{Planck} Collaboration \citep{Planck2018_I}. In the case of the PR4, we have employed the noise simulations described in \citet{NPIPE}.\footnote{Simulations available at NERSC under \texttt{/global/cfs/cdirs/cmb/data/planck2020}.} 
\end{itemize}

While the frequency channels of different experiments are uncorrelated, there might be correlations between channels of a given instrument. This is the case for the 11 and 13\,GHz low-frequency MFI channels. On the other hand, we have assumed no correlations between frequency channels for WMAP and \textit{Planck}. Thus, for a given pixel $p$, the \textit{Planck} and WMAP frequency covariance matrices are diagonal while QUIJOTE's has non-zero off-diagonal terms. For a given configuration, the covariance matrix is obtained as a block matrix, where each block corresponds to the frequency covariance matrix of each instrument included in that configuration. 

To obtain the experiments' frequency covariance matrices, first we pre-process the noise simulations in the same manner as the data maps. Then, for \textit{Planck} and WMAP, the diagonal terms are calculated as the variance of the noise simulations at the corresponding pixel for each frequency. Each pixel covariance matrix between QUIJOTE 11 and 13\,GHz is calculated as the sample covariance matrix using the values of the 11 and 13\,GHz noise simulations at that specific pixel.

One test to verify that our covariance matrices are well estimated is the following. We obtained a distribution of $\chi^2_{n,i}$ values as:
\begin{equation}
    \chi^2_{n,i}= \myvector{n}_{i}^T\mymatrix{C}^{-1}_{i}\myvector{n}_{i}\, ,
    \label{eq:chi2_noise_red} 
\end{equation}
where $\myvector{n}_{i}$ is a noise simulated map\footnote{The noise simulations used in this test are different from the noise simulations used to calculate the noise covariance matrices.} at the frequency $i$ and $\mymatrix{C}_{i}$ is the noise covariance matrix described above.
The $\chi^2_{n,i}$ distributions should have the expected form with $N_{\rm pix}$ degrees of freedom (d.o.f.). This is consistent with the values obtained for \textit{Planck} and WMAP. In the case of QUIJOTE, the distribution deviates slightly from the expected $N_{\mathrm{pix}}$ d.o.f. $\chi^2$-distribution since they are not end-to-end noise simulations and hence not as accurate~\citep[see][for details]{mfiwidesurvey}. However, as subsequent analyses will show, we find that when the astrophysical emission is included, the obtained $\chi^2$ is correct as expected, i.e., in the regions where the model properly explains the data (outside the Galactic plane). Thus, QUIJOTE's noise simulations are accurate enough to perform scientific analyses.

We explored the possibility of including correlations among neighbouring pixels within a 1 degree radius\footnote{The pixels contained within this radius are the ones with the largest correlations induced by the smoothing process.}. The smoothing process of the maps induces noise correlations among different pixels and, although this does not affect our pixel-by-pixel analyses, it can affect analyses where we assume a uniform parameter value within one region. Therefore, for each pixel, we calculated the covariance matrix among its neighbouring pixels from noise simulations. Then we generated a sparse covariance matrix where the only non-zero values in each row were the diagonal element and the correlation with the neighbouring pixels. In this case the distribution does not follow a $N_{\rm pix}$ degrees-of-freedom $\chi^2$ distribution as one would expect. The recovered values were smaller than expected, more notably for \textit{Planck} maps. This is a consequence of not having enough noise simulations, which prevents us from obtaining a good characterization of the noise correlations. Therefore, we use the covariance matrices that do not take into account possible noise correlations among neighbouring pixels in the following.

\begin{figure*}
    \centering
    \includegraphics[width=1\textwidth,trim={2.8cm 1cm 3cm 1cm},clip]{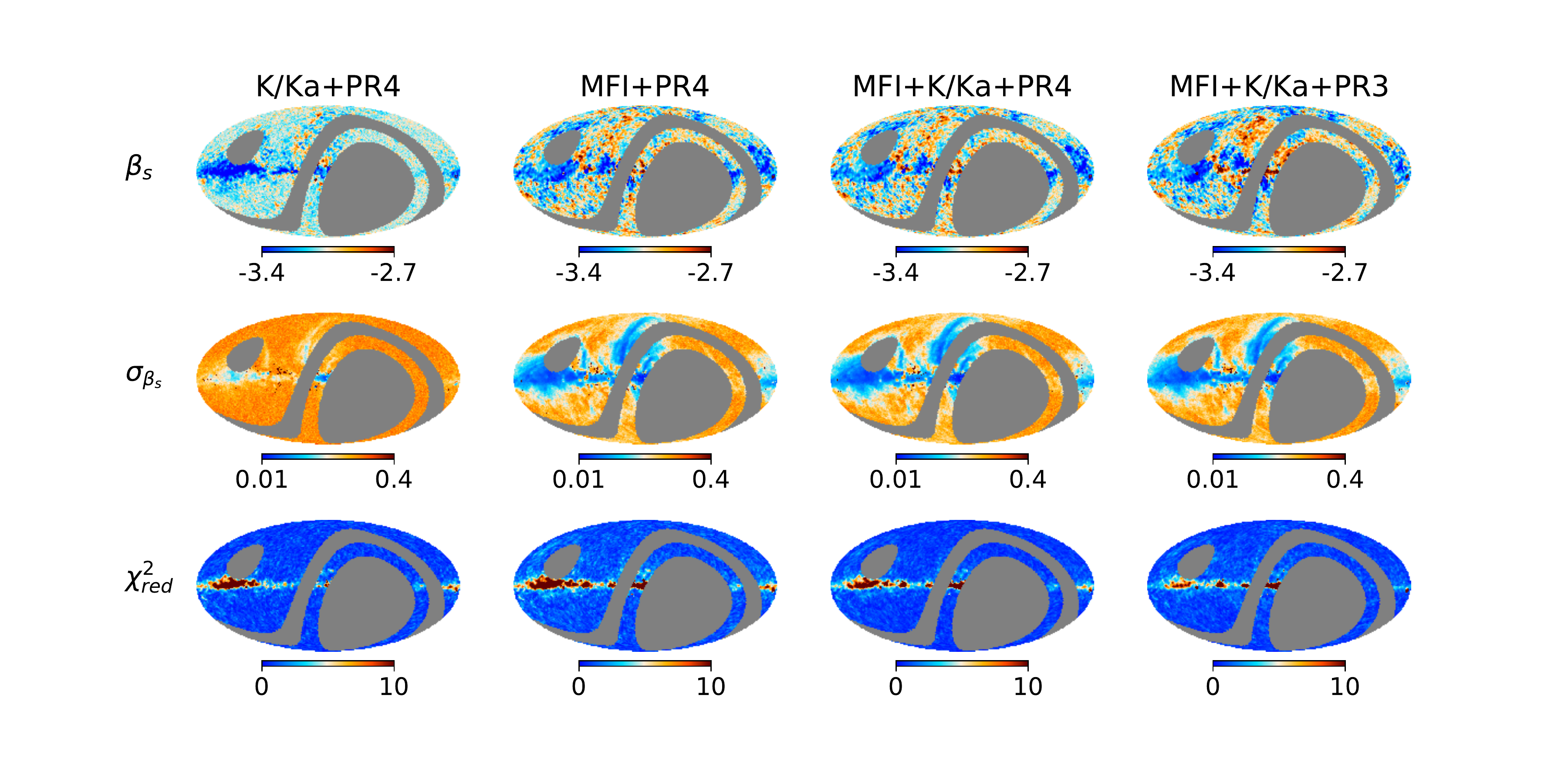}
    \caption{Synchrotron spectral index (top row) and uncertainty maps (middle row) obtained after component separation with four different datasets. The synchrotron emission is modelled with a power law. Bottom row: reduced $\chi^2$ map obtained for each dataset.}
    \label{fig:beta_s_datasets}
\end{figure*}

As explained in \citet{mfiwidesurvey}, in order to correct residual RFI signals emerging after co-adding all data in the map-making process of the QUIJOTE-MFI data, the polarization maps are corrected using a function of the declination (FDEC). This correction is equivalent to applying a filter to QUIJOTE data, which removes the zero mode in lines of constant declination. In Appendix~\ref{sec:appendix_beta_s_simus} we studied whether this correction affects the recovery of foregrounds spectral parameters such as $\beta_s$. We found that if only QUIJOTE maps are filtered with FDEC the recovered $\beta_s$ map is biased in regions such as the North Polar Spur. When all data maps are filtered in the same way this bias disappears. Thus, for this analysis we have filtered all signal maps with their corresponding FDEC function.

Another important instrumental effect arises from detectors having a finite bandwidth. This issue has to be taken into account when dealing with foreground components whose amplitude varies within that frequency band. This effect can be corrected by adding a multiplicative factor, called colour correction, to the signal that depends on the spectral behaviour. We have used the \textsc{fastcc} \textsc{Python} code (\citealt{fastcc}, \citealt{mfipipeline}) to obtain the colour corrections of each experiment considered here. Therefore, our model for the sky signal presented in Section~\ref{sec:sky_model} is corrected as follows:
\begin{equation}
        X_{\nu} = X_{\nu,\notsotiny{\mathrm{cmb}}} + \dfrac{X_{\nu,s}}{C_{s}(\alpha, \nu)} + \dfrac{X_{\nu,d}}{C_{d}(\beta_d, T_d, \nu)},
    \label{eq_sky_model_cc_corrected}
\end{equation}
where $X$ is either $Q$ or $U$,  $C_{s}(\alpha, \nu)$ is synchrotron colour correction whose spectral behaviour is modelled as a power law with  $\alpha=\beta_s + 2$. 
The spectral behaviour of dust colour correction $C_{d}(\beta_d, T_d, \nu)$ is assumed to be a modified black body and it is determined by its $\beta_d$ and $T_d$ parameters. The colour correction values are updated in each MCMC iteration.

\section{Results}
\label{sec:results}

In this Section we present the component separation products obtained using the recently released QUIJOTE low-MFI  data along with the already available \textit{Planck} and WMAP data. We have focused primarily on the synchrotron spectral parameters since those are the parameters where a greater improvement is found, see Section~\ref{subsec:synch_spectral_index} and Section~\ref{subsec:synch_running}. Moreover, we show the recovered amplitudes of the CMB, synchrotron and thermal dust and, compare them with those obtained by \textsc{Commander} using PR4 data  in Section~\ref{subsec:amplitudes}. In Section~\ref{subsec:dust_spectral_index} we present the spectral parameters of the thermal dust. Finally, we evaluate the robustness of these results in Section~\ref{subsec:goodness_of_the_fit}.

\subsection{Synchrotron Spectral Index}
\label{subsec:synch_spectral_index}

The major improvement obtained from including the low-frequency QUIJOTE-MFI channels is having the sufficient sensitivity to study the synchrotron spectral index with great accuracy. Here we have conducted a deep study on several aspects with regard to this parameter. First, we have compared the recovered $\beta_s$ maps using different combinations of the available datasets (Section~\ref{subsec:beta_s_datasets}). Section~\ref{subsec:beta_s_spatial_variabily} studies the spatial variability of $\beta_s$.  Finally,  we compare our results to the available $\beta_s$ models that are often exploited in simulations used in CMB science forecasts in Section~\ref{subsec:beta_s_comparison_pysm}.

\begin{figure*}
    \centering
    \includegraphics[width=1\textwidth,trim={0cm 0cm 0cm 0cm},clip]{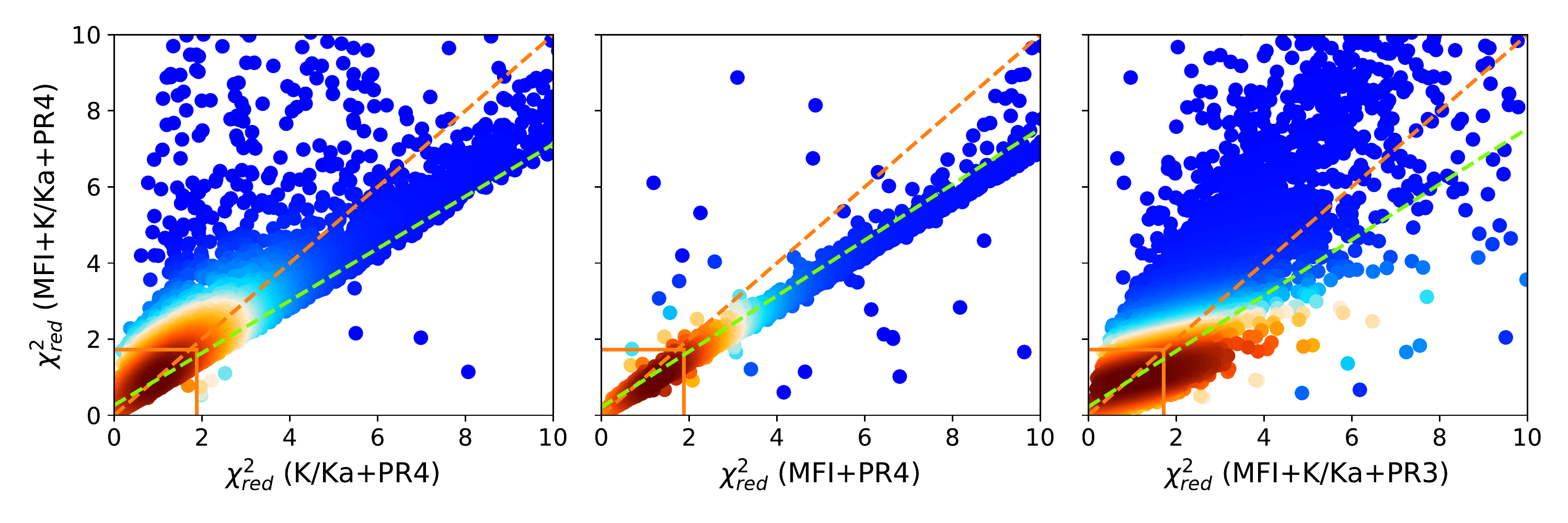}
    \caption{Reduced $\chi^2$, $\chi^2_{\mathrm{red}}$, obtained using the MFI+K/Ka+PR4 dataset vs. the $\chi_{\mathrm{red}}^2$ obtained using K/Ka+PR4 (left), MFI+PR4 (center) and MFI+K/Ka+PR3 (right). The color scale is related to the density of points, redder (bluer) corresponds to denser (sparser) regions. The orange rectangle shows the $\chi^2_{\mathrm{red}}$ within a 95\% confidence region. The slope calculated with the points within this 95\% confidence region is $m = 0.686 \pm 0.004$ (left column), $m=0.732 \pm 0.003$ (center column) and $m=0.731\pm 0.003$ (right column), shown with a green dashed line. The orange dashed line shows the one-to-one line. The synchrotron emission is modelled with a power law.}
    \label{fig:chi2_datasets}
\end{figure*}
\begin{figure}
    \centering
    \includegraphics[width=.5\textwidth,trim={1.2cm 1.6cm .5cm 1.2cm},clip]{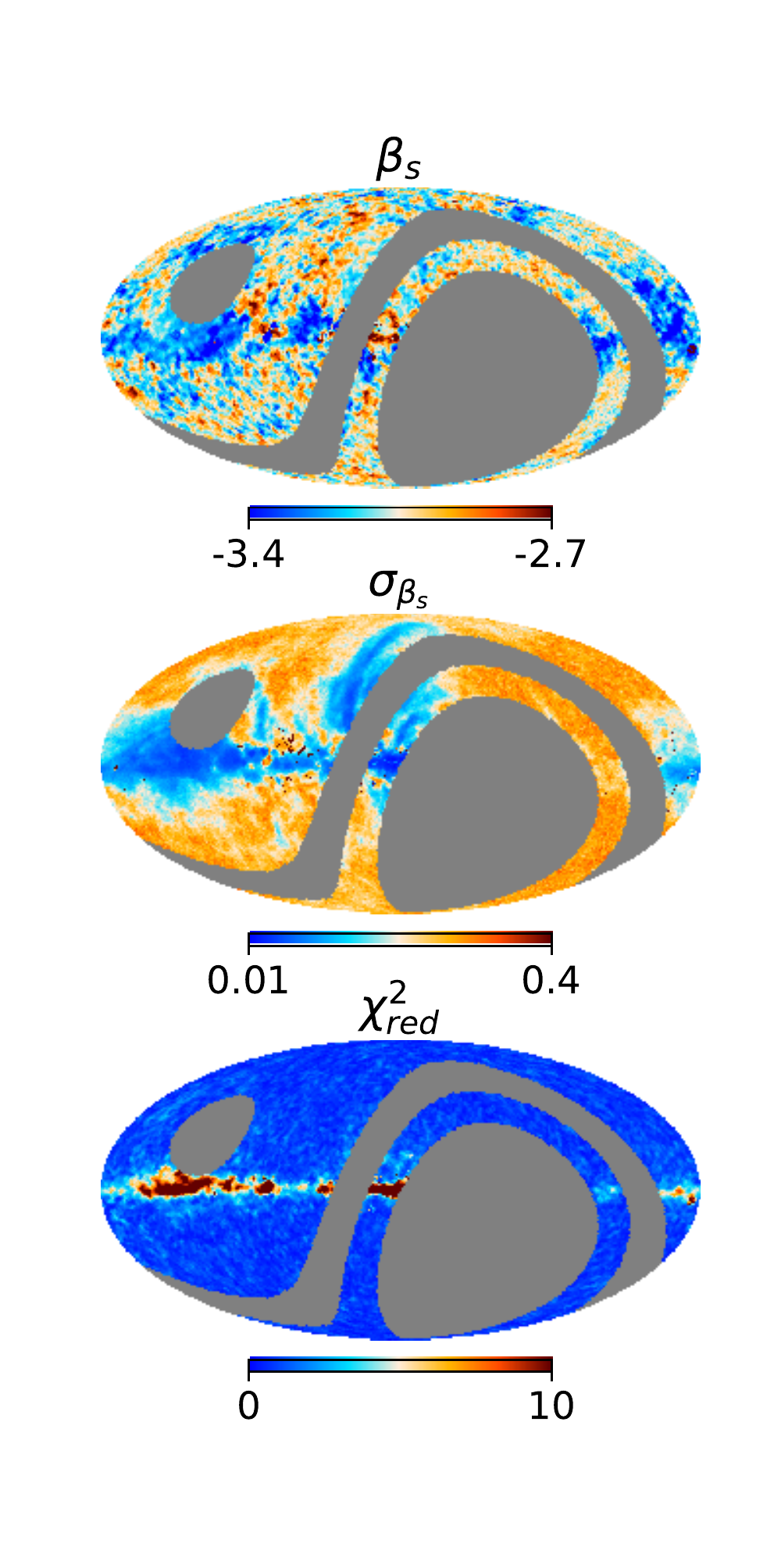}
    \caption{Synchrotron spectral index (top), its uncertainty (middle) and reduced $\chi^2$ (bottom) maps obtained after component separation with the default dataset MFI+K/Ka+PR4. The synchrotron emission is modelled with a power law.}
    \label{fig:beta_s_MFIKKaPR4}
\end{figure}

\subsubsection{Datasets}
\label{subsec:beta_s_datasets}

We have obtained different $\beta_s$ maps from component separation analyses using the four following  datasets: WMAP K and Ka bands with PR4 (K/Ka+PR4); QUIJOTE-MFI 11 and 13\,GHz channels with PR4 (MFI+PR4); QUIJOTE-MFI 11 and 13\,GHz channels, WMAP K and Ka bands and PR4 (MFI+K/Ka+PR4) and   QUIJOTE-MFI 11 and 13\,GHz channels, WMAP K and Ka bands and PR3 (MFI+K/Ka+PR3). The results are shown in Fig.~\ref{fig:beta_s_datasets}. It is clear from the comparison of the synchrotron spectral index uncertainty maps obtained in the K/Ka+PR4 case (first column) with respect to the MFI+K/Ka+PR4 case (third column), that the inclusion of QUIJOTE channels significantly improves the estimation of $\beta_s$. Moreover, we observe that, outside the Galactic plane, the estimation of $\beta_s$ is very close to the mean value of the prior set on this parameter, in this case $-3.1$. In other words, the information contained in that fraction of the data, i.e., the likelihood, is very poor and the estimation is driven by the prior. 

This improvement does not come from the inclusion of more channels, but from channels where the synchrotron contribution is larger. This is evident from the comparison of the results from K/Ka+PR4 with respect to MFI+PR4, where the number of frequency channels is the same but the results are significantly better for the latter.

Finally, we have compared also the results obtained with MFI+K/KA+PR3 and MFI+K/Ka+PR4 (fourth and third column respectively). In this case the recovered uncertainty maps are virtually the same but there are some distinct differences between the $\beta_s$ maps that should be ascribed to changes in \textit{Planck} maps. 

One of the advantages of using a parametric component separation method is that we can evaluate the goodness of the fit with certain estimators. In this work we use the reduced $\chi^2$ estimator, whose value at a given pixel $p$ is calculated as:
\begin{equation}
    \chi^2_{\mathrm{red},p}= \dfrac{1}{N_{\rm{dof}}}\sum\limits_{i \in \{Q, U\}}(\myvector{d}_{p,i}-\myvector{S}_{p,i})\mymatrix{C}^{-1}_{p,i}(\myvector{d}_{p,i}-\myvector{S}_{p,i})\, ,
    \label{eq:chi2_red}
\end{equation}
where the sum is over all $Q$ and $U$ frequency channels, and $N_{\rm{dof}}$ is the number of d.o.f.. The bottom row of Fig.~\ref{fig:beta_s_datasets} shows the $\chi^2_{\mathrm{red}}$ maps obtained for each dataset combination. These maps show that our default model, i.e., a power law and a modified black body to model the synchrotron and thermal dust emission respectively, provides a good fit (low values of $\chi^2_{\mathrm{red}}$) outside the Galactic plane. Within the Galactic plane, this model is not able to capture all the physical complexity and the $\chi^2_{\mathrm{red}}$ values are quite large. 
However, we note that in this analysis we have considered statistical uncertainties but not calibration errors, which in QUIJOTE are of $5\%$. Apart from the higher complexity of the Galactic plane emission, the higher $\chi_{\rm red}^2$ in this region could also be due, in part, to having neglected calibration errors.

We have also used the $\chi^2_{\mathrm{red}}$ estimator to select the dataset that is used as the default for further tests between the MFI+K/Ka+PR3 and the MFI+K/Ka+PR4 datasets, i.e., the only combinations that include all the channels considered. In Fig.~\ref{fig:chi2_datasets} the $\chi^2_{\mathrm{red}}$ obtained using the MFI+K/Ka+PR4 dataset is plotted against the $\chi^2_{\mathrm{red}}$ obtained with K/Ka+PR4, MFI+PR4 and MFI+K/Ka+PR3. The 95\% confidence regions are delimited by orange lines. These lines indicate the $\chi^2$ values, from the reduced $\chi^2$-distribution with $N_{\rm{dof}}$ d.o.f.\footnote{The $\chi^2$-distribution with $N_{\rm{dof}}$ divided by $N_{\rm{dof}}$.}, that satisfy that the normalized area covered to their left is equal to 0.95. We have also fitted the points within this confidence regions to a straight line to determine which dataset has more pixels with smaller $\chi^2_{\mathrm{red}}$. If the slope is larger than unity, the dataset on the horizontal axis has more pixels with smaller $\chi^2$. If the slope is smaller than unity, the dataset on the vertical axis is the one which satisfies that condition.

Although it is not clear from  the left plot of Fig.~\ref{fig:chi2_datasets} which dataset is better, the slope $m=0.686 \pm 0.004$ indicates that the MFI+K/Ka+PR4 dataset provides a better fit. Moreover, the K/Ka+PR4 dataset has larger uncertainties which can mask model inconsistencies. On the other hand, from the middle plot of Fig.~\ref{fig:chi2_datasets}, we observe that the inclusion of the K and Ka WMAP bands to the MFI+PR4 dataset improves the goodness of the fit. Finally, comparing MFI+K/Ka+PR4 with MFI+K/Ka+PR3, we see that PR3 provides a better fit in the Galactic plane, while PR4 fits better outside the Galactic plane (Fig.~\ref{fig:beta_s_datasets}). Since the fit in the Galactic plane is bad in both cases we have chosen the MFI+K/Ka+PR4 as our default dataset as it retrieves better fits within the 95\% confidence regions (pixels outside the Galactic plane, Fig.~\ref{fig:beta_s_MFIKKaPR4}).

\begin{figure}
    \begin{minipage}{\linewidth}
    \centering
    \includegraphics[width=\linewidth,trim={0.55cm 0.5cm 0.55cm .2cm},clip]{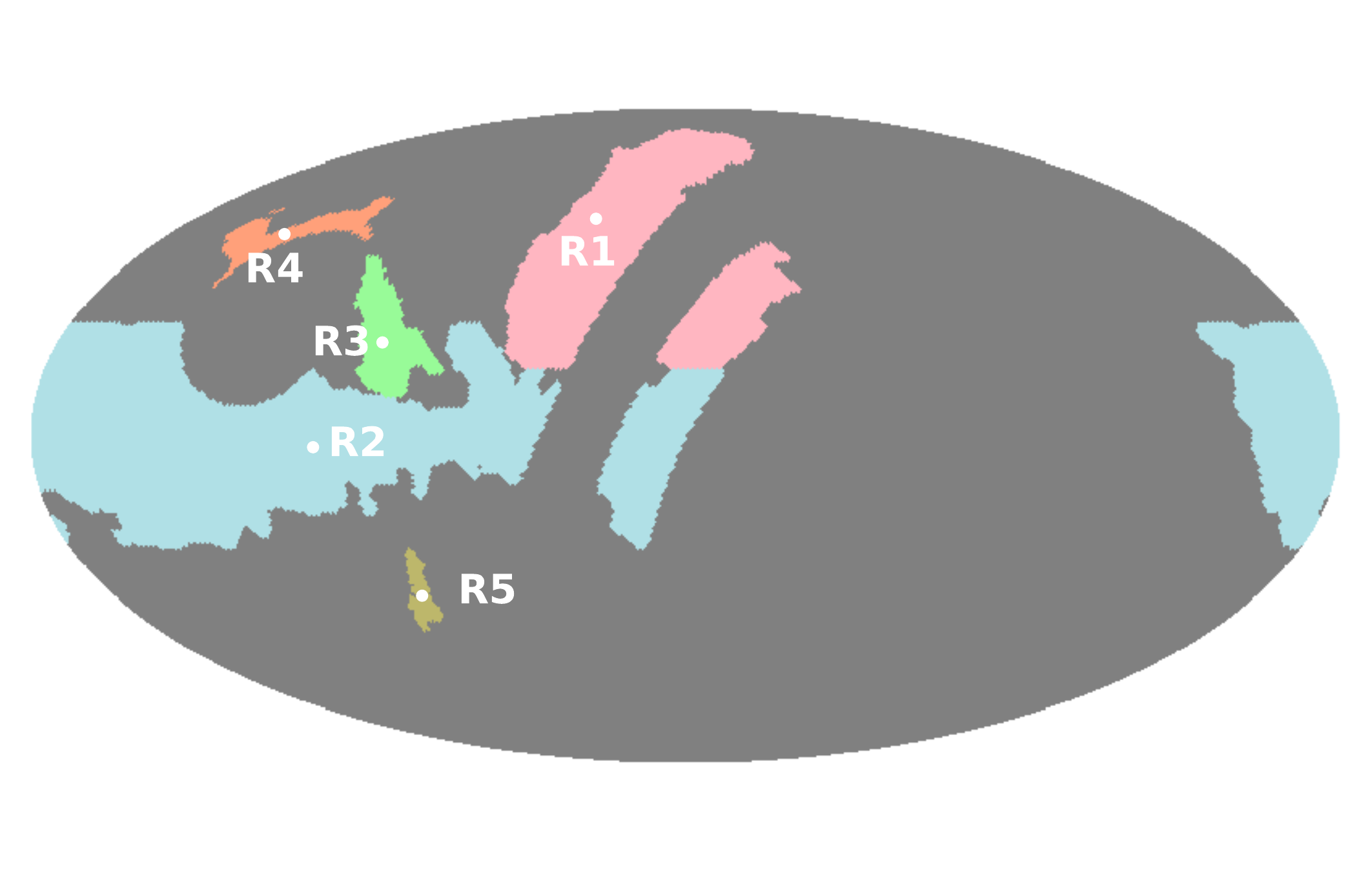}
    \caption{R1, R3, R4 and R5 are sky regions where $\beta_s$ is assumed uniform in Section~\ref{subsec:beta_s_spatial_variabily} and R2, which encompasses the Galactic plane seen by QUIJOTE, is a very heterogeneous region. These regions satisfy that $\beta_s$ is recovered with a signal-to-noise larger than 15.}
    \label{fig:regions_beta_s}
    \end{minipage}
    \vspace{.4cm}
    
    \begin{minipage}{\linewidth}
    \centering
    \captionof{table}{Synchrotron spectral index estimation $\beta_s^R$ and its uncertainty $\sigma({\beta^R_s})$ obtained assuming uniform value across the regions R1, R3, R4 and R5 shown in Fig.~\ref{fig:regions_beta_s}.}
    \begin{tabular}{cccc}
    \toprule
    Region & $f_{\mathrm{sky}}$ (\%) & $\beta_s^R$ & $\sigma({\beta^R_s})$ \\
    \midrule
    R1 & 4.84  & -3.028  &  0.002  \\ 
    R3 & 0.96  & -2.945  &  0.008  \\ 
    R4 & 0.56  & -3.319  &  0.011  \\ 
    R5 & 0.21  & -3.228  &  0.019  \\ 
    \bottomrule
    \end{tabular}
     \label{tab:beta_s_regions}
    \end{minipage}
\end{figure}

\begin{figure*}
    \centering
    \includegraphics[width=1\textwidth,trim={0cm 0cm 0cm 0cm},clip]{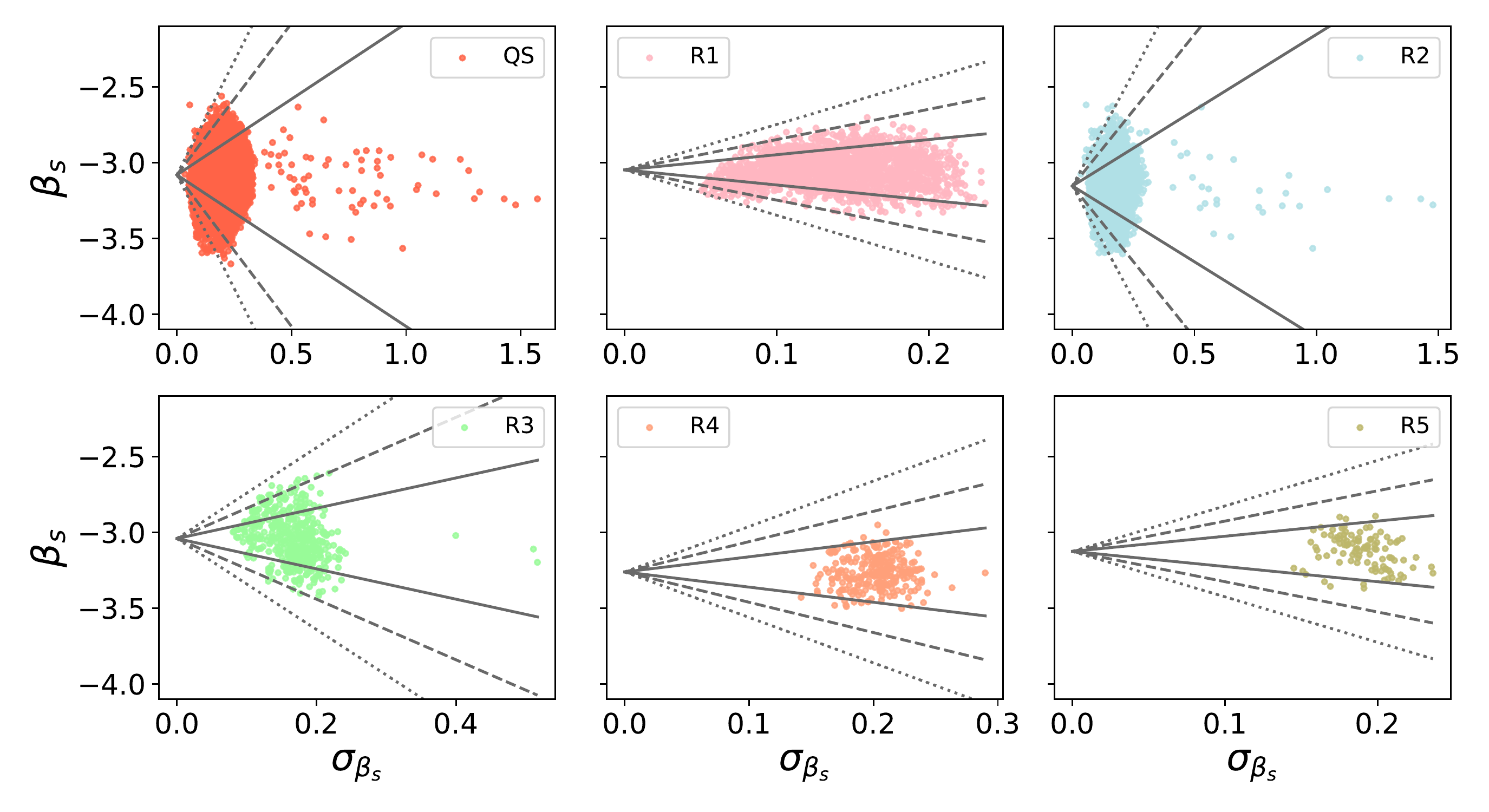}
    \caption{Synchrotron spectral index estimate against its uncertainty within different sky regions: QUIJOTE-MFI sky (QS) (Fig.~\ref{fig:mask_satband}); R1, R2, R3, R4 and R5 are shown in Fig.~\ref{fig:regions_beta_s}. The solid, dashed, and dotted lines enclose the values of $\beta_s$ within 1$\sigma$, 2$\sigma$ and 3$\sigma$ of the weighted mean respectively. The study is limited to those pixels whose $\chi^2_{\mathrm{red}}$ lies within the 95\% confidence region. }
    \label{fig:beta_s_vs_sigma_beta_regions_QKNPIPE}
\end{figure*}

\subsubsection{Spatial Variability}
\label{subsec:beta_s_spatial_variabily}

We have also studied the spatial variability of the synchrotron spectral index in several high signal-to-noise regions of the sky, see Fig.~\ref{fig:regions_beta_s}. These connected regions satisfy the condition that $\beta_s$ is estimated with a signal-to-noise ratio larger than 15. In particular, R1 is associated with the North Polar Spur (NPS), and R2 encompasses the Galactic plane. R3, R4 and R5 are other sky regions where the polarized synchrotron intensity has a large signal-to-noise ratio.

Fig.~\ref{fig:beta_s_vs_sigma_beta_regions_QKNPIPE} shows the estimated synchrotron spectral index against the uncertainty on the estimation of all the pixels within a given region. We have limited this study to those pixels with a $\chi^2_{\mathrm{red}}$ within the 95\% confidence region. The area delimited by the dotted lines contains the values that are consistent within 3$\sigma$ with the weighted mean in each region. The top left panel indicates that $\beta_s$ has a large spatial variability across the whole available QUIJOTE-MFI sky (QS). Therefore,  a constant value of $\beta_s$ is not a good model of the synchrotron emission. 
On the contrary, the R1, R3, R4 and R5 pixels values are well within those lines, i.e., a uniform $\beta_s$ value could be a good model for all pixels within each region. Finally, R2 (top right panel) shows a significant spatial variability which is consistent with the large heterogeneity observed in the $\beta_s$ map.

The study of regions with uniform $\beta_s$ values helps with improving the detectability of primordial $B$-modes. Allowing spatial variations of the spectral parameters at the pixel level  results in a  very robust parametrization of the signal sky. However, this robustness comes at the expense of an increase in the statistical uncertainty of the parameters as less information is provided in the fit \citep{multipatch}. Thus, several approaches have been proposed in the literature to define sky regions with uniform spectral parameters. For example, in \citet{multipatch}, these regions are chosen as super-pixels at a lower HEALPix maps resolution, whereas in \citet{clustering}, the regions are obtained using clustering algorithms such as the mean-shift clustering algorithm. Recently, \citet{spectral_clustering} has presented a new methodology based on spectral  clustering to define geometrical affine regions with similar spectral parameters. It is worth noting that if the assumption of uniform spectral parameters within those regions does not hold, the modelling errors  introduced might bias cosmological parameters measurements obtained from the output CMB map after component separation, as well as foreground model parameters.

We have calculated the value of $\beta_s$ in some of these regions assuming a constant value  within each region. We have performed the fit in the following manner:
\begin{itemize}
    \item First we fix $\beta_s$ to a given value and fit the rest of the model parameters in each pixel of the region.
    \item Then, the rest of the parameters are fixed to the estimation from the previous fit, and we fit $\beta_s$ assuming a unique value in the whole region under study.
    \item $\beta_s$ is fixed to the new obtained value and the process is repeated until it reaches convergence.
\end{itemize}
We have chosen the median of the $\beta_s$ values (obtained pixel-wise) within that region as the initial guess of $\beta_s$. The results  are shown in Table~\ref{tab:beta_s_regions}. Notice that the uncertainty on the recovered $\beta_s$ has dramatically decreased. This is simply a result of having $N^{R}_{\rm pix}$ (the number of pixels contained within the region $R$) times more information to fit the parameter. The $\beta_s$ values recovered in each region (R1, R3, R4 and R5) are not consistent among them. These results further showcase the spatial variability of the synchrotron's spectral parameter.

\begin{figure}
    \centering
    \includegraphics[width=.5\textwidth]{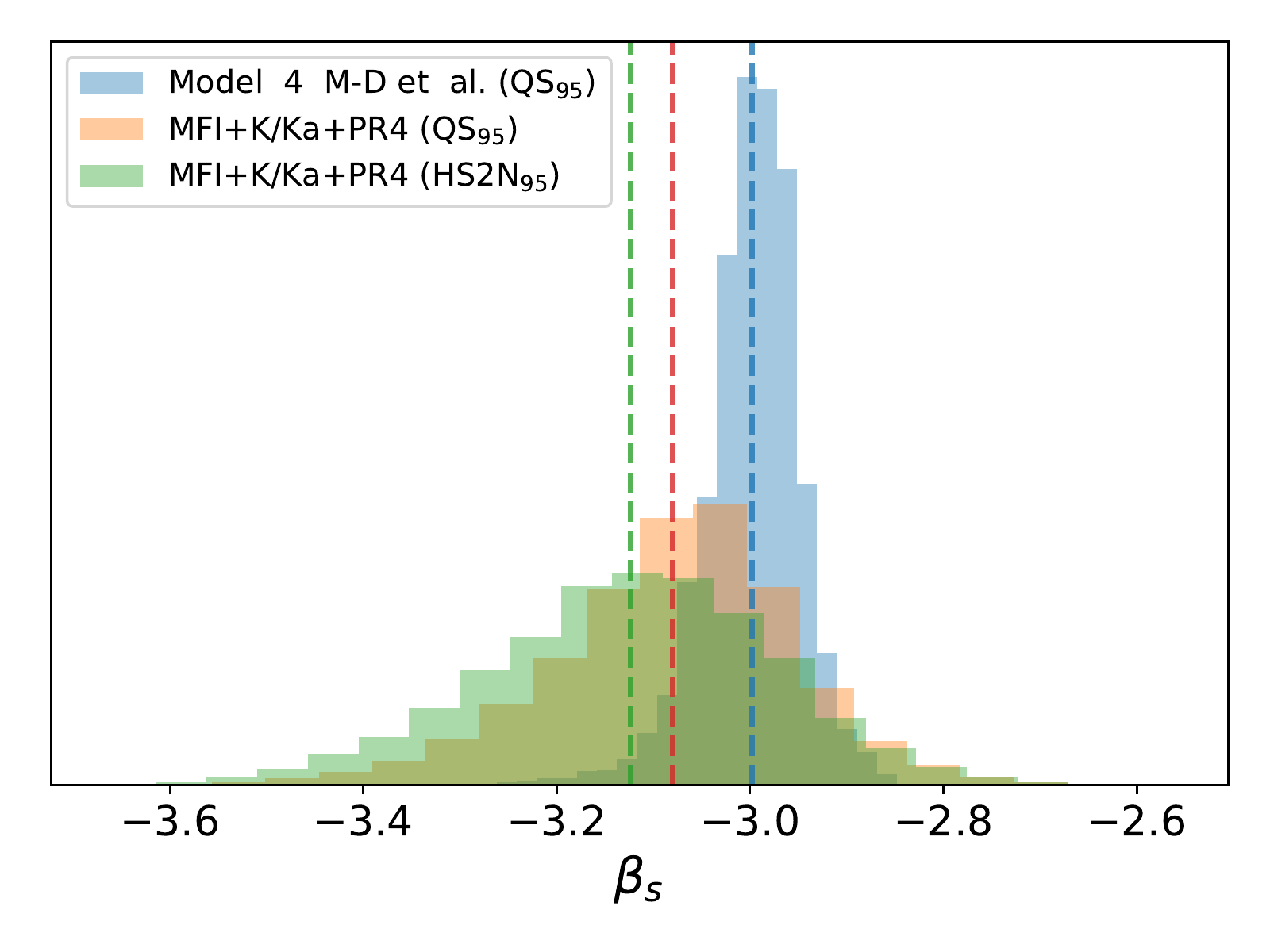}
    \caption{Distribution of the synchrotron spectral index from ``Model 4'' of \citet{2008AAseparation} and from our estimation using the MFI+K/Ka+PR4 dataset. Vertical dashed lines indicate the mean value for each distribution.}
    \label{fig:comparison_pysm_beta_s}
 \end{figure}

\subsubsection{Comparison with current $\beta_s$ models}
\label{subsec:beta_s_comparison_pysm}

In this Section we compare our $\beta_s$ map with the currently most used $\beta_s$ template\footnote{Used for example in the \textit{Planck} Sky Model \citep{PSM2012}, or in the Python Sky Model (\textsc{PySM}) a \textsc{Python} library to simulate foregrounds \citep{pysm}.}, the  ``Model 4'' Miville-Deschenes et al. template, which was constructed with Haslam and WMAP observations in temperature \citep{2008AAseparation}. Fig.~\ref{fig:comparison_pysm_beta_s} shows the distribution of the spectral index value for this model (blue) and for our analysis (orange), considering only those QUIJOTE-MFI pixels that lie within the 95\% confidence region of the $\chi^2$ ($\rm{QS}_{95}$). In the $\rm{QS}_{95}$ region, the mean and the standard deviation from the ``Model 4'' of \citet{2008AAseparation} template are $-3.00\pm0.05$ while those from our estimate are $-3.08\pm 0.13$. It is interesting to note that the variability observed in our analysis is significantly larger. A direct comparison of the dispersion of both maps (using the same mask) indicates an increment of the spatial variability in our study around a factor of 2.6, i.e., $\sigma(\beta_s^{\rm{MFI+K/Ka+PR4}})/\sigma(\beta_s^{\rm{Model\, 4}}) \sim 2.6$.

One may wonder if this result can be affected by the considered prior, since the estimated spectral indices for low signal-to-noise pixels are significantly constrained by it (see Section~\ref{subsubsec:robustness_priors}). In order to test this point, we have repeated the previous analysis considering only those pixels satisfying  that the recovered $\beta_s$ values have a signal-to-noise larger than 15 (i.e., where the synchrotron signal-to-noise is high and thus the results are not driven by the prior) and lie within the 95\% confidence region of the $\chi^2$ ($\rm{HS2N}_{95}$). In this case, we find that the mean value and dispersion of the distribution of $\beta_s$ are $-3.12\pm 0.15$ for our analysis (see green histogram in Fig.~\ref{fig:comparison_pysm_beta_s}) versus $-3.00\pm0.05$ for ``Model 4'' in the same region, confirming our finding. Although our estimations can be affected by the presence of noise, the results show that the variability of the synchrotron spectral index assumed in current templates is underestimated. A similar increment in the variability was also noted by analysing the S-PASS data in the Southern Hemisphere \citep{Nicoletta2018}.

Recently \citet{weiland_beta_s} published a composite map of $\beta_s$ using publicly available data covering approximately 44\% of the sky. In the region covered in our study, they obtained $\beta_s$ estimates in the Galactic Plane and the North Polar Spur using information from WMAP K and Ka band, and estimates at latitudes larger than $40^{\circ}$ using K, Ka and DRAO 1.41\,GHz map \citep{DRAO}. From a visual inspection our results are compatible within the North Polar Spur. We find that our derived spectral indices are steeper at the Galactic plane. \citet{weiland_beta_s} found discrepancies between the $\beta_s$ values obtained in the Fan Region when they performed the analysis using WMAP K and Ka band versus WMAP K band and Planck LFI 30\,GHz channel. In the latter case, the recovered $\beta_s$ were significantly steeper. We repeated our analysis excluding the PR4 30\,GHz channel and did not observe a discrepancy concerning the $\beta_s$ recovered from the default analysis in Fan Region. This results from the fact that the $\beta_s$ recovery is mainly driven by QUIJOTE-MFI data. At high latitudes we cannot make a reasonable  comparison since our $\beta_s$ estimates are driven by the prior. They also show that DRAO data have some unexplained systematics and can be affected by Faraday Rotation depolarization.

\begin{figure*}
    \centering
    \includegraphics[width=.75\textwidth,trim={2cm 1cm 1.5cm 1cm},clip]{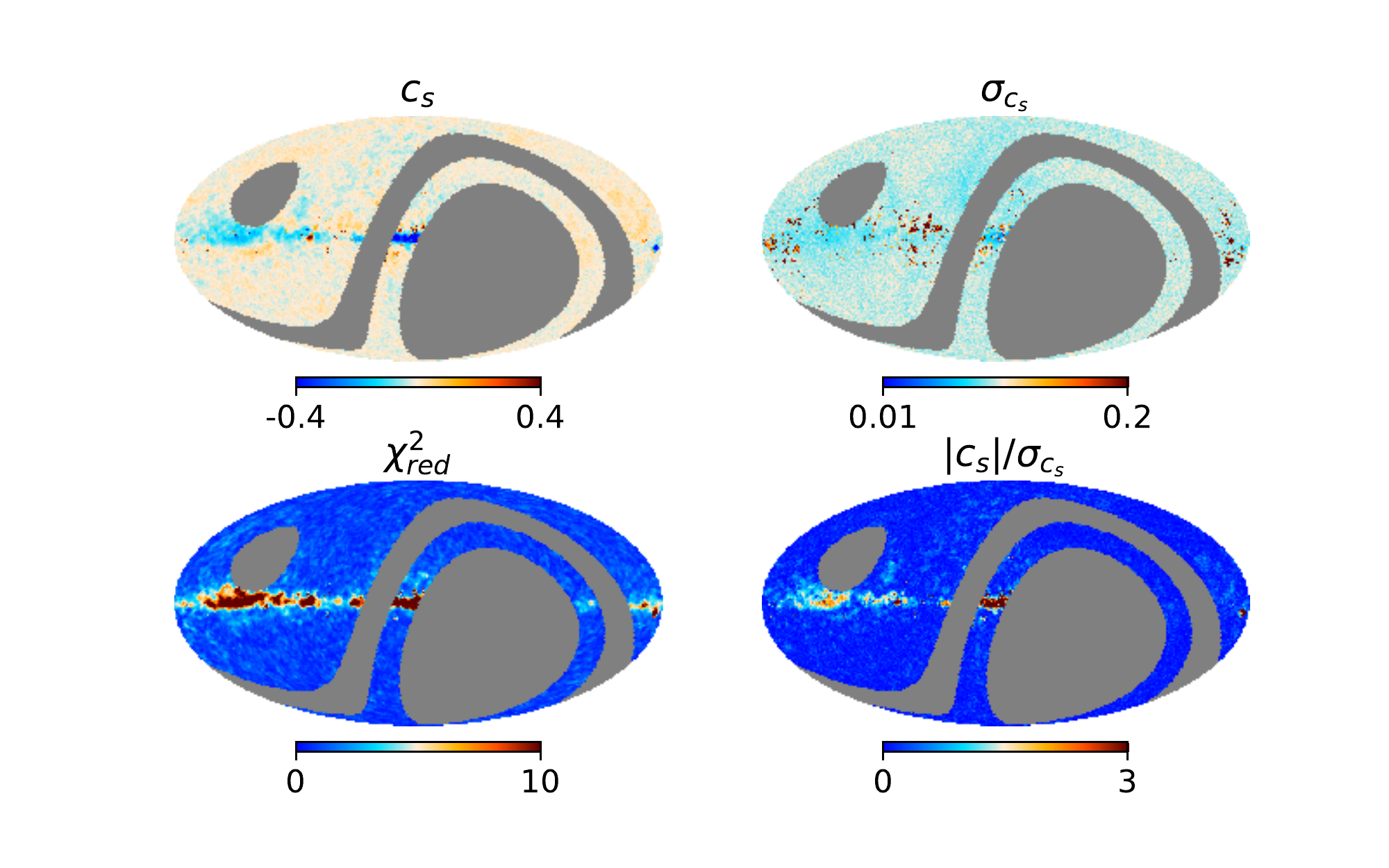}
    \caption{Top row: Synchrotron curvature estimate (left) and uncertainty (right) maps obtained after component separation using the default dataset (MFI+K/Ka+PR4). The synchrotron emission is modelled using a power law with spatially varying curvature (pixel-wise). Bottom row:  reduced $\chi^2$ map (left) and $c_s$ signal-to-noise map (right).}
    \label{fig:c_s_QKNPIPE_QU}
\end{figure*}

\begin{figure}
    \centering
    \includegraphics[width=.5\textwidth]{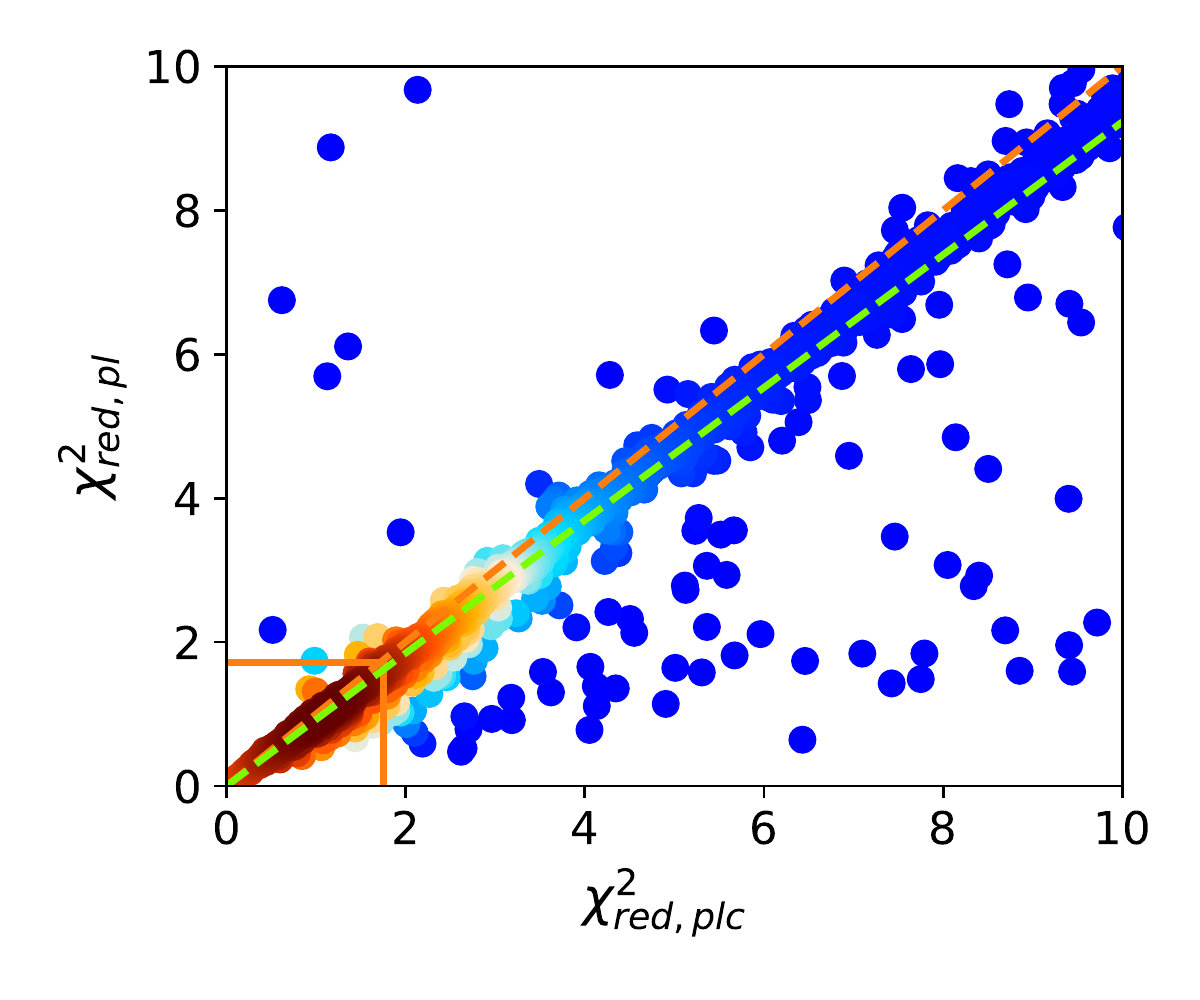}
    \caption{Reduced $\chi^2$ calculated using a power law as a model of the synchrotron emission ($\chi^2_{\mathrm{red,pl}}$) vs. $\chi^2_{\mathrm{red}}$ when the model is a power law with spatially varying curvature ($\chi^2_{\mathrm{red,plc}}$). The color scale is related to the density of points, redder (bluer) corresponds to denser (sparser) regions. The red rectangle shows the $\chi^2_{\mathrm{red}}$ within a 95\% confidence region. The slope at the 95\% confidence region is $m=0.9227\pm 0.0005$, shown with a green dashed line. The orange dashed line shows the one-to-one line.}
    \label{fig:chi2_c}
\end{figure}
Other studies, such as those presented in \citet{Vidal2015, Fuskeland2014,Fuskeland2019,felice_synchrotron}, also find variability of the spectral index analyzing different regions of the sky. However it is difficult to compare the same regions in our map, since they compute a global spectral index for large areas, while we work pixel by pixel. For example, near the center of the Galactic plane we see a fair amount of structure that cannot be accounted for in the T–T scatter plots analyses carried out in some of the cited papers, that use several pixels to obtain a single $\beta_s$ value. In that sense, the methodology followed here is more complete given that we perform a full component separation in each pixel, retrieving information at smaller scales for a large fraction of the sky. 

\citet{mfiwidesurvey} obtain an estimate of the synchrotron spectral index map directly from the comparison of the QUIJOTE-MFI 11\,GHz map with the WMAP K band map. The results obtained there are fully consistent with the ones from this work.

\begin{figure}
    \begin{subfigure}{\linewidth}
        \centering
        \includegraphics[width=.98\linewidth]{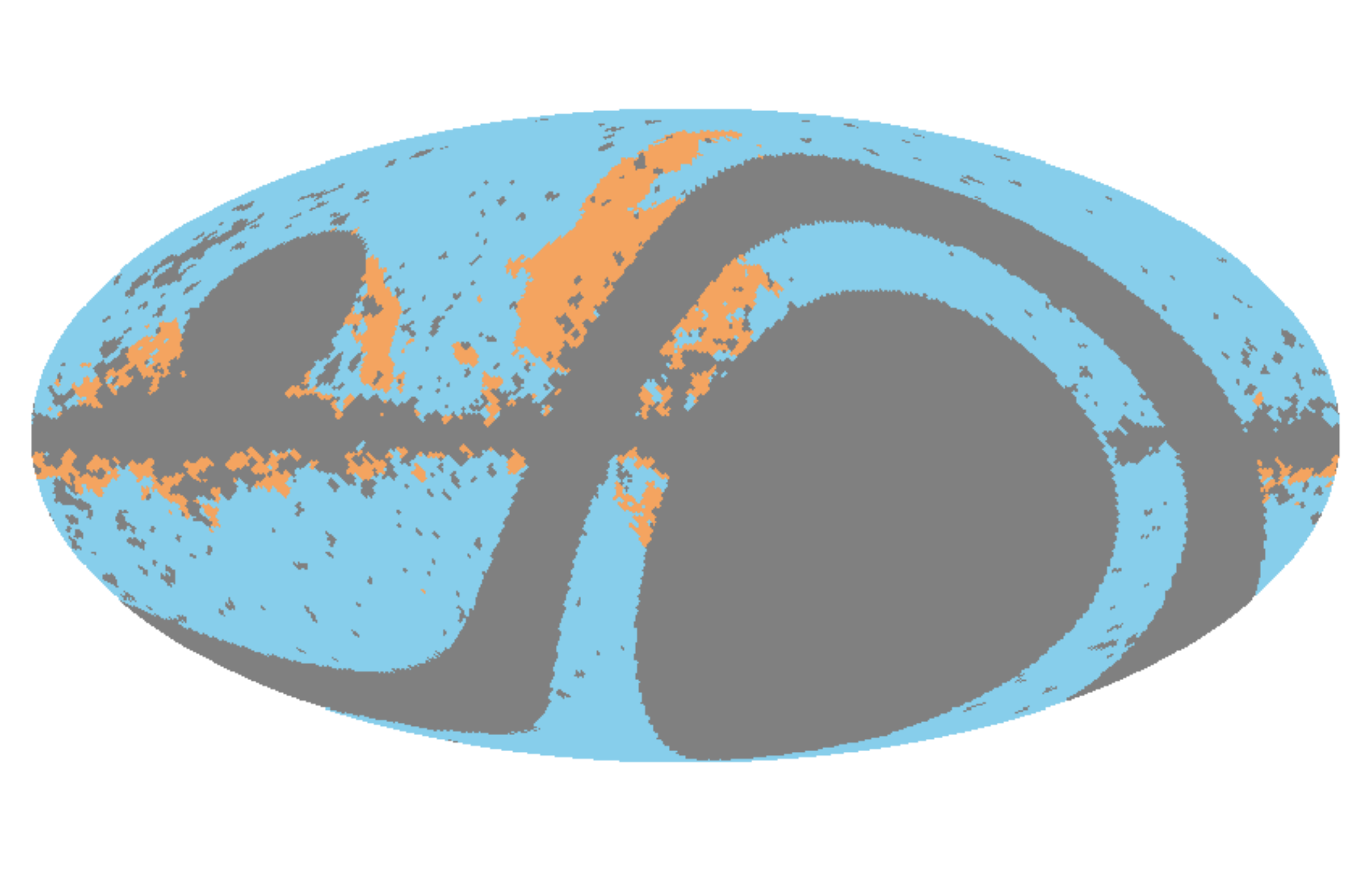}
        \caption{RC1 (coloured) and RC2 (orange) regions.}
        \label{fig:RC1_RC2_regions}
    \end{subfigure}
    \begin{subfigure}{\linewidth}
        \centering
        \includegraphics[width=.98\linewidth]{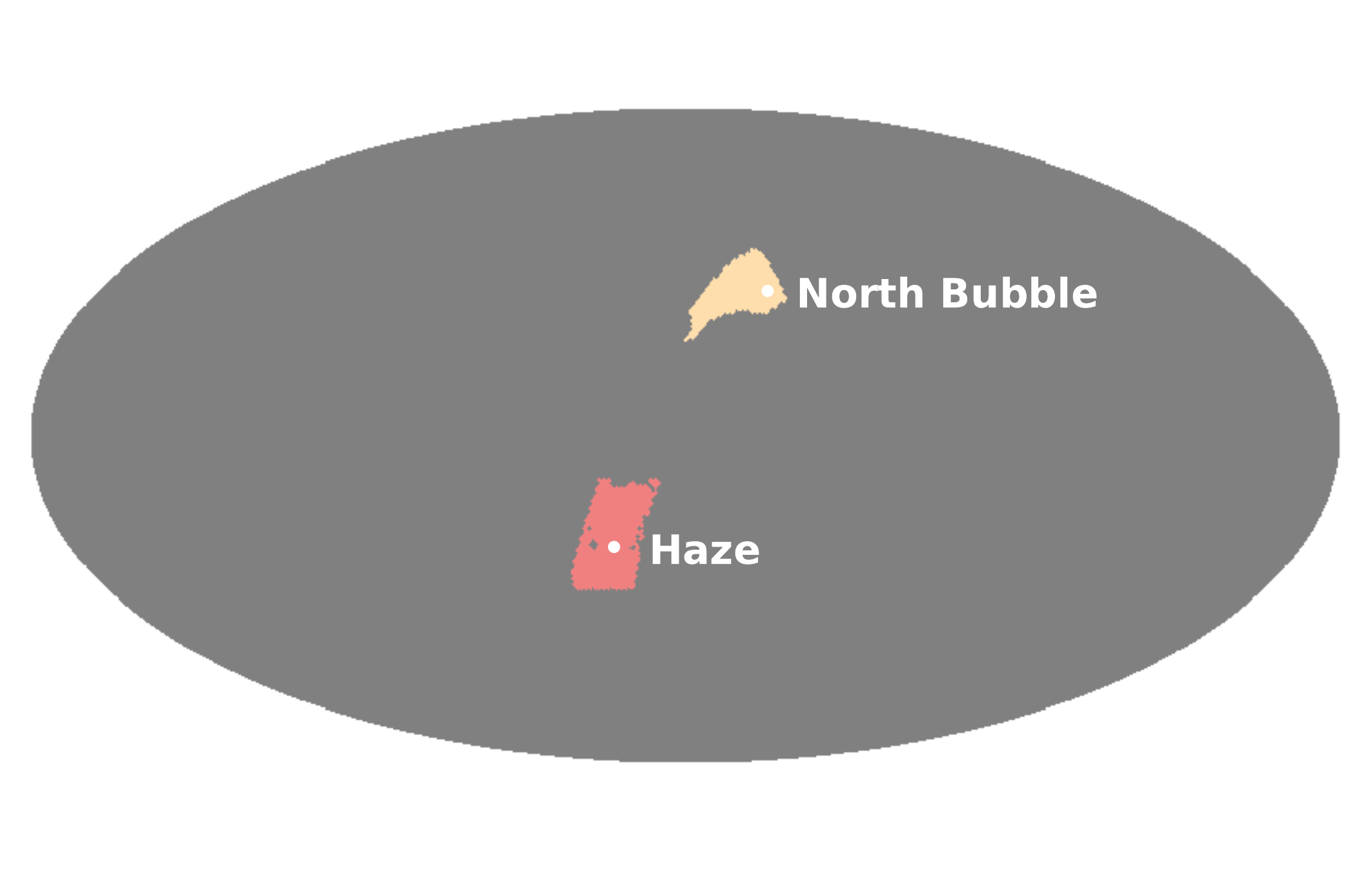}
        \caption{Haze and North bubble regions.}
        \label{fig:fede_regions}
    \end{subfigure}
    \caption{Regions where $c_s$ has been assumed to be uniform.}
    \label{fig:c_s_regions}
\end{figure}
\subsection{Synchrotron Curvature}
\label{subsec:synch_running}

We have also explored a synchrotron model with curvature, i.e., the model presented in equation~(\ref{eq:synchrotron_model_curvature}), using the MFI+K/Ka+PR4 dataset. Fig.~\ref{fig:c_s_QKNPIPE_QU} shows the estimation and uncertainty maps of the curvature parameter as well as the $\chi^2_{\mathrm{red}}$ map and the $c_s$ signal-to-noise map. 

We observe from the signal-to-noise map that curvature is detected at more than 3$\sigma$ in the Galactic plane, in regions where the fit is not good as it can be seen from the $\chi^2_{\mathrm{red}}$ map. Even though the inclusion of a curvature parameter is not able to explain the complexity of this region, this parameter can account for some effects along the Galactic plane, e.g., Faraday rotation.

Outside the Galactic plane the estimated $c_s$ values are close to zero and their uncertainties are around 0.1, which are the expected value and the spread of the prior set on $c_s$. Moreover, the recovered $\beta_s$ map in this case is very similar to the one obtained when the synchrotron is model with a power law. This means that we do not have enough sensitivity to detect a spatially varying curvature. Hopefully, joint analyses with future releases of the Northern Celestial Hemisphere data like the new MFI2 instrument and C-BASS at 5\,GHz \citep{CBASS2018} might elucidate more details on changes of the power law spectrum.

In Fig.~\ref{fig:chi2_c}
we compare the goodness of fit using a power law  versus a power law with curvature as the synchrotron model. We see that there are  more points located below the bisector. Besides, the slope $0.9227 \pm 0.0005$ calculated at the 95\% confidence region, shows that, given the current data, the power law model is slightly preferred over the power law plus curvature model. 

\begin{table}
    \caption{Estimated values of the curvature and its uncertainty obtained assuming the curvature is  uniform within the region.}
    \begin{tabular}{ccccc}
    \toprule
    Region & $f_{\mathrm{sky}}$ (\%) & $c^R_s$ & $\sigma_{c^R_s}$ & $\left|c^R_s\right|/\sigma_{c^R_s}$ \\
    \midrule
    RC1 & 45.48 & -0.0797 & 0.0012  &  63.75\\ 
    RC2 & 5.93  & -0.2768 & 0.0017  & 161.57 \\ 
    Haze & 0.94 & 0.041 & 0.010 & 4.23 \\
    North bubble & 0.63 &  -0.083 & 0.007 & 11.43 \\
    \bottomrule
    \end{tabular}
    \label{tab:c_s_regions}
\end{table}

Furthermore, we have considered modelling the synchrotron emission with a power law with uniform curvature. We have assumed a constant $c_s$ in four regions: RC1, RC2, and the Haze and North bubble (Fig.~\ref{fig:c_s_regions}). The recovered curvature values  are shown in Table~\ref{tab:c_s_regions}. RC1 encompasses all the pixels whose $\chi^2_{\mathrm{red}}$ is within 95\% confidence region. RC2 is composed of the RC1 pixels that also satisfy that the synchrotron polarized intensity signal-to-noise ratio at 30~GHz is larger than 5.
We detect curvature in all regions. The detection is more evident in RC1 and RC2, mostly due to the higher sensitivity (lower $\sigma_C$) in these regions. However, it is important to highlight that there is no physical reasoning behind the definition of RC1 and RC2, and the assumption of uniform curvature in all synchrotron high signal-to-noise regions is arbitrary\footnote{Any curvature will be more easily detected in high signal-to-noise regions than in low signal-to-noise regions.}.  In the Haze and North bubble, we find a  curvature value different from zero at more than 3$\sigma$. These regions are studied in greater detail in \citet{hazewidesurvey}. 

\begin{figure}    
    \centering
    \includegraphics[width=\linewidth]{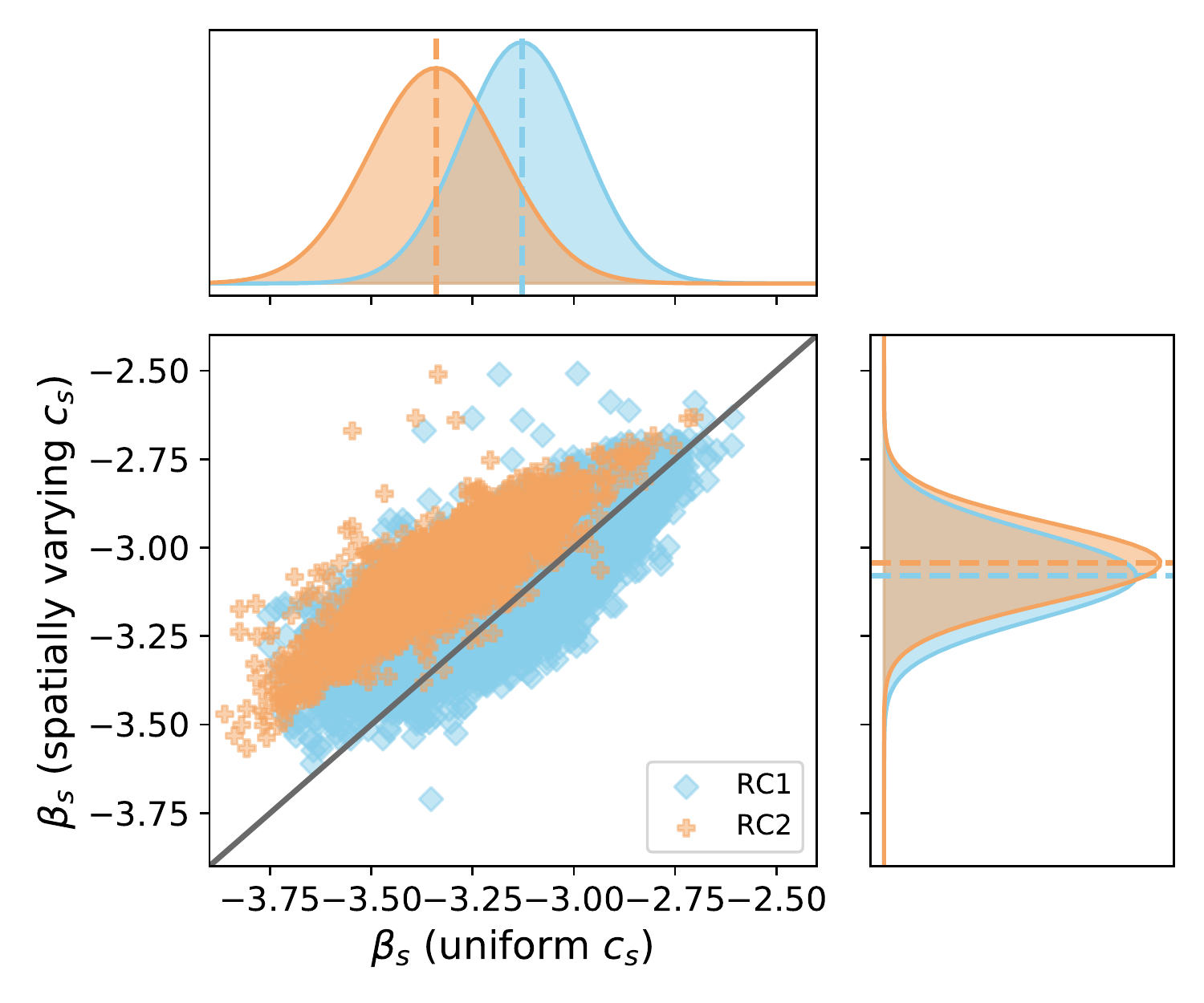}
    \caption{Comparison between the pixel $\beta_s$ values obtained when fitting the synchrotron emission with a spatially varying curvature model  ($y$-axis) versus with a model with uniform curvature ($x$-axis) in the regions RC1 and RC2 using the MFI+K/Ka+PR4 dataset.}
    \label{fig:beta_s_comparison_rc_clusters_pixels}
\end{figure}

We have studied how $\beta_s$ changes when we impose the constraint of having a uniform $c_s$ value within each region. The results are displayed in Fig.~\ref{fig:beta_s_comparison_rc_clusters_pixels}. For RC2, we observe that $\beta_s$ steepens considerably. The weighted mean value of $\beta_s$ in RC2 is $\mleft<\beta_s\mright> = -3.022 \pm 0.011$ in the pixel-wise analysis and, $\mleft<\beta_s\mright> = -3.375 \pm  0.002$ when $c_s$ is imposed to be uniform in RC2. For RC1, this effect is not as considerable. The weighted mean values are  $\mleft<\beta_s\mright> =-3.079 \pm 0.002$ and $\mleft<\beta_s\mright> =-3.1651 \pm 0.0014 $ when $c_s$ varies pixel-wise and is uniform respectively. The steepening of $\beta_s$ leads to values of the exponent $\beta_s + c_s\log(\nu/\nu_s)$ within [-3.04,-3.10] at 11\,GHz which are compatible with the average value of $\beta_s$ when we fit to a power law model. From these results, we infer that the $\beta_s$ and $c_s$ parameters are not independent. More sensitive data at the QUIJOTE frequencies and at lower and/or higher frequencies are required to break the degeneracy.

\begin{figure*}
    \centering
    \includegraphics[width=.99\textwidth,trim={3.5cm .5cm 2.5cm 1cm},clip]{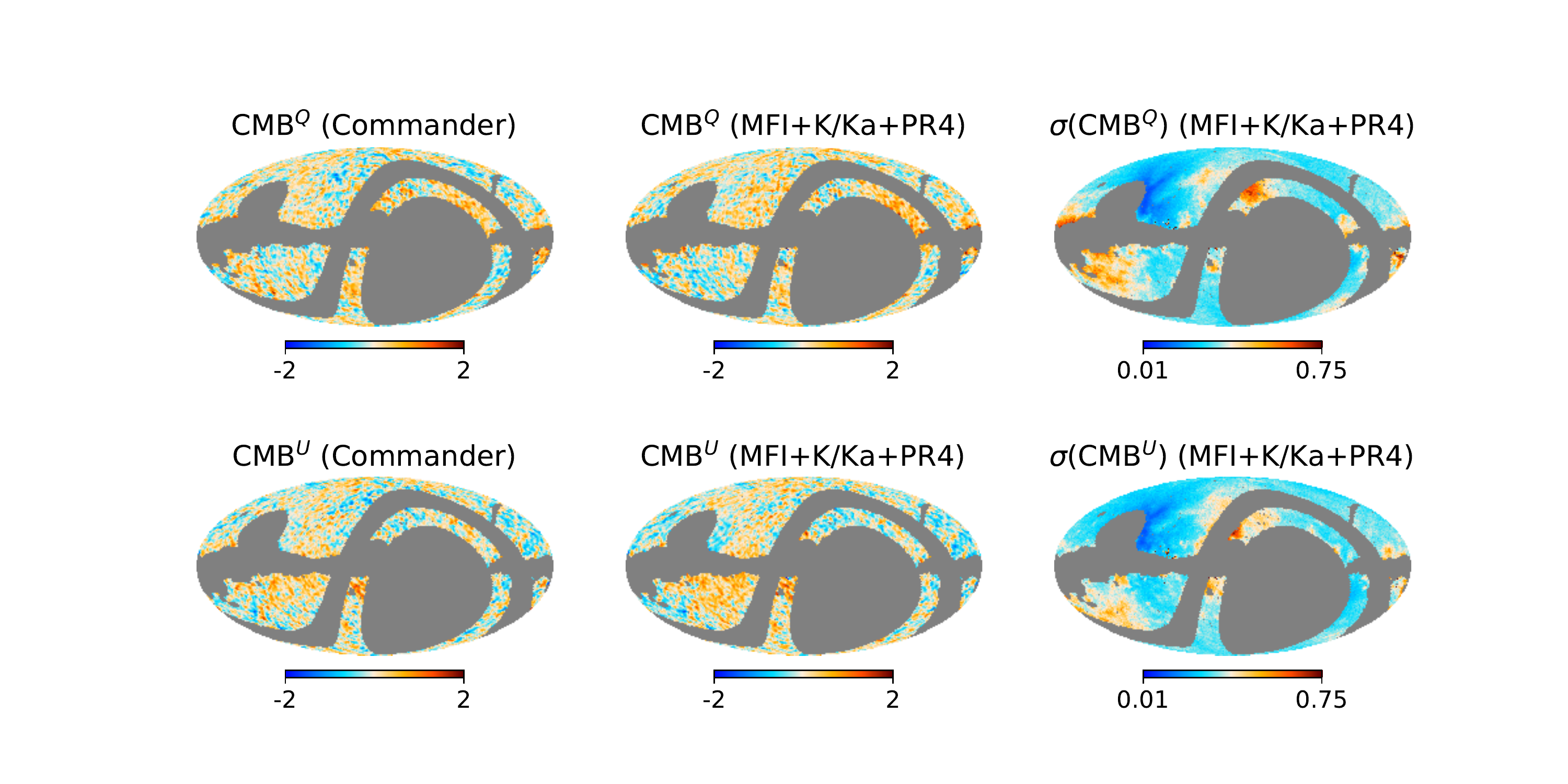}
    \caption{Left column: \textsc{Commander} $Q$ (top) and $U$ (bottom) CMB maps at $N_{\rm side}$ = 64, smoothed with a Gaussian beam to a final resolution of $\mathrm{FWHM} = 2^{\circ}$. Centre column: CMB $Q$ and $U$ maps using the MFI+K/Ka+PR4 dataset. Right column: uncertainty of the CMB maps. Maps are in thermodynamic temperature ($\mu$K). We apply the common polarization confidence mask provided by \textit{Planck}.}
    \label{fig:cmb_QKNPIPE_Commander}
\end{figure*}

\begin{table}
    \centering
    \caption{Reduced $\chi^2$ obtained using either a power law or a power law with curvature model in different regions, $R$. We have considered two curvature models: one where $c_s$ varies spatially (spatial) and other where   $c_s$ is assumed constant in $R$.} 
    \begin{tabular}{cccc}
    \toprule
        Model & Curvature & Region & $\chi^2_{red,R}$ \\
        \midrule
        power law             &  -- &           RC1 &        0.892 \\
        power law + curvature &  spatial  &           RC1 &        0.965 \\
        power law + curvature &  uniform &           RC1 &        0.936 \\
        power law             &  -- &           RC2 &        1.010 \\
        power law + curvature &  spatial  &           RC2 &        1.088 \\
        power law + curvature &  uniform  &           RC2 &        1.081 \\
        power law             &  -- &          Haze &        0.845 \\
        power law + curvature &  spatial     &          Haze &        0.936 \\
        power law + curvature &  uniform  &          Haze &        0.885\\
        power law             &  -- &  North bubble &        0.961 \\
        power law + curvature &  spatial &  North bubble &        1.041 \\
        power law + curvature &  uniform &  North bubble &        0.986\\
    \bottomrule
    \end{tabular}
    \label{tab:chi2_models}
\end{table}

In order to test which model provides a better goodness of fit we calculate the reduced $\chi^2$ of a given region $R$ as follows:
\begin{equation}
    \chi^2_{{\rm{red}},R} = \dfrac{1}{N_{\rm{dof}}}\sum\limits_{p=1}^{N^R_{\rm{pix}}}\sum\limits_{i\in \{Q,U\}}(\myvector{d}_{p,i}-\myvector{S}_{p,i})\mymatrix{C}^{-1}_{p,i}(\myvector{d}_{p,i}-\myvector{S}_{p,i})\, ,
    \label{eq:chi2_red_regions}
\end{equation}
where we sum over all pixels $N^R_{\rm{pix}}$ within $R$. The d.o.f. are given as ${N_{\rm{dof}}} = N^R_{\rm{pix}}(2N-N_{\theta})$ when all model parameters are allowed to vary pixel-wise, and ${N_{\rm{dof}}} = N^R_{\rm{pix}}(2N-(N_{\theta}-1)) -1$ when $c_s$ is assumed uniform in the analysis, where $N_{\theta}$ is the number of model parameters. We calculated the value of this estimator in three cases: i) when the model parameters are allowed to vary spatially using a power law model for the synchrotron component, ii) when the model parameters vary from pixel-to-pixel using a power law with curvature model, iii) when we fit the data assuming uniform curvature using a power law with curvature model. The results are given in Table~\ref{tab:chi2_models}. 

The $\chi^2_{\mathrm{red}}$ results show that the models we used, i.e., power law and power law with curvature, are compatible with the data. However there is not enough statistical significance to discern which model suits better the data. Especially, considering that we have not been able to take into account possible
correlations between pixels and that the power law with curvature model is degenerate\footnote{We considered applying other statistics such as the Bayesian evidence to do model selection. However, since the QUIJOTE-MFI noise simulations are not end-to-end and the Bayesian evidence is very computationally expensive we did not perform any model selection analysis. This is left for future work.}.

\subsection{Recovered Amplitudes and Comparison with \textit{Planck} results}
\label{subsec:amplitudes}

\begin{figure*}
    \centering
    \includegraphics[width=.99\textwidth,trim={3.5cm .5cm 2.5cm 1cm},clip]{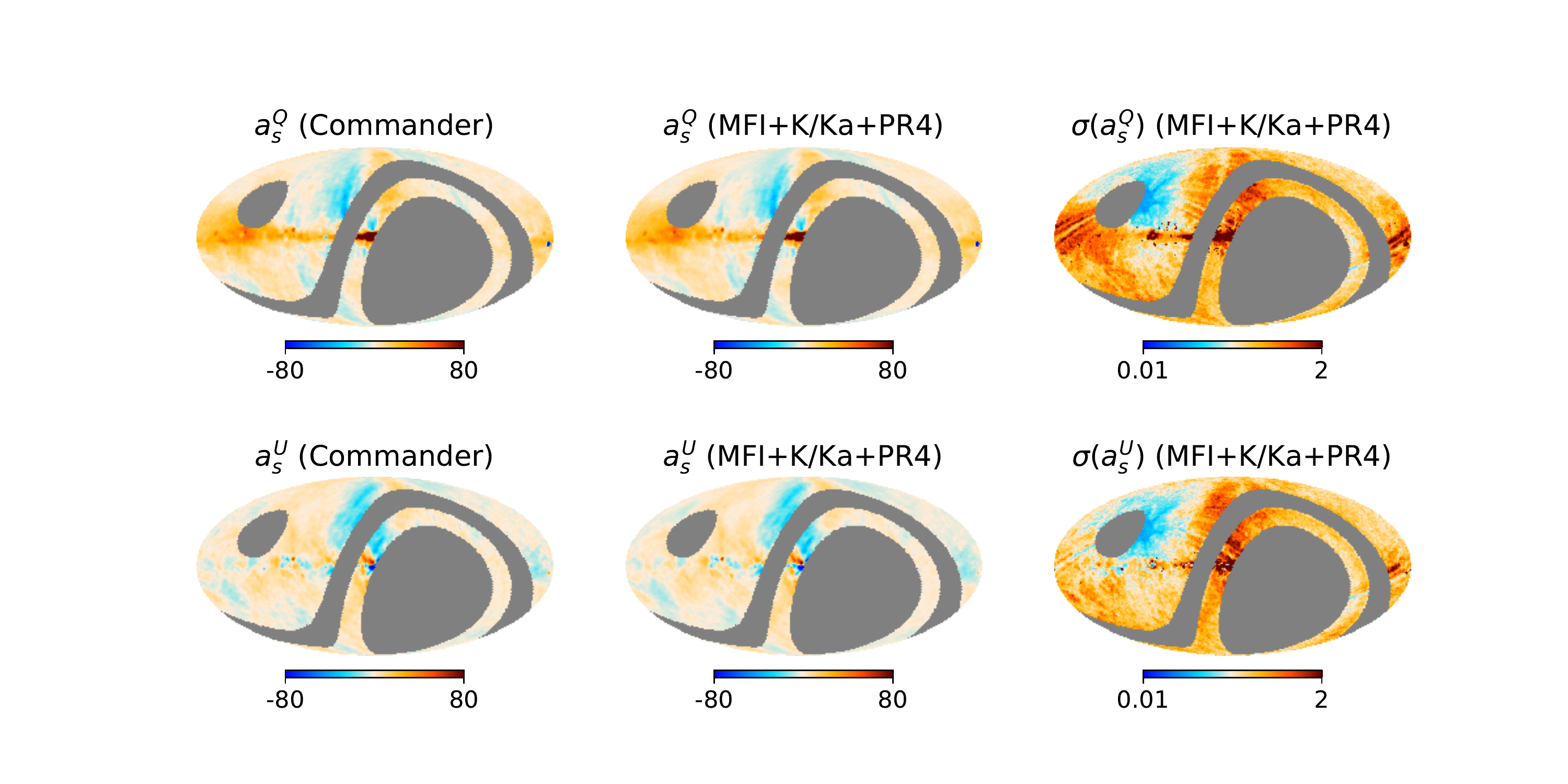}
    \caption{Left column: \textsc{Commander} $Q$ (top) and $U$ (bottom) synchrotron amplitude maps at 30\,GHz at $N_{\rm side}$ = 64, smoothed with a Gaussian beam to a final resolution of $\mathrm{FWHM} = 2^{\circ}$. Centre column: our estimate of the synchrotron amplitude at 30\,GHz, using the MFI+K/Ka+PR4 dataset. Right column: uncertainty of the estimated synchrotron amplitude. Maps are in antenna temperature ($\mu$K).}
    \label{fig:a_s_QKNPIPE_Commander}
\end{figure*}

\begin{figure*}
    \centering
    \includegraphics[width=.99\textwidth,trim={3.5cm .5cm 2.5cm 1cm},clip]{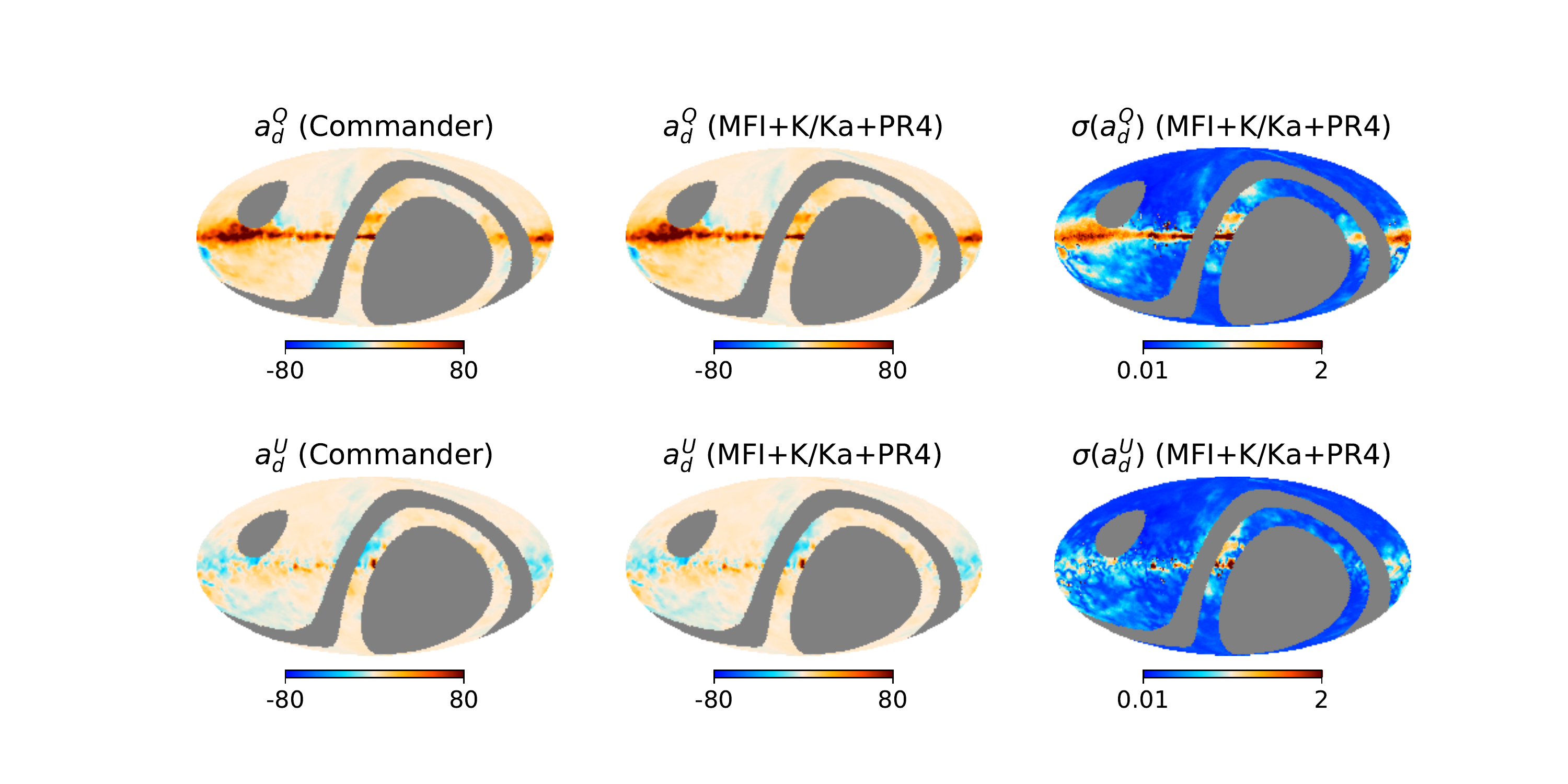}
     \caption{Left column: \textsc{Commander} $Q$ (top) and $U$ (bottom) thermal dust amplitude maps at 353\,GHz at $N_{\rm side}$ = 64, smoothed with a Gaussian beam to a final resolution of $\mathrm{FWHM} = 2^{\circ}$. Centre column: our estimate of the thermal dust amplitude at 353\,GHz,  using the MFI+K/Ka+PR4 dataset. Right column: uncertainty of the estimated thermal dust ampltitude. Maps are in antenna temperature ($\mu$K).}
     \label{fig:a_d_QKNPIPE_Commander}
\end{figure*}

We have compared our baseline results, i.e., using the MFI+K/Ka+PR4 dataset and a power law as the synchrotron model, to those obtained from the \textsc{Commander} pipeline \citep{Commander} applied to PR4 data\footnote{Data available at NERSC under \texttt{/cmb/daa/planck20}.}. We have only considered this pipeline among those used by \textit{Planck}, since it is the reference method with regard to the recovery of foreground components. In Figs.~\ref{fig:cmb_QKNPIPE_Commander} to \ref{fig:a_d_QKNPIPE_Commander}, we show a comparison of the CMB, the synchrotron emission at 30\,GHz, and the thermal dust emission at 353\,GHz  between \textsc{Commander} and our results. In order to perform a direct comparison we have filtered \textsc{Commander} results with FDEC. The left column shows the $Q$ and $U$ \textsc{Commander} amplitudes, the center column our amplitudes and the right column the corresponding uncertainties. A visual inspection shows that both estimates are very similar, especially the synchrotron and thermal dust emissions which are the dominant contributions in polarization.

\begin{figure*}
    \centering
    \includegraphics[width=\textwidth]{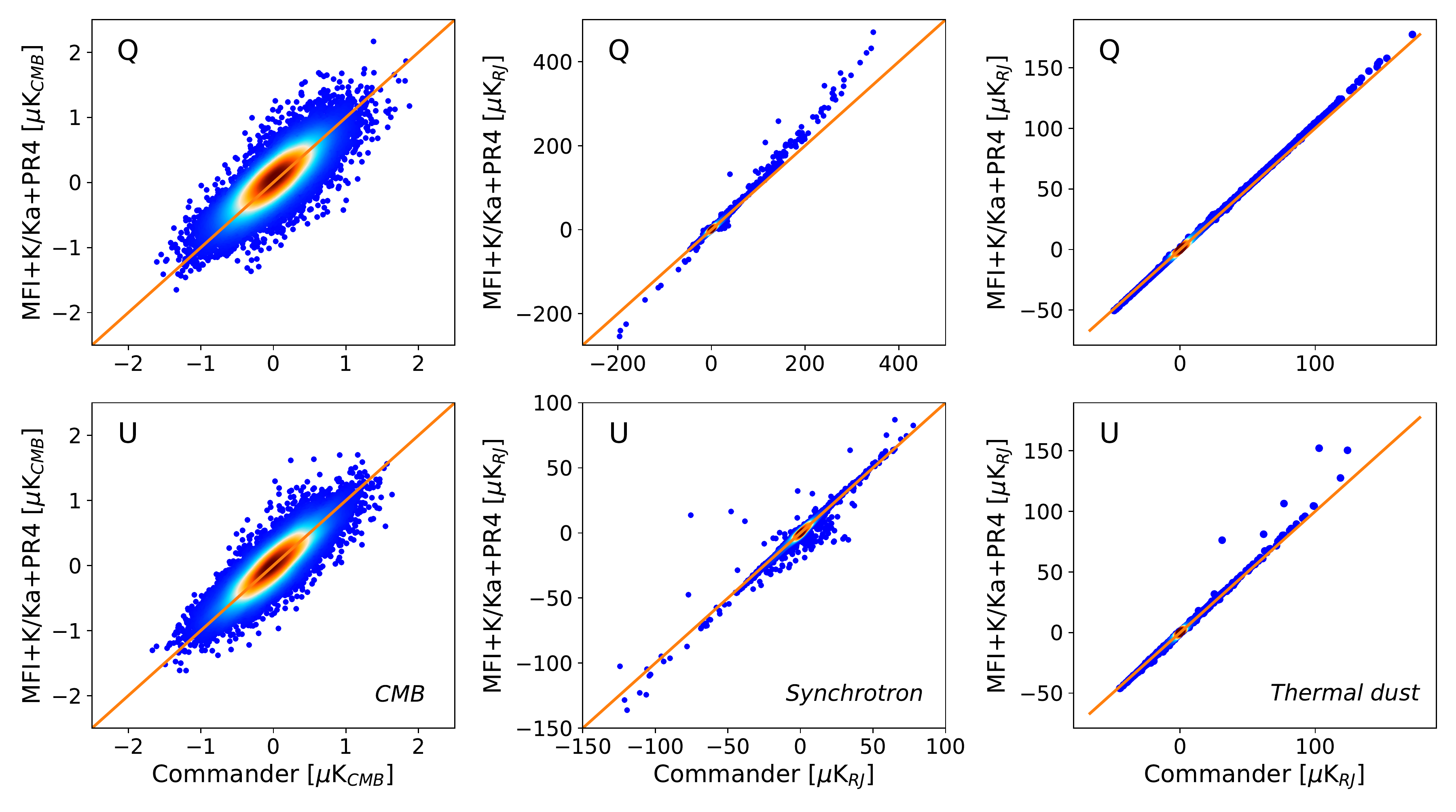}
    \caption{Comparison of CMB (left), synchrotron at 30\,GHz (centre) and thermal dust at 353\,GHz (right) amplitudes recovered using the MFI+K/Ka+PR4 dataset and the ones obtained by \textsc{Commander} using PR4 data. The correlation factors are $\rho^Q = 0.543$ and $\rho^U = 0.817$ (CMB), $\rho^Q = 0.992$ and $\rho^U = 0.973$ (synchrotron) and $\rho^Q = 1.000$ and $\rho^U = 0.997$ (thermal dust).}
    \label{fig:tt_comparison}
\end{figure*}

\begin{figure}
    \centering
    \includegraphics[width=.43\textwidth,trim={2cm 1.4cm 1.7cm 1cm},clip]{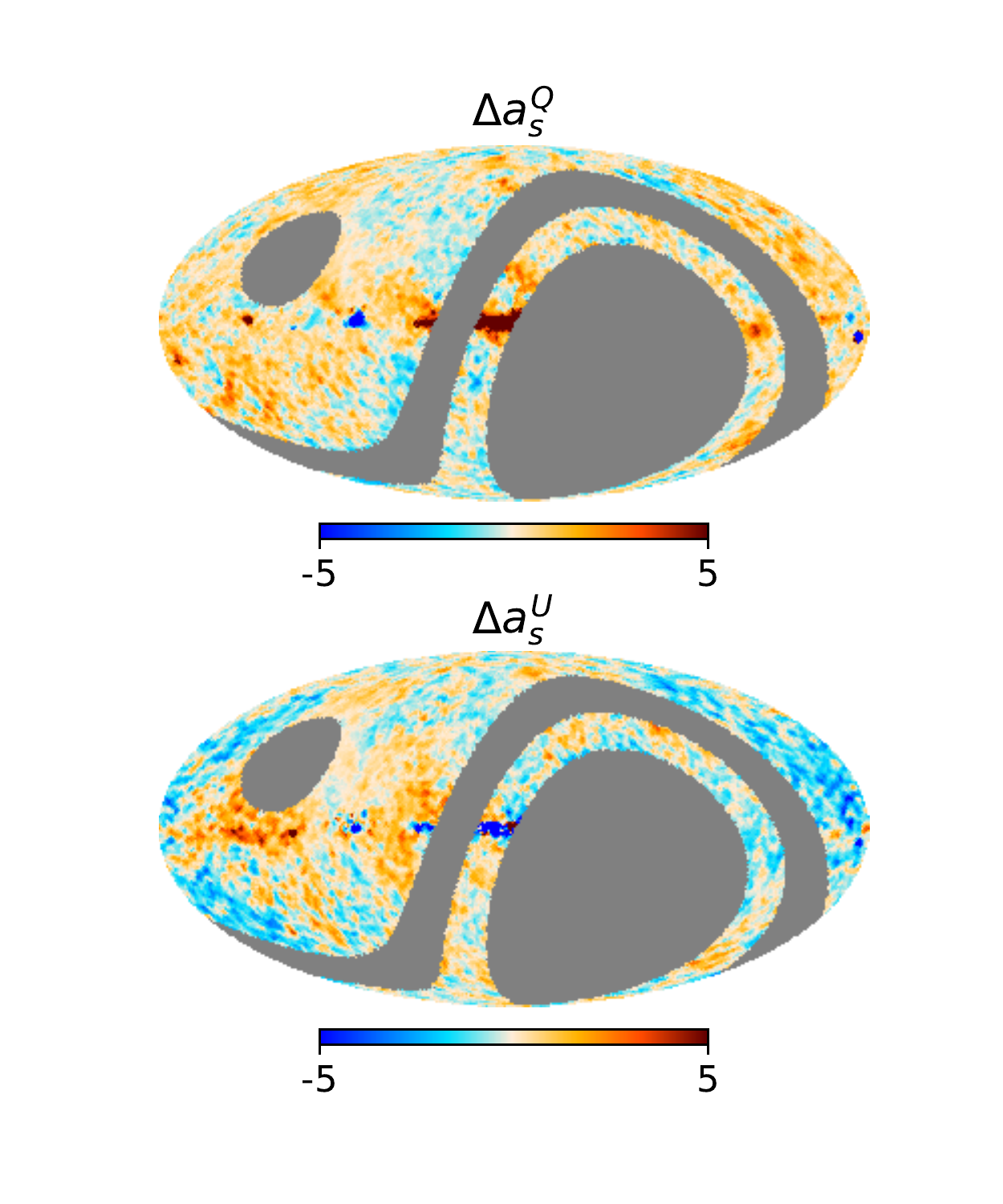}
    \caption{Difference between the synchrotron amplitude $a_s^Q$ ($a_s^{U}$) obtained with the MFI+K/Ka+PR4 and the \textsc{Commander} estimate, top row (bottom row). Maps are in antenna temperature ($\mu$K).}
    \label{fig:difference_a_s_maps_Commander_QKNPIPE}
\end{figure}

\subsubsection{CMB}
\label{subsec:cmb_comparison_Commander}

Regarding CMB, the left column of Fig~\ref{fig:tt_comparison} shows the pixel-to-pixel comparison for the recovered CMB map from our analysis and from \textsc{Commander} both in $Q$ and $U$. We have applied a combination of the QUIJOTE observed sky and the common polarization confidence mask provided by the \textit{Planck} Collaboration\footnote{\label{note:common_mask}Available at \url{https://pla.esac.esa.int/\#maps}.} \citep{Planck2018_IV}. 

We observe from the maps that there is a discrepancy. We found that the application of the FDEC filter, before the component separation process, leads to a decrease of the amplitude in the power spectra of our recovered CMB map. This power reduction appears only when \textit{Planck} and WMAP are filtered with FDEC, since the CMB information is extracted mainly from those channels. Instead of applying the FDEC filter, one could apply a filter that suppresses the large scales. This would be equivalent to applying a linear function to the CMB and there would not be a reduction of power. However, since we want to study all scales we decided to apply the FDEC filter. Since the aim of this work is the study of the foregrounds, we keep the results obtained with all the data filtered with FDEC to recover the $\beta_s$ map without any bias. One can in principle recover the unbiased CMB following one of the approaches described below:
\begin{itemize}
    \item Perform the component separation analysis without filtering the data with FDEC and including the FDEC correction in QUIJOTE-MFI data as part of the model; or
    \item Given the unbiased $\beta_s$ map\footnote{Obtained in the component separation analysis using the data filtered with FDEC.} and \textit{Planck} data, one can construct a template with the modes that QUIJOTE-MFI data is missing after being filtered with FDEC. Then perform the analysis with the reconstructed QUIJOTE-MFI maps. 
\end{itemize}
Since the estimation of the CMB is out of the scope of this paper, we leave this analysis for future works.

\subsubsection{Synchrotron}
\label{subsec:synch_comparison_Commander}

Fig.~\ref{fig:difference_a_s_maps_Commander_QKNPIPE} shows the difference between the synchrotron amplitude maps obtained using the MFI+K/Ka+PR4  and the \textsc{Commander} reconstruction using the PR4 data. The largest differences observed are located in the Galactic plane where the model fails to reproduce the sky signal. We also observe large scale structures in the difference map. These structures can originate from the fact that we have obtained a more accurate estimation of the scaling law as our fit is performed using additional frequencies. However, overall, the correlation between both methods is very good.  

This can also be seen in the centre column of Fig.~\ref{fig:tt_comparison}, where a pixel-to-pixel comparison is given, showing that both methods present a synchrotron amplitude at 30\,GHz highly correlated for $Q$ and $U$ except in some pixels where the synchrotron emission is very large. Those pixels are located primarily in the Galactic plane. These discrepancies are likely to arise from differences in the amplitude of the polarised intensity instead of from differences in the polarization angles. In Fig.~\ref{fig:tt_comparison} we observe that both the slopes, in the Q and U plots, are higher than unity. If the discrepancies were originated from differences in the polarization angle,  one slope would be higher than unity and the other lower.

\subsubsection{Dust}
\label{subsec:dust_comparison_Commander}

Regarding thermal dust emission, this foreground strongly dominates the 353\,GHz \textit{Planck} frequency map and, therefore, the recovered amplitude is very much determined by this channel. This was also the case in the \textsc{Commander} analysis done by the \textit{Planck} Collaboration. Thus, our recovered $Q$ and $U$ components of the thermal dust are strongly correlated with those obtained using \textsc{Commander}, see the right column of Fig.~\ref{fig:tt_comparison}.

\begin{figure*}
    \centering
    \includegraphics[width=.9\textwidth,trim={0cm 0cm 0cm 0cm},clip]{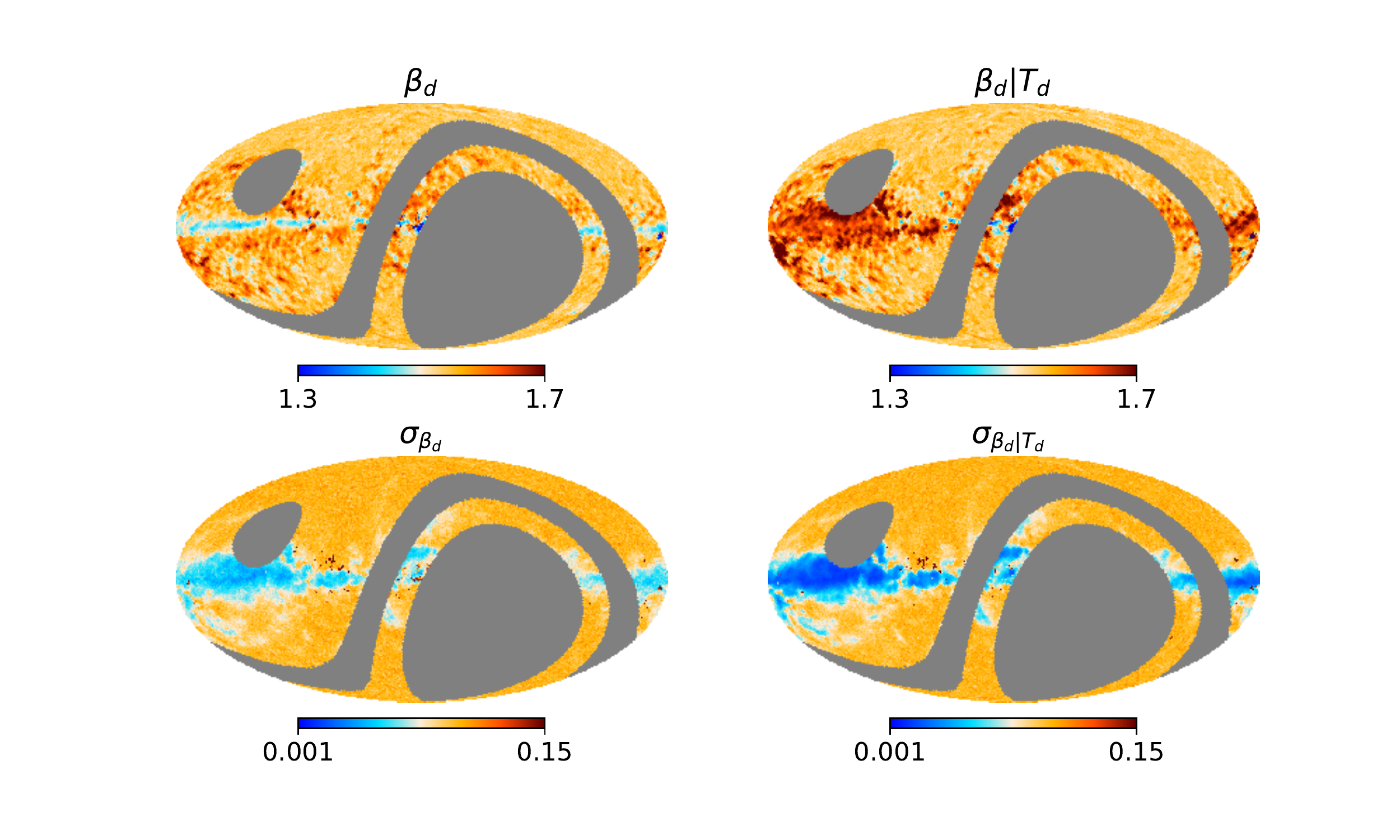}
    \caption{Left column: estimate (top) and uncertainty (bottom) of thermal dust spectral index obtained when $T_d$ is included as a model parameter. Right column: estimate (top) and uncertainty (bottom) of thermal dust spectral index obtained when the $T_d$  template obtained by \textsc{Commander} in the intensity analysis is used to fix $T_d$ in the component separation process.}
    \label{fig:beta_d_fixed_T}
\end{figure*}

\subsection{Dust Spectral Parameters}
\label{subsec:dust_spectral_index}

\begin{figure}
    \centering
    \includegraphics[width=.41\textwidth,trim={2cm 1.35cm 1.5cm 1cm},clip]{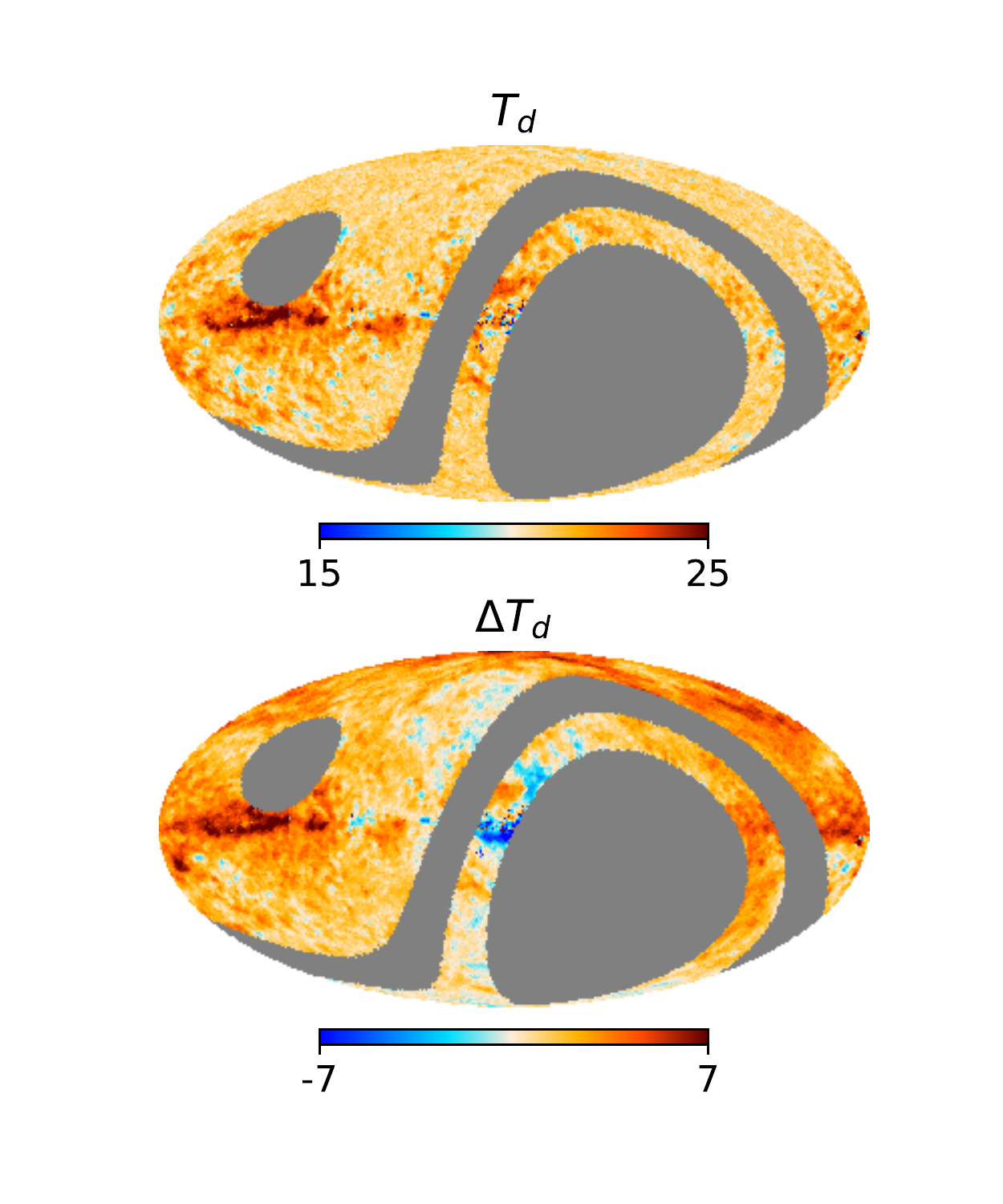}
    \caption{Top row: thermal dust temperature map recovered in the default case. Bottom row: difference map between the top row map and the $T_d$ template used in the analysis. Maps are in Kelvin.}
    \label{fig:T_d_comparison_commander}
\end{figure}

Although the frequencies of QUIJOTE-MFI do not overlap with the spectral range where the thermal dust is more dominant, we have studied whether the inclusion of this data set in the analysis can help with the thermal dust characterization due to an improvement on the determination of the rest of the polarized foreground parameters. Fig.~\ref{fig:beta_d_fixed_T} shows the thermal dust spectral index $\beta_d$ recovered with the default data set, modelling the synchrotron emission  as a power law, in two cases:
\begin{itemize}
    \item $T_d$ is included as a model parameter.
    \item $T_d$ is fixed to \textsc{Commander}'s estimation of the thermal dust temperature from the component separation analysis in intensity \citep{Planck2016_X} like \textsc{Commander} did in their polarization analysis. Fixing $T_d$ helps breaking its degeneracy with $\beta_d$ in the Rayleigh-Jeans part of the thermal dust spectrum, which is the one observed with \textit{Planck} in polarization.
\end{itemize}
\begin{figure}
    \centering
    \includegraphics[width=.41\textwidth,trim={2cm 1.35cm 1.5cm 1cm},clip]{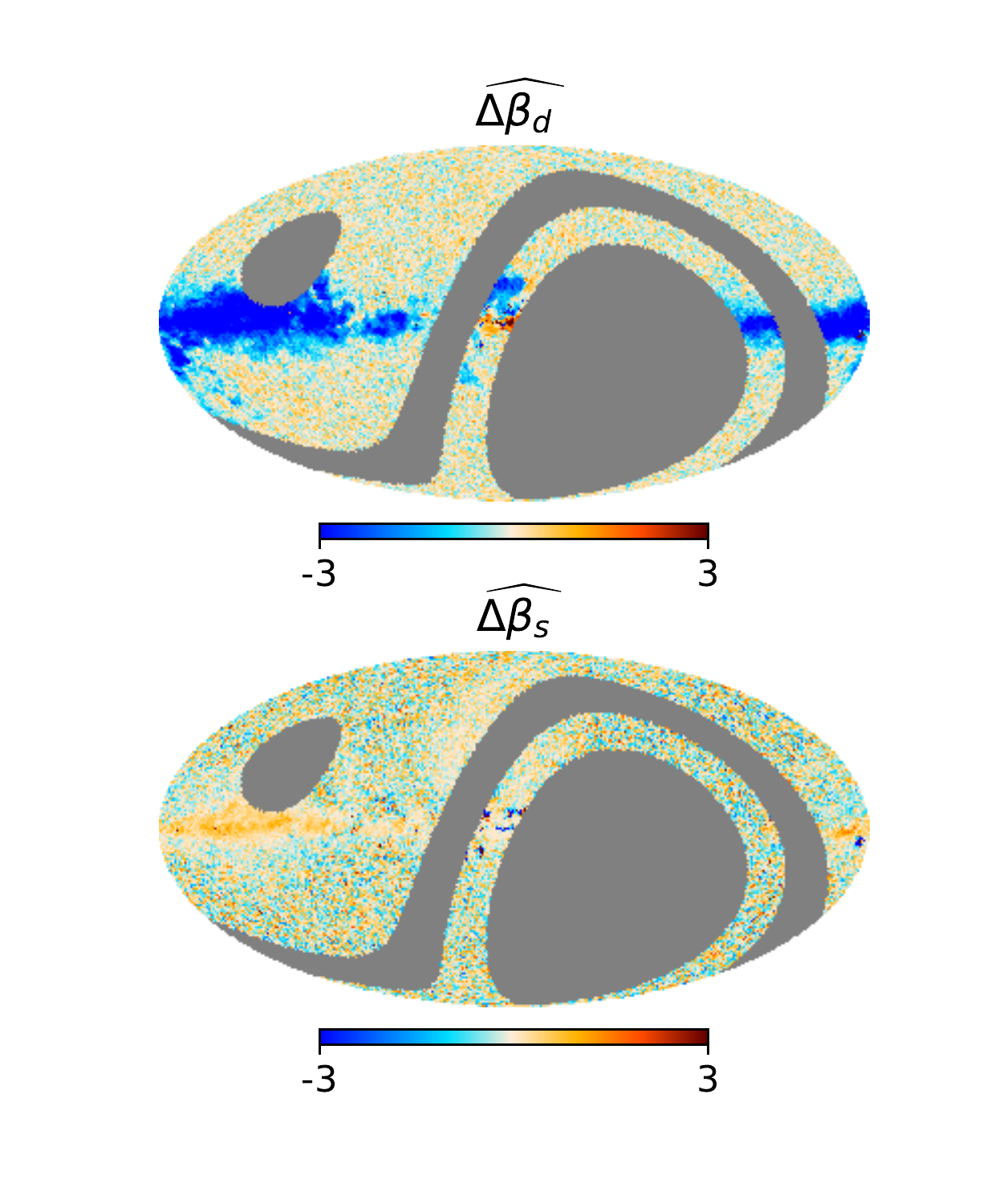}
    \caption{$\beta_d$ (top row) and $\beta_s$ (bottom row) relative difference map between the maps obtained when we include $T_d$ as a model parameter and when we fix it.}
    \label{fig:relative_difference_spectral_par}
\end{figure}
In both maps we find that the recovered $\beta_d$ values are close to the expected value of the prior, i.e., 1.55, except close to the Galactic plane where the thermal dust signal is larger\footnote{Notice that the uncertainty does not improve in the regions where the $\beta_d$ values are close to the mean value of the prior when we fix one parameter. The uncertainty in those pixels is the spread of the prior.}. The results differ significantly along the Galactic plane, see Fig.~\ref{fig:relative_difference_spectral_par}. This difference originates since our recovered $T_d$ map does not resemble the used $T_d$ template as shown in Fig.~\ref{fig:T_d_comparison_commander}. We remark that although in the first case $T_d$ is estimated from the polarization analysis, the $T_d$ recovered values lie close to the expected value of the prior (22 K) except along the Galactic plane where the fit is not good. Moreover, it is very difficult to fit $T_d$ from polarization data only, as the highest frequency is 353 GHz, and thus we are not able to trace the thermal dust peak.

\begin{figure}
    \centering
    \includegraphics[width=.41\textwidth,trim={2cm 1.35cm 1.5cm 1cm},clip]{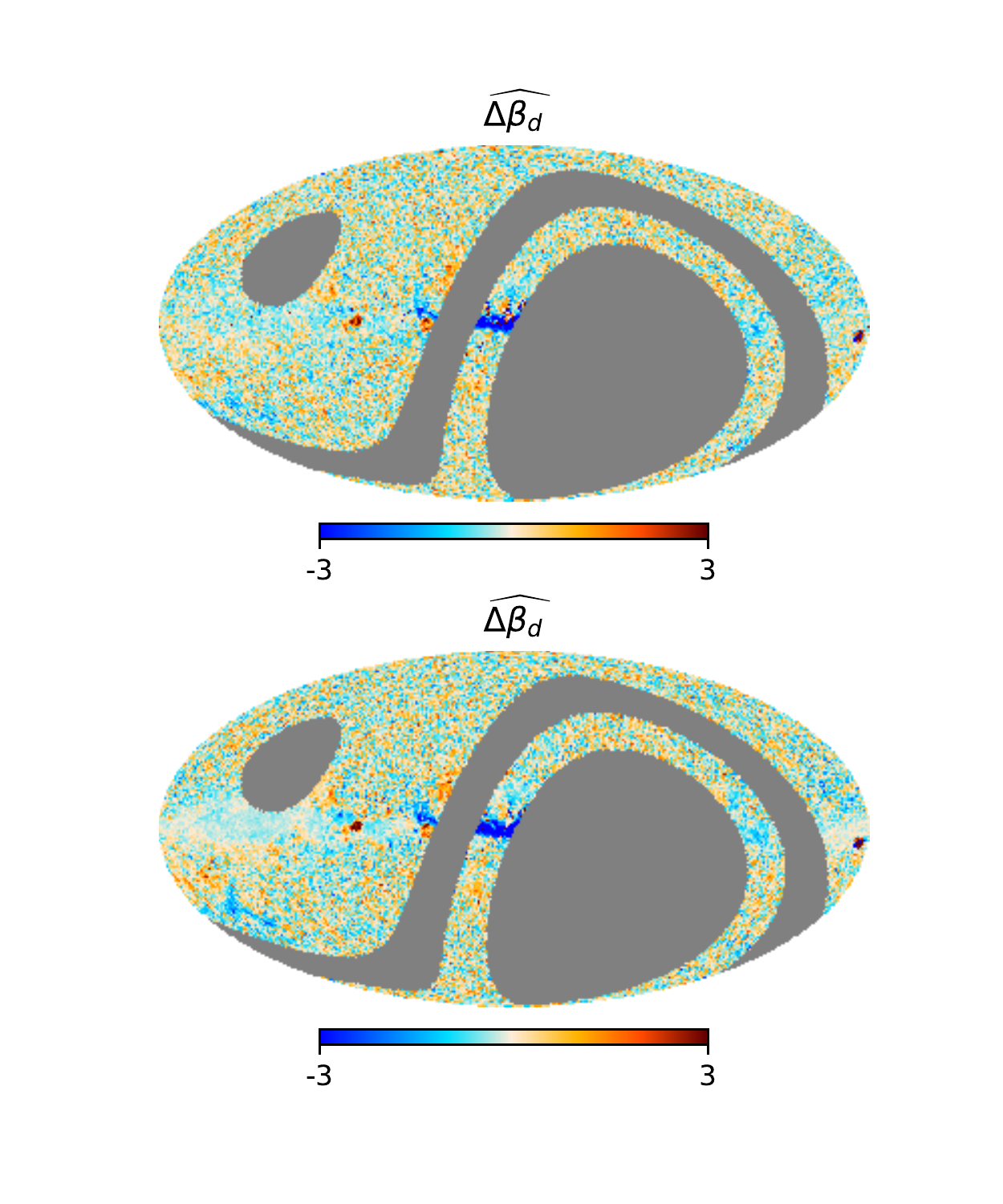}
    \caption{$\beta_d$ relative difference map between the map obtained using the MFI+K/Ka+PR4 and the one obtaiened with K/Ka+PR4 datasets when we include $T_d$ as a model parameter  (top row) and when we fix it (bottom row).}
    \label{fig:relative_difference_beta_d_KN_QKN}
\end{figure}

In Fig.~\ref{fig:relative_difference_spectral_par} we show the relative difference between spectral index map of the thermal dust and synchrotron obtained when $T_d$ is included as a model parameter and when it is fixed. The relative difference is calculated as follows:
\begin{equation}
    \widehat{\Delta\beta}_{1,2} = \dfrac{\beta_1-\beta_2}{\sqrt{\sigma^2_{\beta_1} + \sigma^2_{\beta_2} - 2\sigma_{\beta_1,\beta_2}}} \, ,
    \label{eq:relative_difference}
\end{equation}
where $\sigma^2_{\beta_1}$ ($\sigma^2_{\beta_2}$) is the variance of the $\beta_1$ ($\beta_2$) map, and $\sigma_{\beta_1,\beta_2}$ is the covariance between the $\beta_1$ and $\beta_2$ maps that are being compared. As expected from Fig.~\ref{fig:beta_d_fixed_T} the differences close to the Galactic plane are significantly large in the case of $\beta_d$. On the other hand, we find that, the $\beta_s$ maps recovered in both cases are compatible and the differences resemble Gaussian noise except along the Galactic plane where the model fails. 

We also studied the relative difference between the $\beta_d$ map obtained with the MFI+K/Ka+PR4 and K/Ka+PR4 data sets in Fig.~\ref{fig:relative_difference_beta_d_KN_QKN}. The top panel  shows the relative difference when $T_d$ is included as a model parameter and the bottom panel when $T_d$ is fixed. We observed that both maps are compatible except in regions where the fit is not good. Moreover, when we compare the uncertainty maps we find that there is not a significant improvement when we include QUIJOTE-MFI channels. Thus, we conclude that the improvement in the characterization of low-frequency foregrounds does not help necessarily with the estimation of thermal dust spectral parameters.

\begin{figure}
    \centering
    \includegraphics[width=.47\textwidth,trim={.5cm .4cm .5cm .5cm},clip]{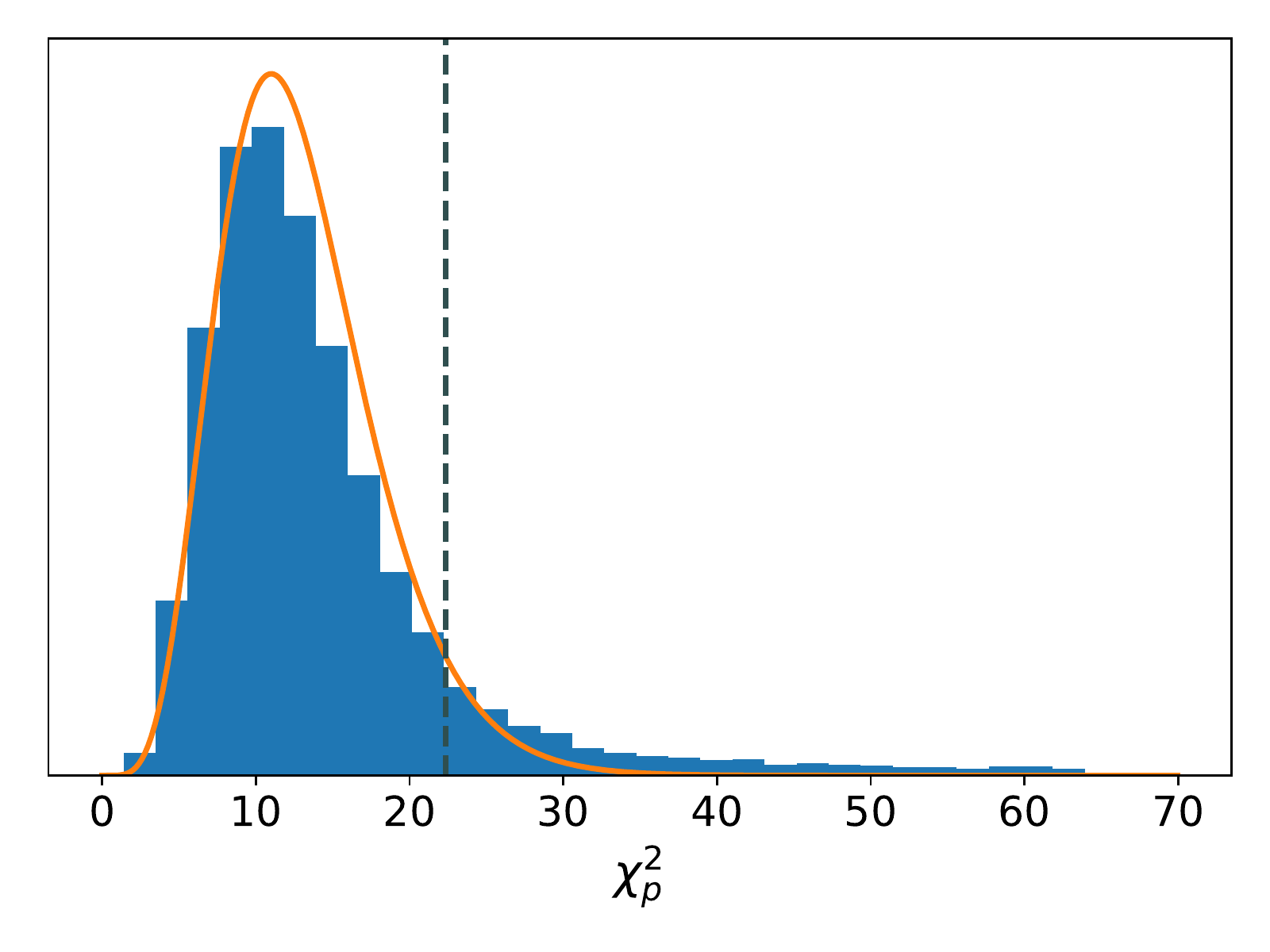}
    \caption{$\chi^2_p$ distribution obtained using the default dataset. The orange curve shows the theoretical $\chi^2$ probability density function with $N_{\rm{dof}} =13$. The area to the left of the gray dashed line shows values within the 95\% confidence region.}
    \label{fig:chi2_distribution_default}
\end{figure}

\begin{figure*}
    \centering
         \begin{subfigure}[b]{0.33\textwidth}
         \centering
         \includegraphics[width=1\linewidth,trim={1.75cm 1.75cm 1cm 1.25cm},clip]{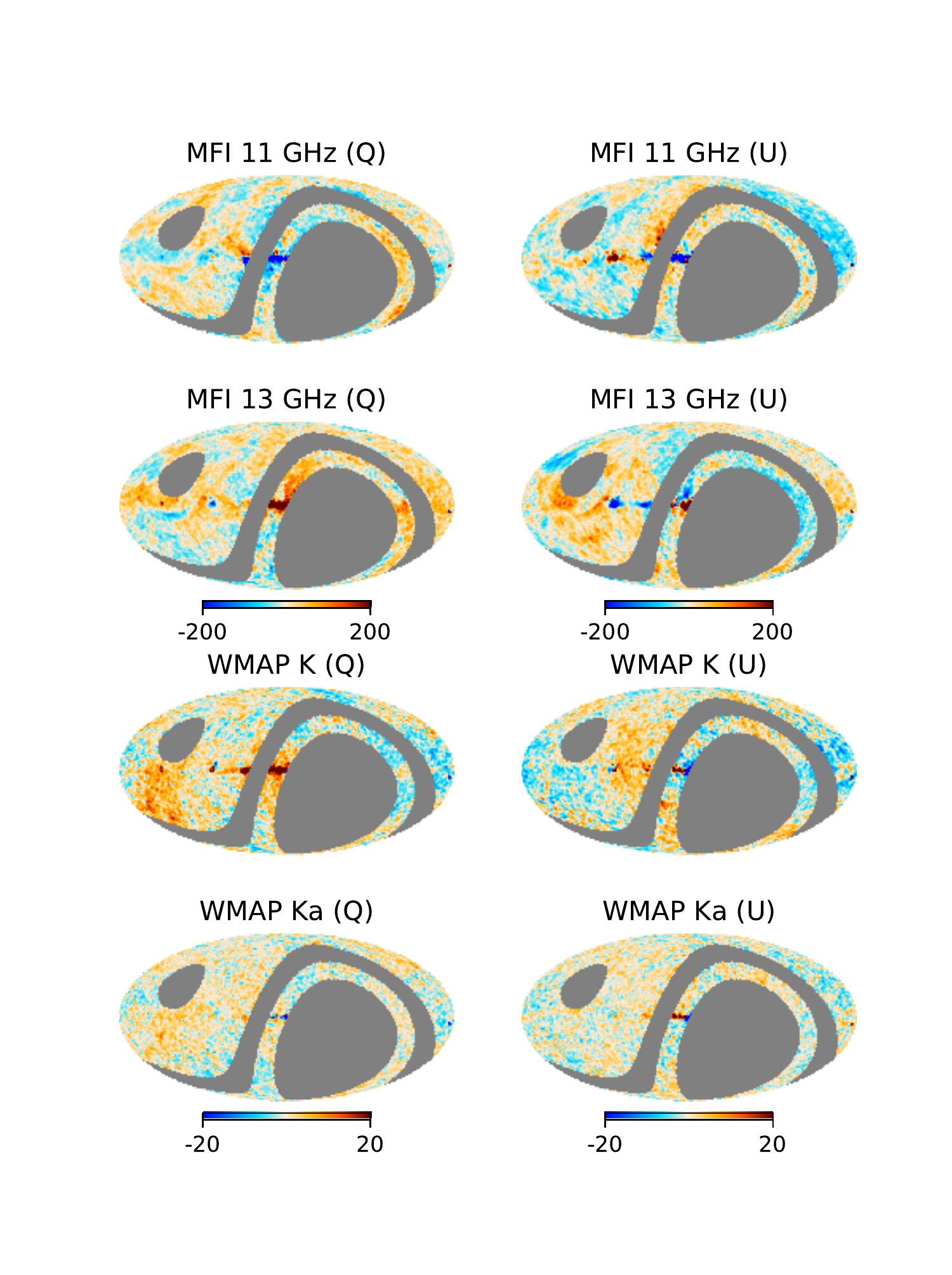}
     \end{subfigure}
     \hfill
     \begin{subfigure}[b]{0.33\textwidth}
         \centering
         \includegraphics[width=1\linewidth,trim={1.75cm 0.25cm 1cm 1.25cm},clip]{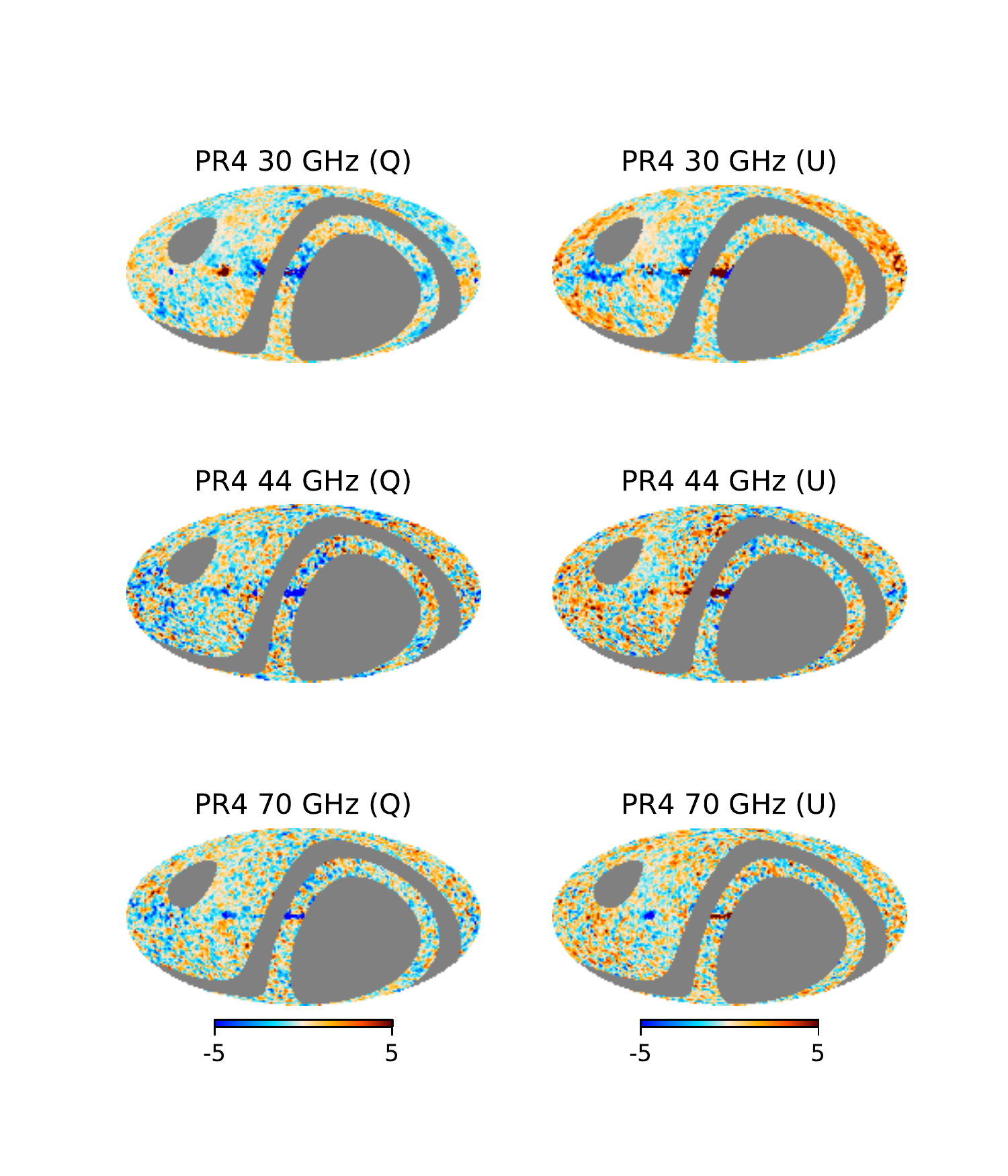}
     \end{subfigure}
     \hfill
     \begin{subfigure}[b]{0.33\textwidth}
         \centering
         \includegraphics[width=1\linewidth,trim={1.75cm 1.75cm 1cm 1.25cm},clip]{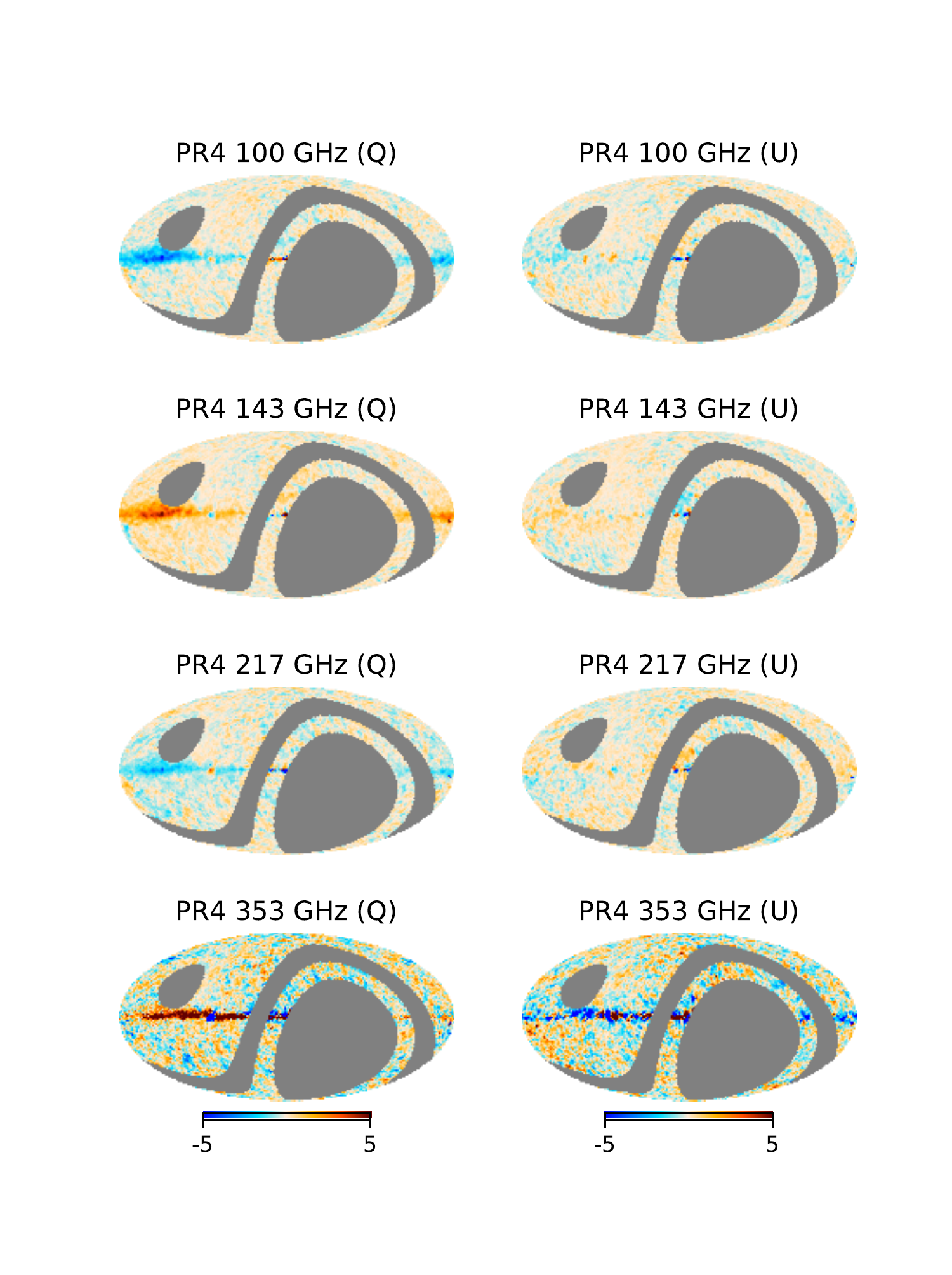}
     \end{subfigure}
    \caption{$Q$ and $U$ residual maps for each frequency channel at $N_{\rm{side}} = 64$. Maps are displayed in thermodynamic temperature ($\mu$K).}
    \label{fig:residuals_maps}
\end{figure*}

\subsection{Goodness of fit}
\label{subsec:goodness_of_the_fit}

In this Section we study in depth the quality of the results obtained using the default dataset. In Section~\ref{subsubsec:chi2_distribution} we analyze the $\chi^2$ distribution of the results as well as the $Q$ and $U$ residuals of each channel. Section~\ref{subsubsec:robustness_priors} investigates the robustness of our results regarding the estimation of the synchrotron spectral index with respect to the prior applied to this parameter.

\subsubsection{$\chi^2$ distribution and residuals}
\label{subsubsec:chi2_distribution}

We have studied the pixel $\chi^2$ distribution obtained from the fit using MFI+K/Ka+PR4 (see Fig.~\ref{fig:chi2_distribution_default}):
\begin{equation}
    \chi^2_{p} = N_{\rm{dof}}\cdot\chi^2_{\mathrm{red}} \, .
    \label{eq:chi2}
\end{equation}
Moreover, we have also calculated the residuals per channel involved in the analysis:
\begin{equation}
    r_{p,\nu}=\dfrac{(d_{p,\nu} - S_{p\nu})}{\sigma_{p,\nu}} \, .
    \label{eq:residuals}
\end{equation}
In the perfect scenario, residuals maps are consistent with instrumental noise alone. Therefore, they are a valuable tool to look for either systematic effects or mismatches in the foreground modelling.

First of all, we recall that the number of d.o.f. for this analysis is 13 (11 channels $\times$ 2 ($Q$ and $U$) minus 9 free parameters). We find $\langle\chi^2_{p}\rangle=14.3$ and $\sigma=8.9$ slightly larger than what is expected for the theoretical number of d.o.f.. Fig.~\ref{fig:chi2_distribution_default} shows that the $\chi^2_p$ values follow a $\chi^2$-like distribution, whose peak lies close to $N_{\rm{dof}}=13$. However, there is an excess of pixels at large values of $\chi^2$ with respect to the $\chi^2_{N_{\rm{dof}}}$-distribution. That excess appears since there are pixels where the model is not able to track the true sky emission, mainly in the Galactic plane. Thus, those pixels are highly inconsistent with this $\chi^2$ distribution. 

Fig.~\ref{fig:residuals_maps} shows the $Q$ and $U$ residuals maps of every frequency channel from the MFI+K/Ka+PR4 dataset. We find that  \textit{Planck} and WMAP residuals maps are reasonably consistent with the expected noise except along the Galactic plane. The residuals in this region are a consequence of an incorrect modelling of the sky as we saw in the $\chi^2_{\rm{red}}$ maps. For the MFI channels we observe that the largest residuals are located in compact regions along the Galactic plane. We observe in the 11\,GHz $U$ channel a redder region in the NPS's closest part to the Galactic centre. This region overlaps with the area where we obtain a better goodness-of-fit if Faraday Rotation effects are taken into account, see Fig.~\ref{fig:diff_chi2_FR} in Appendix~\ref{sec:appendix_FR}. Furthermore, artefacts that resemble the FDEC morphology are present in MFI 13\,GHz.

In light of these tests, we are confident of the results obtained in those pixels, that are properly modelled by our assumed parametric model. The pixels outside the confidence region are located mainly in the Galactic plane, probably because our model fails to account for the complexity of this region. It would be convenient to study these regions in more detail with more complex models. However, the aim of this work is to study the diffuse components and the study of specific regions has been conducted in other works \citep{NPSwidesurvey,FANwidesurvey,hazewidesurvey}.

\begin{figure*}
    \centering
    \includegraphics[width=1\textwidth,trim={1.5cm 1cm 2.5cm 1cm},clip]{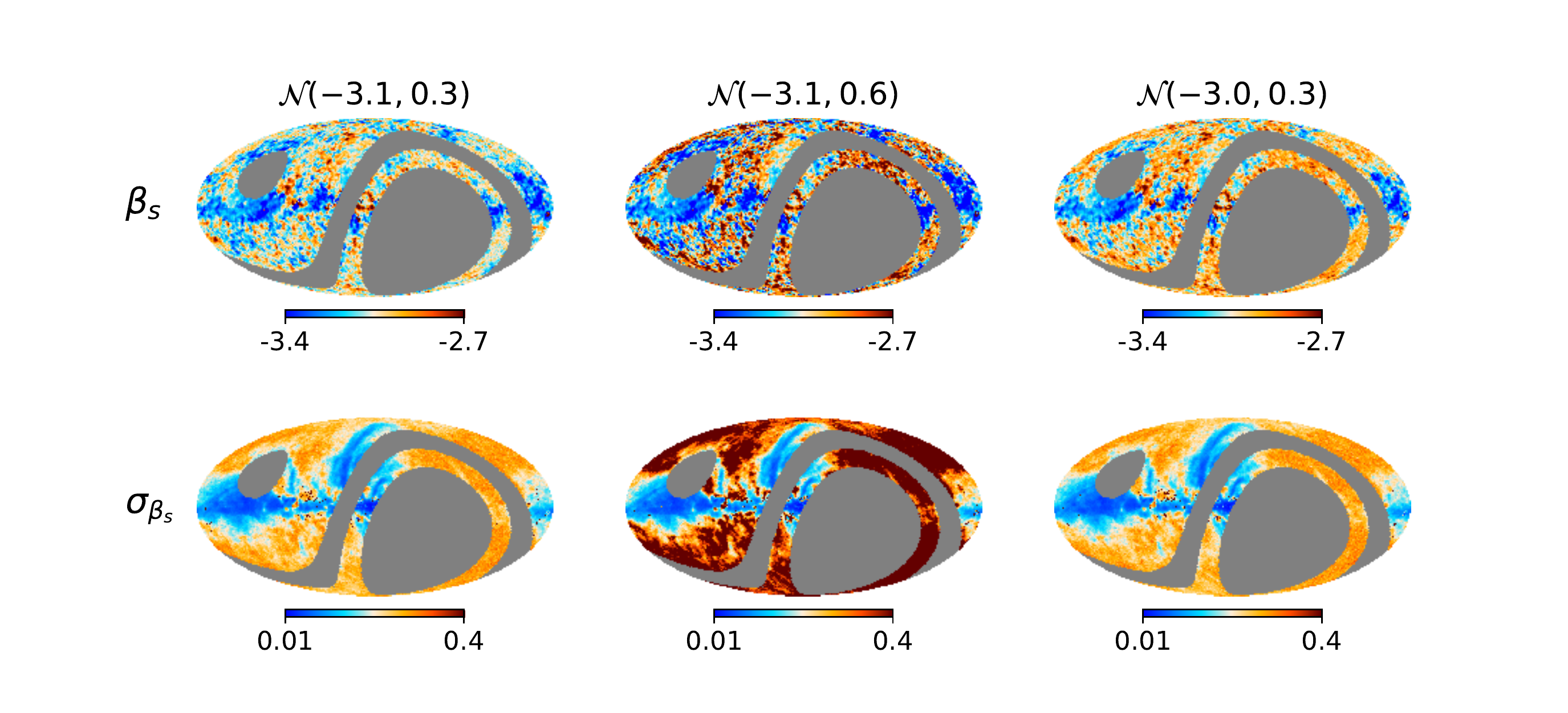}
    \caption{Synchrotron spectral index estimate (top row) and uncertainty (bottom row) obtained using different Gaussian prior distributions and the default dataset (MFI+K/Ka+PR4). The synchrotron emission is modelled as a power law.}
    \label{fig:beta_s_priors}
\end{figure*}

\begin{figure}
    \centering
    \includegraphics[width=.43\textwidth,trim={2cm 1.25cm 1.5cm 1cm},clip]{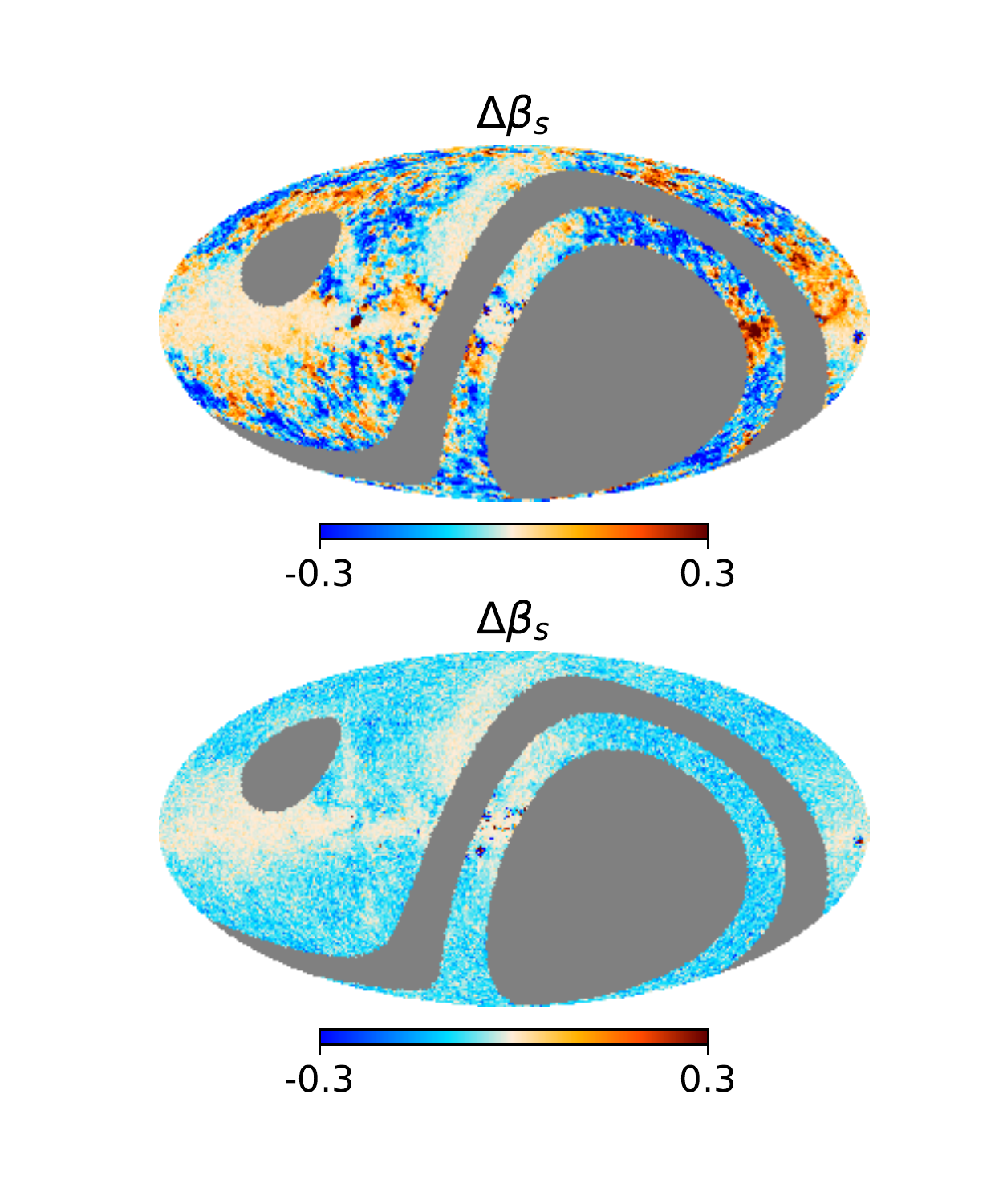}
    \caption{Difference map between the estimated $\beta_s$ using the default prior, i.e., $\mathcal{N}\mleft(-3.1,0.3\mright)$, and the one obtained using an alternative prior, see Fig.~\ref{fig:beta_s_priors}. Top:  $\mathcal{N}\mleft(-3.1,0.6\mright)$ Bottom: $\mathcal{N}\mleft(-3.0,0.3\mright)$.}
    \label{fig:diff_priors_beta_s}
\end{figure}

\subsubsection{Robustness  with respect to the prior}
\label{subsubsec:robustness_priors}

As previously stated, the use of prior information is essential in Bayesian analysis, helping with convergence and computational time reduction. Besides, when the data do not have enough sensitivity, i.e., there is not enough information to obtain a reliable estimation of the spectral index, the prior tends to provide a value close to the mean value of the distribution. In other words, a conservative value is assigned to the spectral index in those pixels. Thus, in order to detect which pixels are prior-dependent we have also performed component separation using two additional Gaussian priors on $\beta_s$,  $\mathcal{N}\mleft(-3.1,0.6\mright)$ and $\mathcal{N}\mleft(-3.0,0.3\mright)$. The $\beta_s$ estimation and uncertainty maps with these new priors are shown in Fig.~\ref{fig:beta_s_priors} together with those obtained with the prior used in the default analysis (left columns).

Comparing the results using the default prior, i.e., $\mathcal{N}\mleft(-3.1,0.3\mright)$, versus a less restrictive prior, i.e., $\mathcal{N}\mleft(-3.1,0.6\mright)$, we observe that the uncertainty on the recovered $\beta_s$ increases at the prior-dominated pixels. On the other hand, in those regions where the synchrotron emission is very intense the uncertainty remains the same. Likewise, the estimated $\beta_s$ in the latter pixels are very similar whereas the other pixels are visually different. The $\beta_s$ distribution of the pixels outside the low-uncertainty regions are compatible with the prior distribution. This is the reason why the estimated values are different and the spread is larger when the prior is relaxed.

When we use a prior with a different expected value, i.e., $\mathcal{N}\mleft(-3.0,0.3\mright)$, but equal standard deviation we obtain a similar uncertainty map. The estimated $\beta_s$ is almost the same in the low-uncertainty regions, i.e., the high-intensity synchrotron regions. However, a flatter spectrum (closer to $-3.0$ instead of $-3.1$) is recovered outside those areas. This is more evident from the bottom panel of Fig.~\ref{fig:diff_priors_beta_s} where the difference between the $\beta_s$ map estimated with the default prior and the $\mathcal{N}\mleft(-3.0,0.3\mright)$ prior is shown. Outside the regions where the synchrotron emission is the largest, the difference is close to $-0.1$ which is the difference between the expected value of the priors. In other words, when there is not enough information from the data the recovered $\beta_s$ is close to the expected value of the prior. This is an advantage of using prior information, since it assigns a conservative value to the spectral index instead of unphysical values or simply failing to perform the fit.

\section{Conclusions}
\label{sec:conclusions}

In this work, we have presented the component separation products in polarization obtained from combining the QUIJOTE-MFI data at 11 and 13\,GHz, with the WMAP K and Ka bands and all \textit{Planck} polarized channels. We have seen that the inclusion of the QUIJOTE-MFI data is crucial to improve the parameter estimation of the low frequency foregrounds, in particular for the estimation of the synchrotron spectral index. 

We have obtained the first detailed $\beta_s$ map of the Northern Celestial Hemisphere at a scale of 2$^{\circ}$ assuming the synchrotron emission is modelled as a power law. This model represents well the data except in the Galactic plane where the physics might be more complex. We find, using the pixels whose $\chi^2_{\rm{red}}$ lies within the 95\% confidence region, an average value of $-3.08$ and a dispersion of $0.13$. The latter is broader than the dispersion of commonly used $\beta_s$ templates. Moreover, we have found that the spectral index is not compatible with a uniform value, i.e., there are statistical significant differences of $\beta_s$ across the observable sky.
 
We have also modelled the synchrotron emission as a power law with curvature. The pixel-based analysis of the curvature shows that $c_s$ is only detected in some regions in the Galactic plane where the fit is bad. When we assume a model with uniform curvature in RC1 (the region that includes all pixels whose $\chi^2_{\rm{red}}$ is within the  95\% confidence region for the power law with curvature model) we found a $c_s = -0.0797 \pm 0.0012$. We found that both models, i.e., power law and power law with uniform curvature, provide a good fit given the available data. However there is not enough statistical significance to distinguish which model is better. A more thorough study is left for future work. 

We found that our recovered synchrotron and thermal dust maps are highly correlated with the maps presented by the \textit{Planck} collaboration using \textsc{Commander}, even though we found some large scale difference between the synchrotron emission maps which arise from better estimation of the SED due to the addition of more frequency channels. On the other hand, we recovered a CMB with less power when we use the filtered K, Ka and PR4  with FDEC. Since our analysis focuses on the characterization of foregrounds we keep the results obtained with the filtered maps. However, as commented in Section~\ref{subsec:cmb_comparison_Commander} an unbiased CMB map can be recovered following other approaches.

We have also performed different analyses to test the validity of our results. First, we found that our results are compatible with a $\chi^2$ distribution in those pixels where the power law model fits well the data. Furthermore, we have calculated the normalized residuals of the pixels with an acceptable goodness of fit of all frequency channels and they are all consistent within the $3\sigma$ level. Finally, we have evaluated the robustness of the estimated $\beta_s$ varying the prior imposed in this parameter. We found that the estimations in the high signal-to-noise synchrotron areas are prior-independent, while outside these regions the prior governs the $\beta_s$ estimation.

\section*{acknowledgments}

We thank the staff of the Teide Observatory for invaluable assistance in the commissioning and operation of QUIJOTE.
The {\it QUIJOTE} experiment is being developed by the Instituto de Astrofisica de Canarias (IAC),
the Instituto de Fisica de Cantabria (IFCA), and the Universities of Cantabria, Manchester and Cambridge.
Partial financial support was provided by the Spanish Ministry of Science and Innovation 
under the projects AYA2007-68058-C03-01, AYA2007-68058-C03-02,
AYA2010-21766-C03-01, AYA2010-21766-C03-02, AYA2014-60438-P,
ESP2015-70646-C2-1-R, AYA2017-84185-P, ESP2017-83921-C2-1-R,
AYA2017-90675-REDC (co-funded with EU FEDER funds),
PGC2018-101814-B-I00, 
PID2019-110610RB-C21, PID2020-120514GB-I00, IACA13-3E-2336, IACA15-BE-3707, EQC2018-004918-P, the Severo Ochoa Programs SEV-2015-0548 and CEX2019-000920-S, the
Maria de Maeztu Program MDM-2017-0765, and by the Consolider-Ingenio project CSD2010-00064 (EPI: Exploring
the Physics of Inflation). We acknowledge support from the ACIISI, Consejeria de Economia, Conocimiento y 
Empleo del Gobierno de Canarias and the European Regional Development Fund (ERDF) under grant with reference ProID2020010108.
This project has received funding from the European Union's Horizon 2020 research and innovation program under
grant agreement number 687312 (RADIOFOREGROUNDS).
EdlH acknowledges financial support from the \textit{Concepci\'on Arenal Programme} of the Universidad de Cantabria. 
DT acknowledges the support from the Chinese Academy of Sciences (CAS) President's International Fellowship Initiative (PIFI) with Grant N. 2020PM0042. 
FP acknowledges support from the Spanish State Research Agency (AEI) under grant number PID2019-105552RB-C43. 
The authors acknowledge the computer resources, technical expertise and assistance provided by the Spanish Supercomputing Network (RES) node at Universidad de Cantabria. 
Some of the presented results are based on observations obtained with \textit{Planck} (http://www.esa.int/Planck), an ESA science mission with instruments and contributions directly funded by ESA Member States, NASA, and Canada.
We acknowledge the use of the Legacy Archive for Microwave Background Data Analysis (LAMBDA) and the Planck Legacy Archive (PLA). Support for LAMBDA is provided by the NASA Office of Space Science. 
Some of the results in this paper have been derived using the HEALPix package \citep{HEALPix}, and the \textsc{healpy} \citep{healpy}, \textsc{numpy} \citep{numpy},  \textsc{emcee} \citep{foreman2013emcee}, and \textsc{matplotlib} \citep{matplotlib} \textsc{Python} packages.

\section*{Data Availability}
\label{sec:data_availability}
The parameter maps obtained from the component separation analysis in the default case, i.e., with the MFI+K/Ka+PR4 dataset using a power law to model the synchrotron emission, are included in the released data products associated to the QUIJOTE-MFI wide survey.

These data products as well as the maps can be freely downloaded from the QUIJOTE web page\footnote{\url{http://research.iac.es/proyecto/cmb/quijote}.}, as well as from the RADIOFOREGROUNDS platform\footnote{\url{http://www.radioforegrounds.eu/}.}. They include also an Explanatory Supplement describing the data formats. 
Any other derived data products described in this paper are available upon request to the QUIJOTE collaboration.

\bibliographystyle{mnras}
\bibliography{bibliography,quijote}

\appendix
\section{Independent \textit{Q} and \textit{U}  Synchrotron spectral index}
\label{sec:appendix_independent_Q_U}
\begin{figure*}
    \centering
    \includegraphics[width=1\textwidth,trim={2cm 1.25cm 2cm 1cm},clip]{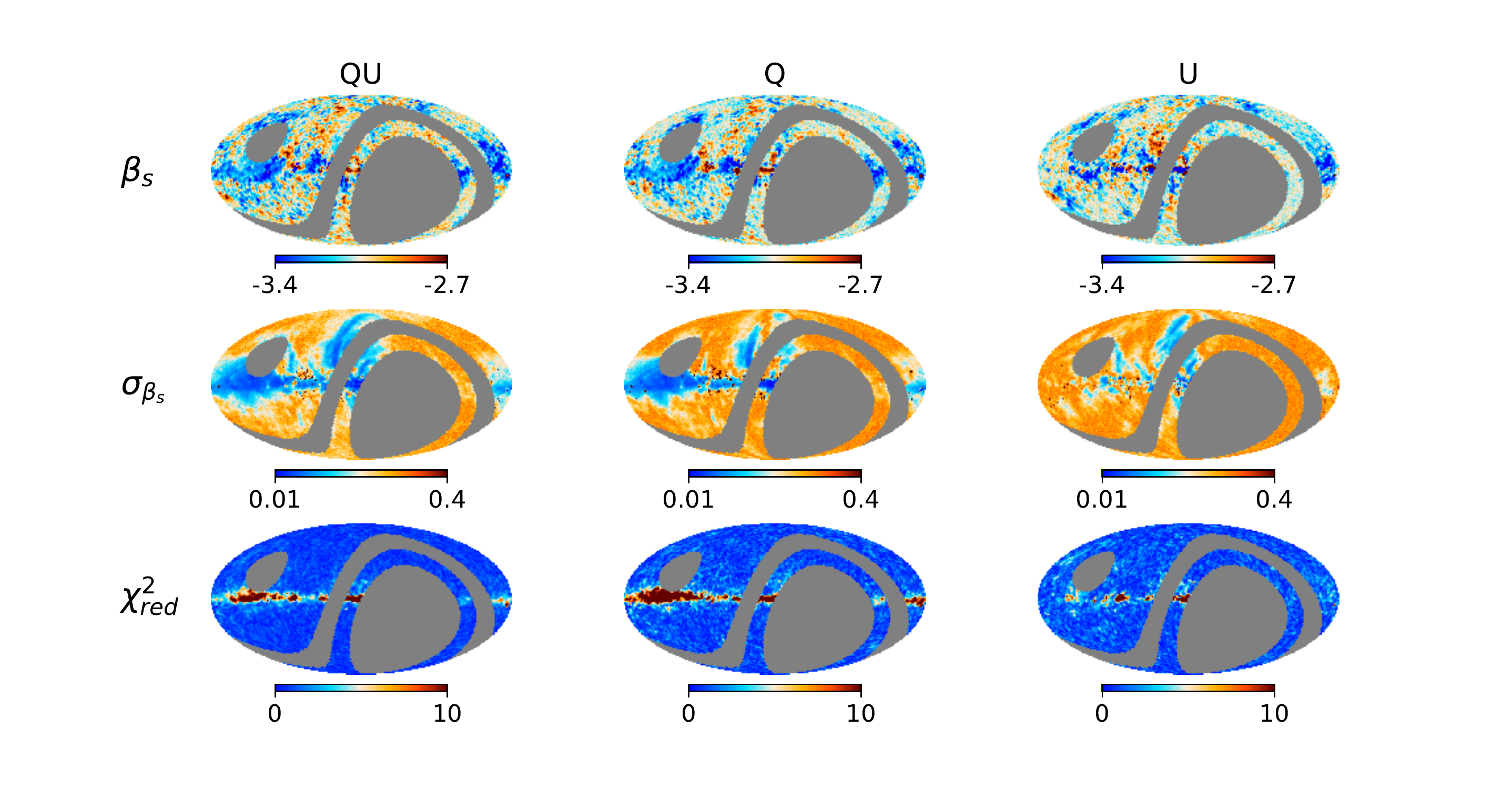}
    \caption{Synchrotron spectral index estimate (top row) and uncertainty maps (second row) obtained after component separation using the MFI+K/Ka+PR4 dataset. The left column shows the $\beta_s$ recovered when we assume that $Q$ and $U$ share the same spectral index, while the centre and right columns depict the $Q$ and $U$  $\beta_s$ when they are assumed to be independent. Bottom row: reduced $\chi^2$ map for each case study considered. The synchrotron emission is modelled as a power law.}
    \label{fig:beta_s_QKNPIPE_QU_Q_U}
\end{figure*}

\begin{figure}
    \centering
    \includegraphics[width=.41\textwidth,trim={2cm 1.35cm 1.5cm 1cm},clip]{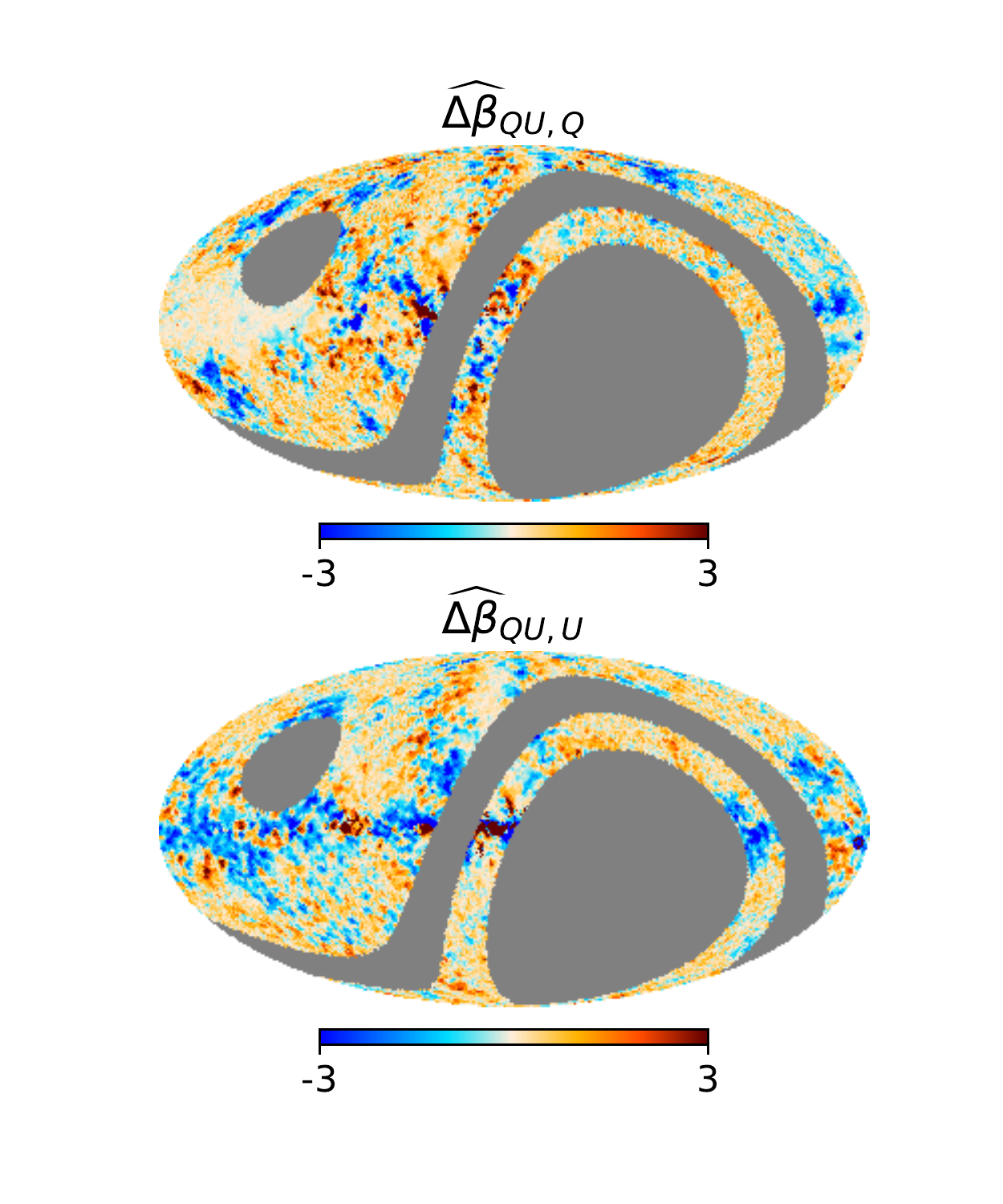}
    \caption{Relative difference map between the $\beta_s$ map obtained when we assume the same $\beta_s$ in both $Q$ and $U$, and $\beta_s$ recovered from the fit with just $Q$ (top) and just $U$ (bottom). The synchrotron emission is modelled as a power law.}
    \label{fig:beta_s_QU_vs_Q_U}
\end{figure}

In order to test the assumption of having the same $\beta_s$ in both $Q$ and $U$, we fit $Q$ and $U$ signals independently. Fig.~\ref{fig:beta_s_QKNPIPE_QU_Q_U} shows the spectral index, the uncertainty of the spectral index as well as the reduced $\chi^2$ maps obtained from the three independent fits using the MFI+K/Ka+PR4 dataset. We infer from the $\chi^2_{\rm red}$ maps that the fit outside the Galactic plane is better when $Q$ and $U$ are fitted together. When we fit just $U$ we observe that the goodness of fit improves significantly in the Galactic plane. However, this effect is due to the low signal-to-noise ratio in that area, not due to a better modelling of the signal. 

The $\beta_s^{Q}$ and $\beta_s^{U}$ maps are distinctly different. The $\beta_s^{QU}$ map resembles more the $\beta_s^{Q}$ map. This is expected, since $Q$ has more signal than $U$ in Galactic coordinates. That is also the reason why the uncertainty on the recovered $\beta_s$ is smaller when we fit just $Q$ compared to $U$. However, in those regions where $\sigma_{\beta_s^U}$ is smaller than $\sigma_{\beta_s^Q}$, i.e., regions where $U$ has more signal than $Q$, the $\beta_s^{QU}$ values obtained are closer to those of $\beta_s^{U}$. This is clearly seen in Fig.~\ref{fig:beta_s_QU_vs_Q_U} where the relative difference between $\beta_s^{QU}$ with respect to $\beta_s^{Q}$ (top row) and $\beta_s^U$ (bottom row) is shown. 
The largest differences shown in the top (bottom) panel are located in regions where the signal-to-noise is larger in $U$ ($Q$). On the other hand the relative difference decreases significantly in the regions where the uncertainty on $\beta_s^Q$ (top) or $\beta_s^U$ (bottom) is smaller.

\section{Faraday Rotation}
\label{sec:appendix_FR}
\begin{figure}
    \centering
    \includegraphics[width=.41\textwidth,trim={2cm 1.35cm 1.5cm 1cm},clip]{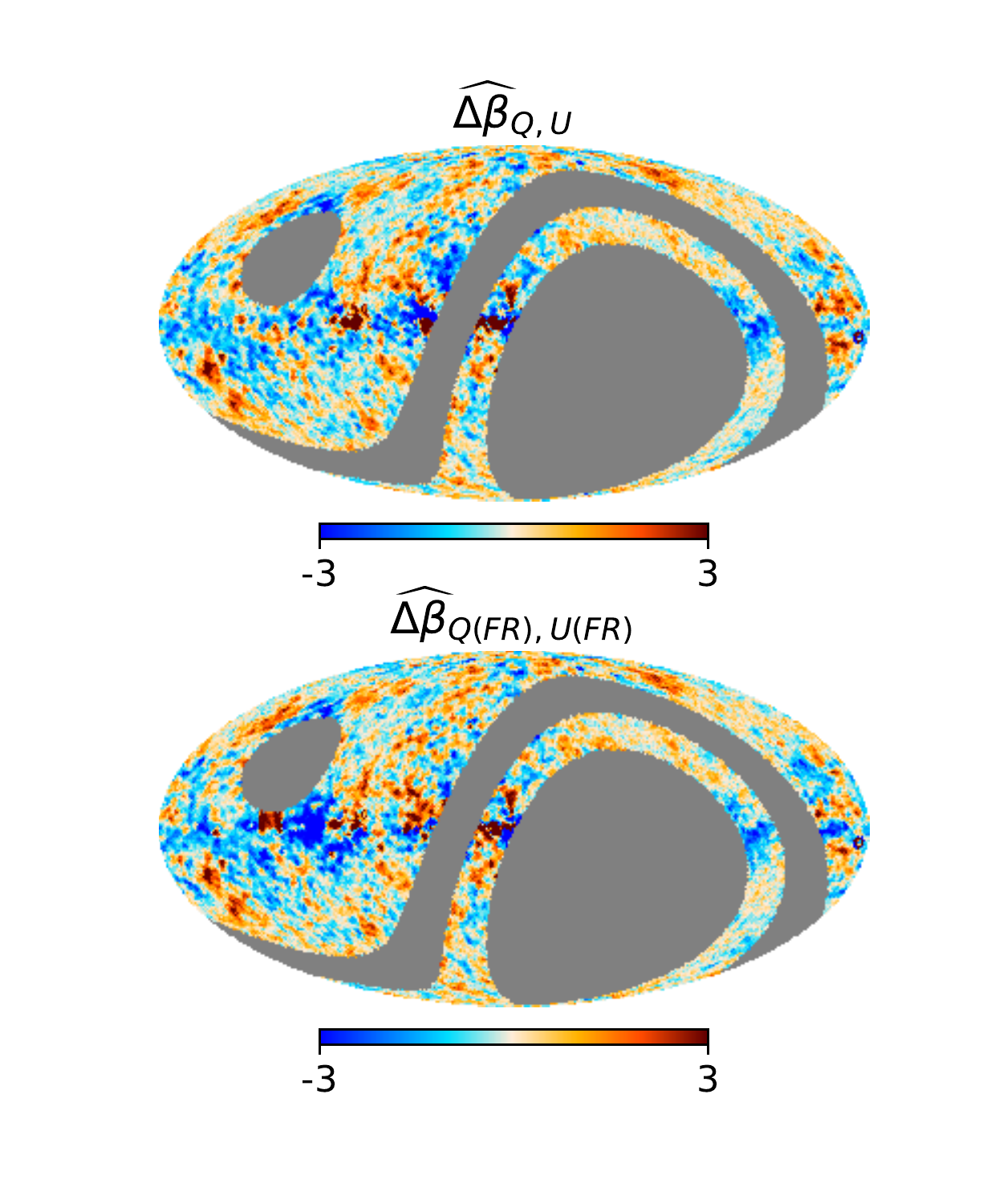}
    \caption{Relative difference map between the $\beta_s$ map from the independent $Q$ and $U$ fit using the MFI+K/Ka+PR4 dataset (top), and using the MFI(FR)+K/Ka+PR4 dataset (bottom). The synchrotron emission is modelled as a power law.}
    \label{fig:beta_s_Q_vs_U_FR}
\end{figure}

\begin{figure}
    \centering
    \includegraphics[width=.43\textwidth,trim={1.6cm 0.5cm 1.2cm .2cm},clip]{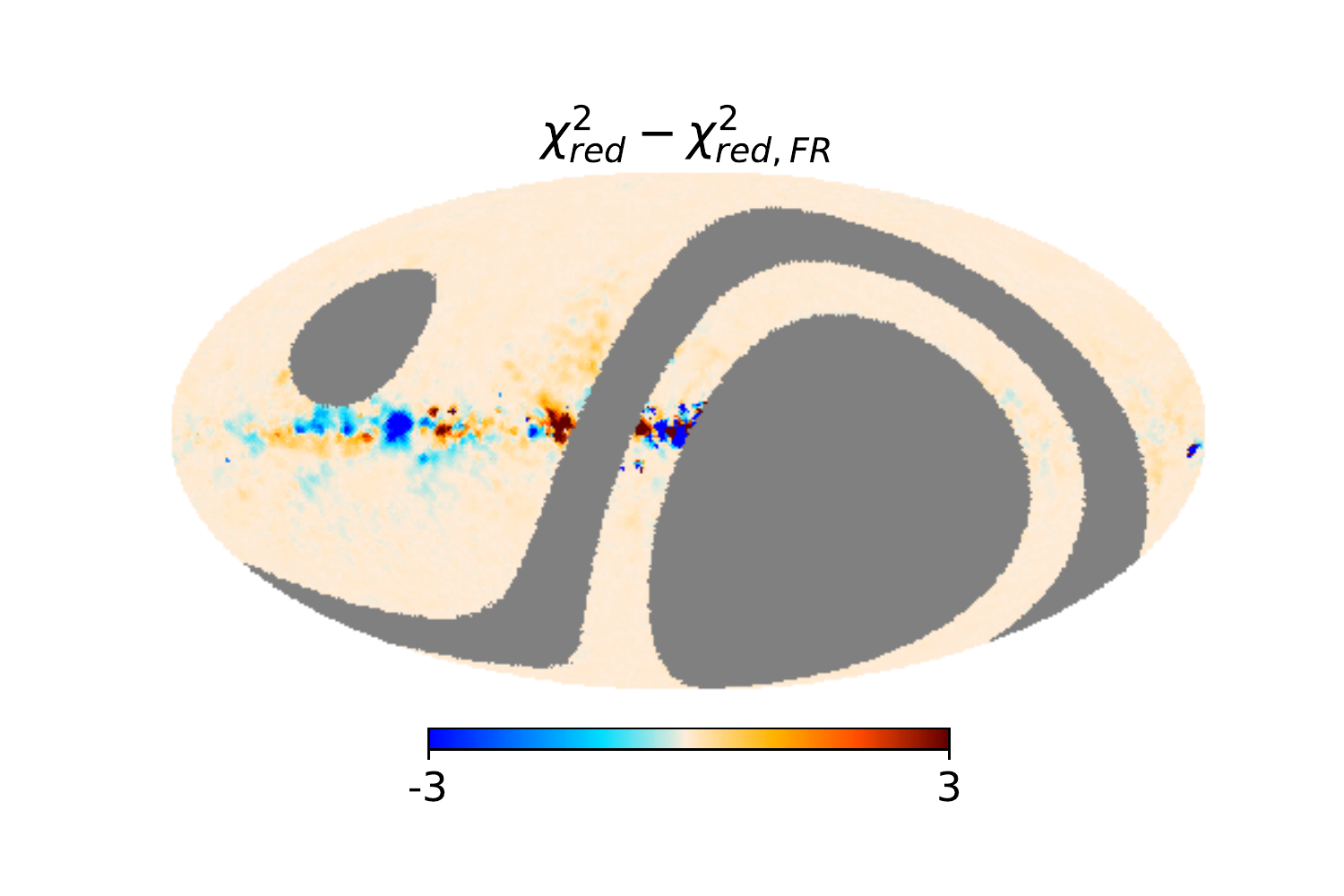}
    \caption{Difference map between the $\chi^2_{\rm red}$ obtained with the MFI+K/Ka+PR4 dataset with respect to the $\chi^2_{\rm red,FR}$ obtained with MFI(FR)+K/Ka+PR4 dataset. In both fits we have assumed that $Q$ and $U$ share the same spectral indices. The synchrotron emission is modelled as a power law.}
    \label{fig:diff_chi2_FR}
\end{figure}

We have also studied the significance of the difference between the $\beta_s^Q$ and $\beta_s^U$ maps, see top row of Fig.~\ref{fig:beta_s_Q_vs_U_FR}. The  discrepancies larger than 3$\sigma$ are concentrated in the Galactic plane, close to the Galactic centre. This could be a tracer of  Faraday rotation. If Faraday rotation  is non-negligible at QUIJOTE frequencies there will be a difference between the polarization angles at QUIJOTE frequencies and those at WMAP/\textit{Planck} frequencies. This yields a $\beta_s^Q$ map different from $\beta_s^U$ due to the bias introduced by the change in angle. That bias is reasonably cancelled out when combining both $Q$ and $U$ to obtain a single index. 

We have studied the possibility of correcting the Faraday rotation effect in the QUIJOTE MFI maps using the model from \citet{Faraday_Rotation}. The rotation of the polarization plane experienced due to the Faraday Rotation effect can be described by:
\begin{equation}
    \Delta\phi = \textrm{RM}\lambda^2 \, ,
    \label{eq:angle_faraday_rotation}
\end{equation}
where $\lambda$ is the wavelength, and RM is the rotation measure. We use the RM map estimated by \citet{Faraday_Rotation} to calculate the rotation angle maps at 11 and 13\,GHz QUIJOTE frequencies. Then, QUIJOTE $Q$ and $U$ maps at a given frequency $\nu$ are de-rotated as follows:
\begin{equation}
    \begin{pmatrix}
        Q_{\rm FR} \\
        U_{\rm FR}
    \end{pmatrix}_{\nu}
    =
    \begin{pmatrix}
        \cos(2\Delta\phi_{\nu}) & -\sin(2\Delta\phi_{\nu}) \\
        \sin(2\Delta\phi_{\nu}) & \cos(2\Delta\phi_{\nu})
    \end{pmatrix}_{\nu}
    \begin{pmatrix}
        Q \\
        U
    \end{pmatrix}_{\nu} \, ,
    \label{eq:faraday_derotation}
\end{equation}
The variance of the de-rotated $Q_{\rm FR}$ and $U_{\rm FR}$ is:
\begin{align}
    \sigma_{Q_{\rm FR}}^2  = & \cos^2(2\Delta\phi_{\nu})\sigma_Q^2 + \sin^2(2\Delta\phi_{\nu})\sigma_{U}^2 \\
    & +
    4\left[\sin(2\Delta\phi_{\nu})Q+\cos(2\Delta\phi_{\nu})U\right]^2\sigma_{\Delta\phi}^2 \nonumber \\
    \sigma_{U_{\rm FR}}^2  = & \sin^2(2\Delta\phi_{\nu})\sigma_Q^2 + \cos^2(2\Delta\phi_{\nu})\sigma_{U}^2 \\
    & +
    4\left[\cos(2\Delta\phi_{\nu})Q-\sin(2\Delta\phi_{\nu})U\right]^2\sigma_{\Delta\phi}^2\nonumber
\end{align}

Therefore, we have repeated the same analysis but using the MFI(FR)+K/Ka+PR4 dataset, where MFI(FR) indicates that the QUIJOTE 11 and 13\,GHz maps have been de-rotated using the angle obtained from the \citet{Faraday_Rotation} model, to correct any possible mismatch due to the Faraday Rotation effect, see bottom row of Fig.~\ref{fig:beta_s_Q_vs_U_FR}. 

We compare these maps (Fig.~\ref{fig:beta_s_Q_vs_U_FR}) with the difference map between the reduced $\chi^2$ map ($\chi^2_{\rm{red}}$) obtained with the MFI+K/Ka+PR4 dataset with respect to the reduced $\chi^2$ ($\chi^2_{{\rm{red}},{\rm FR}}$) obtained with MFI(FR)+K/Ka+PR4 dataset shown in Fig.~\ref{fig:diff_chi2_FR}. The sky regions where the absolute value of the relative difference $\widehat{\Delta \beta}_{Q({\rm FR}),U({\rm FR})}$ is smaller than $\widehat{\Delta \beta}_{Q,U}$ are correlated to those regions where the $\chi^2_{{\rm{red}},\rm{FR}}$ is smaller than $\chi^2_{\rm red}$ (reddish regions) and vice versa (bluish regions). This result suggests that Faraday rotation might be playing a role in some of the significant differences areas observed between $\beta_s^Q$ and $\beta_s^U$.

\section{Function-of-declination Correction Simulations}
\label{sec:appendix_beta_s_simus}
\begin{figure}
    \centering
    \includegraphics[width=.43\textwidth,trim={1.6cm 0.5cm 1.2cm .2cm},clip]{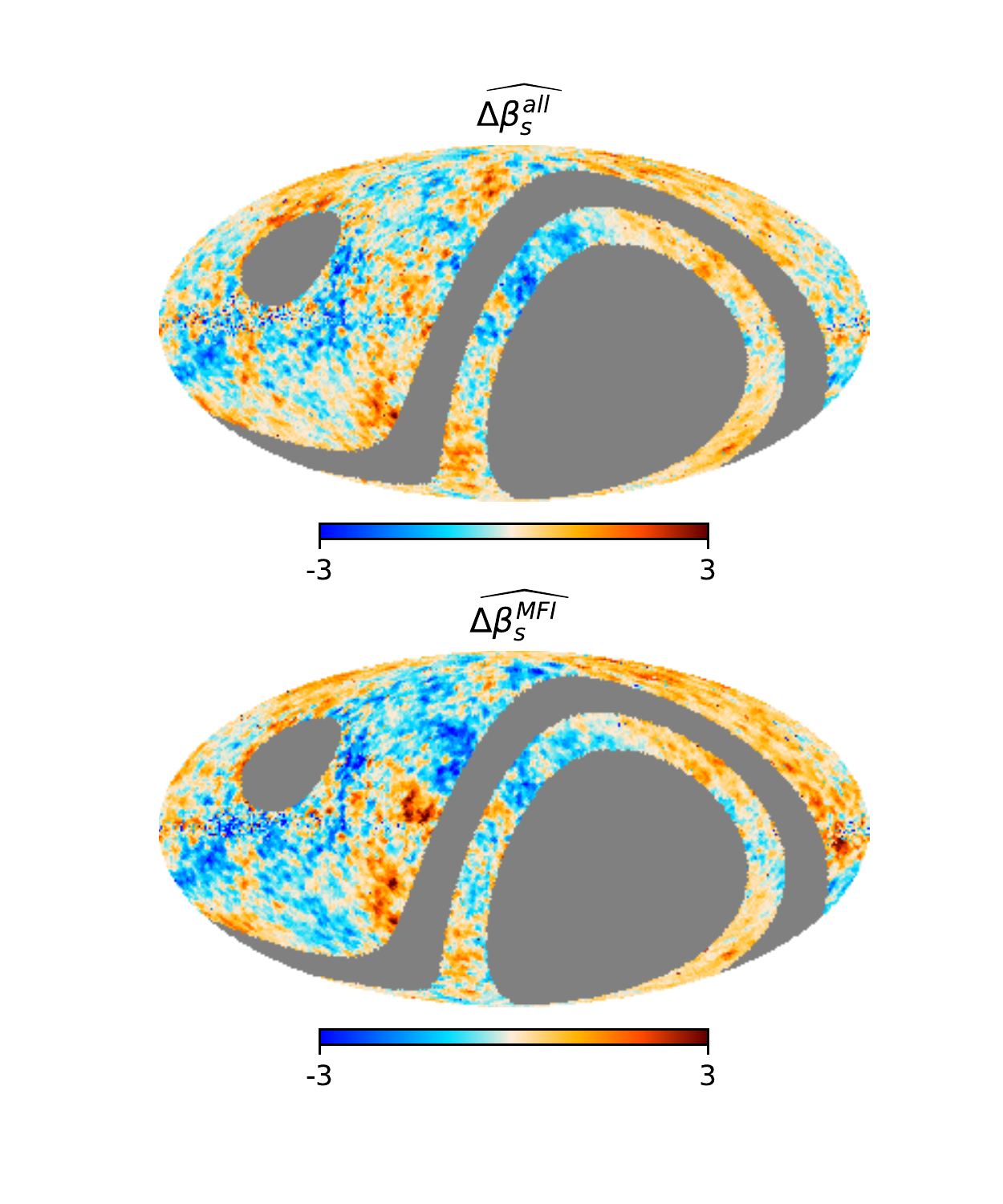}
    \caption{Relative difference map between the $\beta_s$ template used in the simulation and the $\beta_s$ map from the fit using the simulated data with an FDEC filter applied to all maps (top), and an FDEC filter applied only to QUIJOTE-MFI frequencies (bottom).}
    \label{fig:beta_s_simus}
\end{figure}

We studied using simulations if the application of a function-of-declination (FDEC) filter to QUIJOTE-MFI maps biases the $\beta_s$ map obtained from component separation. We generated sky simulation maps with the following components at the QUIJOTE-MFI 11 and 13\,GHz, K and Ka, and PR4 frequencies:
\begin{itemize}
    \item CMB. Generated as Gaussian random samples using the power spectra obtained from CAMB \citep{CAMB} with the latest \textit{Planck} cosmological parameters \citep{Planck2018_VI}.
    \item Synchrotron. Generated using the s1 model of the Python Sky Model (PySM) \citep{pysm}.
    \item Thermal dust. Generated using the d1 model of the PySM.
    \item Realistic noise simulations. For each experiment we use the ones described in Section~\ref{sec:data}.
\end{itemize}
All components are either generated or downgraded to $N_{\rm side}=512$. Then the components maps are added and we apply the corresponding FDEC filter to each signal map. Finally all maps are downgraded to $N_{\rm side}=64$ and smoothed with a Gaussian beam of $\mathrm{FWHM}=2$ deg following the procedure described in Section~\ref{sec:data}. 

We perform the component separation analysis in two scenarios: i) when only the QUIJOTE-MFI frequency signal maps are filtered, and ii) when all maps are filtered. Fig.~\ref{fig:beta_s_simus} shows the relative difference (equation \ref{eq:relative_difference}) between the $\beta_s$ map recovered from the component separation analysis and the $\beta_s$ template (equation \ref{eq:relative_difference} taking into account that the uncertainty of the template map is set to zero, $\sigma_{\beta_{s}} = 0$). We find that when only QUIJOTE-MFI channels are filtered (bottom panel) the relative differences are larger in regions such as the North Polar Spur or the R3 region than when all maps are filtered. Moreover, in those regions the $\beta_s$ relative differences are larger than 3$\sigma$ with respect to the template 
In the case when all maps are filtered (top panel), these biases are reduced significantly.

\vspace{1cm}
\noindent \textit{$^{1}$Instituto de F\'isica de Cantabria (IFCA), CSIC-Univ. de Cantabria, Avda. los Castros s/n, E-39005 Santander, Spain.\\
$^{2}$Dpto. de F\'isica Moderna, Universidad de Cantabria, Avda. de los Castros s/n, E-39005 Santander, Spain. \\
$^{3}$Instituto de Astrof\'{\i}sica de Canarias, E-38205 La Laguna, Tenerife, Spain.\\
$^{4}$Departamento de Astrof\'{\i}sica, Universidad de La Laguna,
E-38206 La Laguna, Tenerife, Spain.\\
$^{5}$Institut d'Astrophysique de Paris, UMR 7095, CNRS \& Sorbonne Universit\'e, 98 bis boulevard Arago, 75014 Paris, France.\\
$^{6}$Astrophysics Group, Cavendish Laboratory, University of Cambridge, J J Thomson Avenue, Cambridge CB3 0HE, U.K.\\
$^{7}$Kavli Institute for Cosmology, University of Cambridge, Madingley Road, Cambridge CB3 0HA, U.K.\\
$^{8}$Dpto. de Ingenieria de COMunicaciones (DICOM), Edif. Ingenieria de Telecomunicacion, Pl. de la Ciencia s/n, E-39005 Santander, Spain.\\
$^{9}$Dpto. de Matem\'aticas, estad\'istica y computaci\'on, Univ. de Cantabria, Avda. de los Castros s/n, E-39005 Santander, Spain. \\
$^{10}$Aurora Technology for the European Space Agency (ESA), European Space Astronomy Centre (ESAC), Camino Bajo del Castillo \\
s/n, 28692 Villanueva de la Cañada, Madrid, Spain.\\
$^{11}$Universidad Europea de Madrid, 28670, Madrid, Spain.\\
$^{12}$Jodrell Bank Centre for Astrophysics, Alan Turing Building, Department of Physics and Astronomy, School of Natural Sciences, \\
University of Manchester, Oxford Road, Manchester M13 9PL, U.K.\\
$^{13}$Consejo Superior de Investigaciones Cientificas, E-28006 Madrid, Spain\\
$^{14}$Dpto. de F\'{\i}sica. Facultad de Ciencias. Univ. de C\'ordoba.  Campus de Rabanales, 
Edif. C2. Planta Baja.  E-14071 C\'ordoba, Spain.\\
$^{15}$Purple Mountain Observatory, CAS, No.10 Yuanhua Road, Qixia District, Nanjing 210034, China. \\
$^{16}$NAOC-UKZN Computational Astrophysics Center (NUCAC), University of Kwazulu-Natal, Durban 4000, South Africa.}

\end{document}